\documentclass[prb,aps,preprint,floatfix]{revtex4}
\usepackage[T1]{fontenc}
\usepackage{lmodern}
\usepackage{graphicx}
\usepackage{wrapfig}
\usepackage{amsmath}
\usepackage{color}
\usepackage{amssymb}
\usepackage[T1]{fontenc}
\usepackage[ansinew]{inputenc}
\usepackage[english]{babel}
\usepackage[left=2.5cm,right=2.5cm,top=2cm,bottom=3.0cm,a4paper]{geometry}
\usepackage{tabularx}
\usepackage{color}
\usepackage{dcolumn}
\usepackage{bm}

\newcommand{\Fs}{{FeSCs}}

\def\k{{\bf k}}
\def\p{{\bf p}}
\newcommand{\be}{\begin{equation}}
\newcommand{\ee}{\end{equation}}
\newcommand{\bea}{\begin{eqnarray}}
\newcommand{\eea}{\end{eqnarray}}
\newcommand{\beq}{\begin{equation}}
\newcommand{\eeq}{\end{equation}}

\newcommand{\e}{{\varepsilon}}
\def\k{{\bf k}}

\newcommand{\KFA}{KFe$_2$As$_2$}

\newcommand{\AFS}{A$_x$Fe$_{2-y}$Se$_2$~}

\begin{document}

\title{Itinerant electron scenario for Fe-based  superconductors}

\author{Andrey Chubukov}

\affiliation {Department of Physics, University of Wisconsin,
Madison, Wisconsin 53706, USA}
\date{\today}

\date{December 15, 2013}

\begin{abstract}
I review  works on Fe-based superconductors which depart from a metal with well defined Fermi surfaces and Fermi liquid-type quasiparticles.
  I  consider normal state instabilities -- SDW magnetism and nematic order, and superconductivity, all three as  the consequences of the instability of a Fermi surface due to interactions between low-energy fermionic quasiparticles. This approach assumes that renormalizations coming from fermions from high energies, of order bandwidth, modify but do not destroy  Fermi liquid behavior
  in the normal state and can be absorbed into the effective low-energy model of interacting fermions located near hole and electron-type Fermi surfaces. I argue that the interactions between these fermions are responsible for (i) a stripe-type SDW magnetic order (and, in some special cases, a checkerboard order ) , (ii) a pre-emptive nematic-type instability, in which magnetic fluctuations break $C_4$ lattice rotational symmetry down to $C_2$, but magnetic order does not yet develop, and (iii) a superconductivity, which competes with these two orders.
   The experimental data on superconductivity show very rich behavior 
  with potentially different symmetry of a superconducting state even for different compositions of the same material. I argue that, despite all this,
 the physics of superconductivity in the itinerant scenario for Fe-based materials is governed by a single underlying pairing mechanism.
 \end{abstract}

\maketitle

\section{Introduction}
\label{sec:1}
The discovery  of superconductivity in $Fe$-based pnictides [\onlinecite{bib:Hosono}] (Fe-based compounds with elements from the 5th group: N, P, As, Sb, Bi)
 was, arguably, among the most significant breakthroughs in condensed matter physics during the past decade. A lot of efforts by the condensed-matter community have
been devoted in the few years  after the discovery
to  understand  normal state properties
 of these materials, the pairing mechanism, and the symmetry and the structure
 of the pairing gap.

The family of $Fe$-based superconductors (FeSCs)  is already quite large and
keeps growing. It includes various Fe-pnictides such as $1111$ systems RFeAsO
($R=$rare earth element)
[\onlinecite{bib:Hosono,bib:X Chen,bib:G Chen,bib:ZA Ren}], $122$ systems XFe$_2$As$_2$(X=alkaline earth
metals) [\onlinecite{bib:Rotter,bib:Sasmal,bib:Ni}],  111
systems like LiFeAs [\onlinecite{bib:Wang}], and
 also Fe-chalcogenides (Fe-based compounds with elements from the 16th group:
 S, Se, Te) such as FeTe$_{1-x}$Se$_x$
~~[\onlinecite{bib:Mizuguchi}] and A$_x$Fe$_{2-y}$Se$_2$
($A = K, Rb, Cs$)~~[\onlinecite{exp:AFESE,exp:AFESE_ARPES}].

 Superconductivity (SC) in FeSCs  emerges upon either hole or electron doping  (see Fig. 1), but can also be induced  by pressure or by
isovalent replacement of one pnictide element by another, e.g., As by
 P (Ref.~[\onlinecite{nakai}]). In some systems, like LiFeAs~~[\onlinecite{bib:Wang}], LiFeP [\onlinecite{bib_mats_1}] and
LaFePO~~[\onlinecite{bib:Kamihara}], SC emerges already at zero doping, instead of a magnetic order.

Parent compounds of nearly all FeSCs are  metals, in distinction to cuprate superconductors for which all parent compounds are Mott
 insulators. Still, in similarity with the cuprates, in most cases these parent compounds are antiferromagnetically ordered~~[\onlinecite{Cruz}].
 Some researchers~[\onlinecite{localized,loc_sel,loc_it}] used this analogy to argue that FeSCs are at short distance from Mott transition, and at least some elements of Mott physics
  must be included into the description of these systems. A rather similar point of view is~[\onlinecite{loc_it}] that fermionic excitations in FeSCs display both localized and itinerant properties and the interplay between the two depends  on the type of the orbital (one set of ideas of this kind
  lead to the notion of "orbital selective Mott transition on FeSCs~[\onlinecite{loc_sel,loc_it}]).  An alternative point of view, which I will present in this review, is that low-energy properties of most of FeSCs can be fully captured in a itinerant approach, without invoking Mott physics.

   In itinerant approach, electrons, which carry magnetic moments, travel relatively freely from site to site. The magnetic order of such electrons
    is often termed as a spin-density-wave (SDW), by analogy with e.g., antiferromagnetic  $Cr$, rather than "Heisenberg antiferromagnetism" -- the latter term is reserved for systems in which electrons are "nailed down" to particular lattice sites by very strong Coulomb repulsion.  From experimental perspective, the majority of FeSCs display a rather small ordered moment in the normal state,  consistent with SDW scenario~[\onlinecite{review}].  There are notable exceptions -- Fe-chalcogenide FeTe 
    (the parent compound of  FeTe$_{1-x}$Se$_x$, which superconduct at $x$ around $0.5$) displays magnetic properties consistent with the  Heisenberg antiferromagnetism
     of localized spins [\onlinecite{Zaliznyak12}]. However, the properties of this material vary quite substantially between $x=0$ and $x=0.5$, and magnetic fluctuations
       at $x \sim 0.5$ are similar to those of other FeSCs.  Another example where magnetism is strong and probably involves localized carriers is  A$_x$Fe$_{2-y}$Se$_2$ (Ref. [\onlinecite{exp:AFESE}]).
     However, in this material, localized carriers and itinerant carriers are most likely phase separated,
        with superconductivity coming primarily from itinerant carriers.

\begin{figure}[tbp]
\includegraphics[angle=0,width=\linewidth]{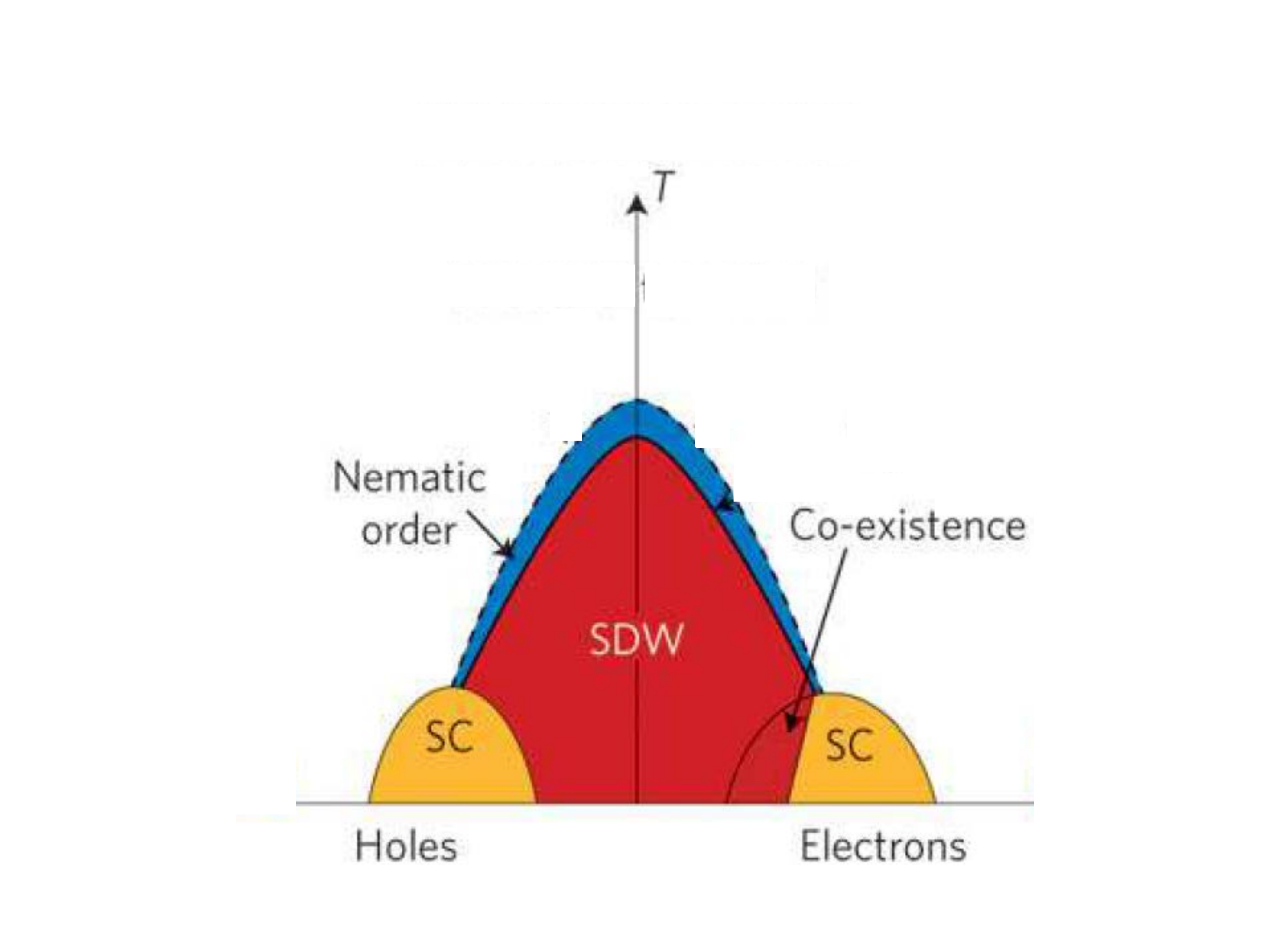}
\caption{Schematic phase diagram of Fe-based pnictides upon hole or electron doping.  In the shaded region, superconductivity and  antiferromagnetism co-exist. Not all details/phases are shown.
Superconductivity can be initiated not only by doping but also by pressure and/or isovalent replacement of one pnictide element by another~~[\onlinecite{nakai}].
 Nematic phase at $T > T_N$ is subject of debates.
 Superconductors at large doping are  KFe$_2$As$_2$ for hole doping~
~[\onlinecite{KFeAs_ARPES_QO,KFeAs_exp_nodal}] and A$_x$Fe$_{2-y}$Se$_2$ (A = K, Rb, Cs) for electron doping~~[\onlinecite{exp:AFESE,exp:AFESE_ARPES}].
 Whether superconductivity in pnictides exists at all
 intermediate dopings is not clear yet.  From Ref.~[\onlinecite{review_we}].}
\label{fig1}
\end{figure}

The itinerant approach to magnetism and superconductivity in FeSCs and the comparative analysis of Fe- and Cu-based superconductors
 have been reviewed in several recent publications~~[\onlinecite{review,review_2,review_3,review_4,Graser,peter,Kuroki_2,rev_physica,mazin_schmalian,review_we,korshunov,ch_annual,maiti_rev}]. This review is an attempt to summarize our current understanding of the phase diagram, the origin of SDW and nematic orders, the pairing mechanism for superconductivity, and the symmetry and the structure of the
 pairing gap at various hole and electron dopings.

\begin{figure}
\includegraphics[angle=0,width=0.8\columnwidth]{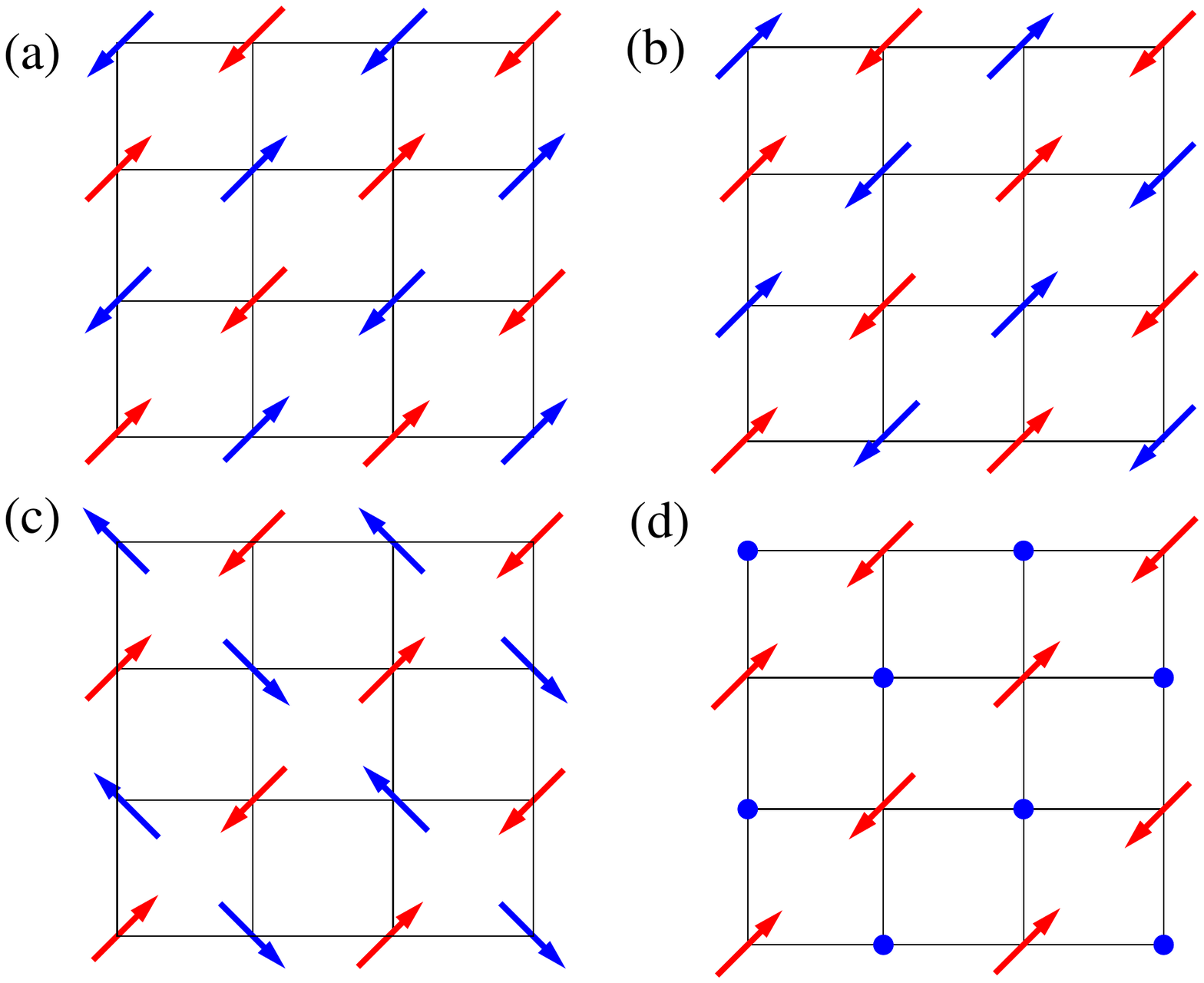}
\caption{(color online) Various SDW spin configurations
  described by $\vec{\Delta}_1 e^{i{\bf Q}_1 R}+\vec{\Delta}_2 e^{i{\bf Q}_2 R}$.
For a perfect nesting only  $\vec{\Delta}^2_1 +
\vec{\Delta}^2_2$ is fixed. Panel (a) --  $\vec{\Delta}_1 = 0$,
panel (b) --  $\vec{\Delta}_2 = 0$, panel (c) --   $\vec{\Delta}_1 \perp \vec{\Delta}_2$, and panel (d) -- $\vec{\Delta}_1 = \vec{\Delta}_2$.
From [\onlinecite{Eremin_10}]}. \label{fig2_sdw}
\end{figure}

Like I said, the very idea of itinerant approach is that magnetism and superconductivity  come from the interactions
  between fermionic states located very near the Fermi surfaces. These interactions  originate from a Coulomb interaction, which is obviously a repulsive one.

 A repulsive  interaction between itinerant carriers is well known to lead to Stoner-type magnetic instability,
   and the presence of the SDW-ordered phase on the phase diagram of FeSCs should not come as a surprise. Less obvious issue is what kind of magnetism is present in FeSCs.
   Experiments show that most of undoped and weakly doped Fe-pnictides display the stripe spin-density wave order at $T<T_{sdw}$, with ordering vectors
$(0,\pi)$ or $(\pi,0)$ in the 1-Fe Brillouin zone (1FeBZ), Ref.[\onlinecite{Yildirim08,Xiang08}] (see Fig. \ref{fig2_sdw}).
 Such an order not only breaks $O(3)$ spin symmetry, but also breaks lattice rotational symmetry from $C_4$ down to $C_2$ (the stripes run either along $X$ or along $Y$ direction).  Stripe, order, however, does not emerge in all cases. Neutron scattering data on more heavily doped
  Ba$_{1-x}$Na$_x$Fe$_2$As$_2$ (Ref. [~\onlinecite{osborn}]) and on Ba(Fe$_{1-x}$Mn$_x$)$_2$As$_2$ ~(Ref. [\onlinecite{rafael_last}])
 show that the magnetic order there does not break $C_4$ symmetry (examples are shown in  Fig. \ref{fig2_sdw}).
  I will argue that both types of magnetic order (the one which breaks $C_4$ symmetry and the one which doesn't)
   emerge in the itinerant scenario for FeSCs.

Another interesting aspect of the normal state phase diagram is that  in weakly doped Fe-pnictides, the stripe SDW order
is often preceded by a {}``nematic'' phase with broken $C_{4}$
tetragonal symmetry but unbroken $O(3)$ spin rotational symmetry. The
emergence of such a phase is not only manifested by a tetragonal to
orthorhombic transition at $T_{n}\geq T_{sdw}$, but also by the onset
of significant anisotropies in several quantities [\onlinecite{Fisher11}],
such as dc resistivity [\onlinecite{Chu10,Tanatar10}], optical conductivity
[\onlinecite{Duzsa11,Uchida11}], local density of states [\onlinecite{Davis10}],
orbital occupancy [\onlinecite{Shen11}],  susceptibility [\onlinecite{Matsuda11}],
and the vortex core in the mixed superconducting state [\onlinecite{Song11}].
The fact that the SDW and structural transition lines
follow each other across all the phase diagrams of 1111 and 122 materials,
 even inside the superconducting dome [\onlinecite{FernandesPRB10,Nandi09}],
prompted researchers to propose that SDW and nematic orders are intimately connected.
The interplay between magnetic and structural transitions in FeSCs
is also quite rich: while in 1111 materials the two transitions are
second-order and split ($T_{n}>T_{cdw}$), in most of the 122 materials
they seem to occur simultaneously or near-simultaneously at small
dopings, but clearly split above some critical doping - $x\approx0.022$
in $\mathrm{Ba\left(Fe_{1-\mathit{x}}Co\mathit{_{x}}\right)_{2}As_{2}}$,
see [\onlinecite{Kim11,Birgeneau11}], and $x\approx0.039$ in $\mathrm{Ca\left(Fe_{1-\mathit{x}}Co\mathit{_{x}}\right)_{2}As_{2}}$,
see [\onlinecite{Prokes11}].

For superconductivity, the central issue is what  causes the attraction between fermions.
 The BCS theory of superconductivity attribute the attraction between fermions
 to the underlying interaction between electrons and phonons~~[\onlinecite{BCS}] (the two electrons effectively interact with each other  by emitting and absorbing the same phonon which then serves as a glue which binds electrons into pairs).
 Electron-phonon mechanism has been successfully applied to explain SC in a large variety of materials, from $Hg$ and $Al$ to recently discovered and extensively studied $MgB_2$ with the transition temperature $T_c =39K$~~[\onlinecite{mgb2}].  However, for FeSCs, early first-principle study of superconductivity due to electron-phonon interaction placed $T_c$ at around $1K$, much smaller that the actual $T_c$ in most of FeSCs.  This  leaves an electron-electron interaction
  as the more likely source of the pairing.

 Pairing due to electron-electron interaction has been discussed even before high $T_c$ era, most notably in connection with
  superfluidity in $^3He$~~[\onlinecite{3he,3he2}], but became the mainstream after the  discovery of SC in the
   cuprates~[\onlinecite{bednortz}].
 This discovery signaled the beginning of the new
era of ``high-temperature superconductivity'' to which FeSCs added a new avenue with  quite high traffic over the last five years.

A possibility to get superconductivity from nominally repulsive electron-electron interaction
 is based on two fundamental principles. First,  in isotropic systems the analysis of superconductivity factorizes~[\onlinecite{stat_phys}]
  between pairing channels with different
    angular momenta $l =0, 1,2,3$, etc [in spatially isotropic systems $l=0$ component is called $s-$wave, $l=1$ component is called $p-$wave, $l=2$ component is called $d-$wave, and so on]. If just one component with some $l$ is attractive, the system undergoes a SC transition at some temperature $T=T_c$.
      Second, the screened Coulomb interaction $U(r)$ is constant and repulsive at short distances, but oscillates at large distances and may develop an attractive component at some $l$.  Kohn and Luttinger (KL) have explicitly proven back in 1965 (Ref. ~[\onlinecite{KL}]) that the combination of these two effects
       necessary leads to a pairing instability, at least at large odd $l$,  no matter what the form of $U(r)$ is.

In lattice systems, angular momentum is no longer a good quantum number, and the  equation for $T_c$ only factorizes between different
 irreducible representations of the lattice space group. In tetragonal systems, which include both cuprates and FeSCs , there are four one-dimensional irreducible representations $A_{1g}$, $B_{1g}$, $B_{2g}$, and $A_{2g}$ and one two-dimensional representation $E_{2g}$.
Each representation has infinite set of eigenfunctions. The eigenfunctions from $A_{1g}$ are invariant under symmetry transformations in a tetragonal lattice: $x \to -x, ~y \to -y,~ x \to y$, the eigenfunctions from $B_{1g}$ change sign under $x \to y$, and so on.
If a superconducting gap has $A_{1g}$ symmetry, it is often called $s-$wave because the first eigenfunction from $A_{1g}$ group is just a constant in momentum space (a $\delta-$function in real space).  If the gap has $B_{1g}$ or $B_{2g}$ symmetry, it is called $d-$wave ($d_{x^2-y^2}$ or $d_{xy}$, respectably),
because in momentum space the leading eigenfunctions in $B_{1g}$ and $B_{2g}$ are
 $\cos k_x - \cos k_y$ and $\sin k_x \sin k_y$, respectively, and these two reduce to $l=2$ eigenfunctions $\cos 2 \theta$ and $\sin 2 \theta$ in the isotropic limit.

\begin{figure}[tbp]
\includegraphics[angle=0,width=\linewidth]{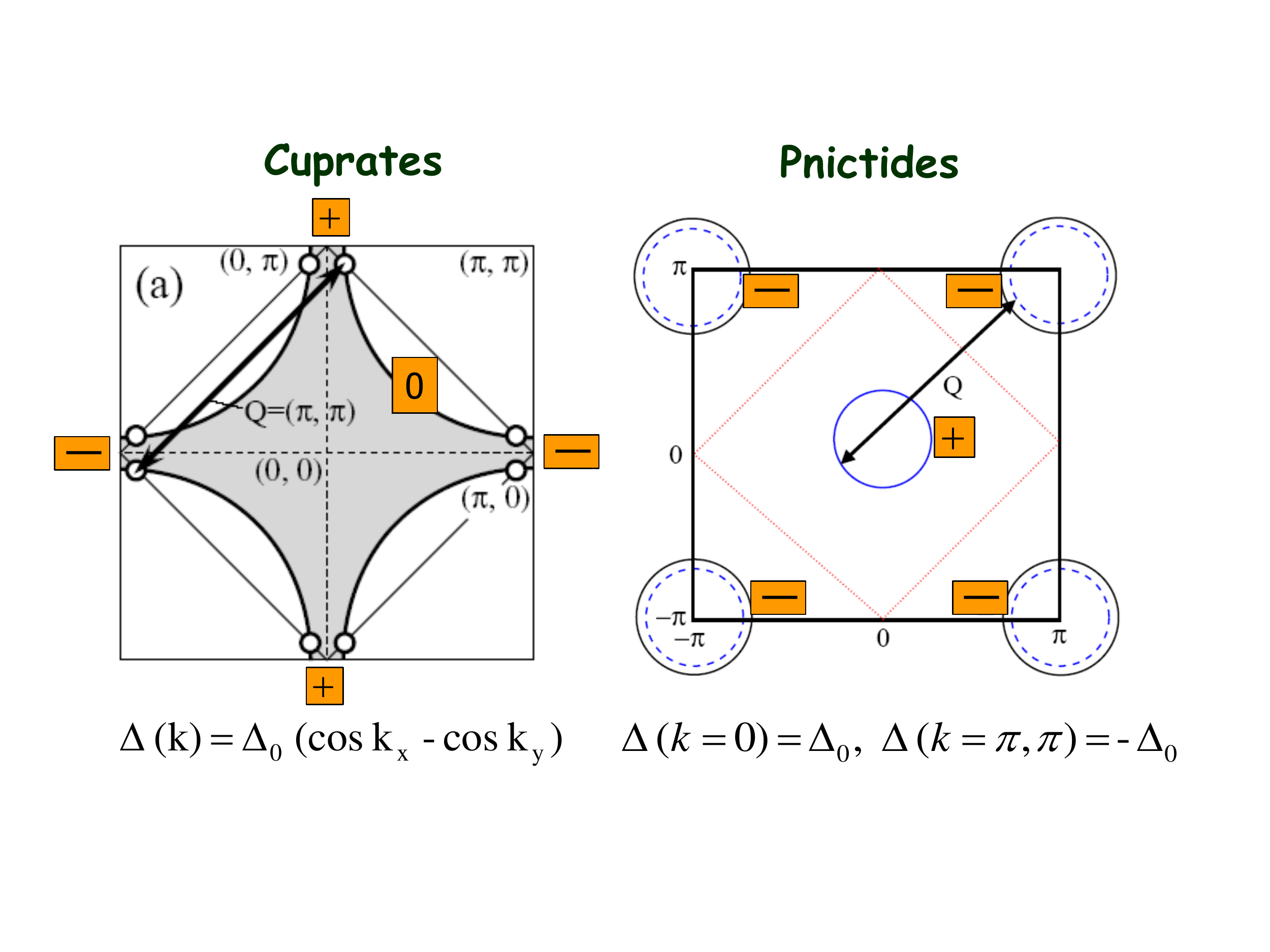}
\caption{A comparison of the pairing state from spin-fluctuation exchange in cuprate SCs and in FeSCs. In the cuprates (left panel) the FS is large, and antiferromagnetic ${\bf Q} = (\pi,\pi)$ connects points on the same FS. Because spin-mediated interaction is positive (repulsive), the
 gap must change sign between FS points separated by ${\bf Q}$. As the consequences, the gap changes sign twice along the FS. This implies a $d-$wave gap symmetry. In FeSCs (left panel) scattering by ${\bf Q}$ moves fermions from one FS to the other. In this situation, the gap must change sign between different FS, but to first approximation remains a constant on a given FS. By symmetry, such a gap is an $s-$wave gap. It is called $s^{+-}$ because it changes sign between different FSs}
\label{fig:comparison}
\end{figure}

In the cuprates, the superconducting gap has been proved experimentally to have $B_{1g}$ symmetry~~[\onlinecite{d-wave}].
 Such a gap appears quite naturally in the doping range where the cuprates  are metals, because KL-type consideration
  shows that $B_{1g}$ interaction becomes attractive if the fully dressed repulsive interaction between fermions near different corners of the Brillouin zone
  (the one at momentum transfer near $(\pi,\pi)$) exceeds the repulsion at small momentum transfer.  The enhancement of $(\pi,\pi)$ interaction is a sure thing if the system displays strong antiferromagnetic spin fluctuations (see Fig.\ref{fig:comparison}).
  That $B_{1g}$ gap is selected is not
 a surprise because such gap $\Delta (k) \propto cos k_x - \cos k_y$ changes sign not only under $k_x \to k_y$ but also
  between ${\bf k}$  and ${\bf k}' = {\bf k} + {\bf Q}$ where ${\bf Q} = (\pi,\pi)$.   This sign change is the crucial element for any electronic mechanism
  of superconductivity because one needs to extract an attractive  component from repulsive  screened Coulomb interaction.

In FeSCs, magnetism and superconductivity are also close neighbors on the phase diagram, and it has been proposed~~[\onlinecite{Mazin,Kuroki}]
 at the very beginning of the $Fe$ era that
  the pairing  mechanism in FeSCs is also a spin-fluctuation exchange.  However,  the geometry of low-energy states in FeSCs and in the
  cuprates is different, and in most FeSCs the momentum ${\bf Q}$ connects low-energy fermionic states near the center and the corner of the Briilouin zone (see Fig.\ref{fig:comparison}). A simple experimentation with trigonometry then tell us that the SC order parameter (the gap)  $\Delta (k)$ must be symmetric with respect to $k_x \to k_y$ and $k_x \to -k_x$, but still must change sign under ${\bf k} \to  {\bf k} + {\bf Q}$.
Such order parameter belongs to  $A_{1g}$ representation, but it only has contributions from a particular subset of $A_{1g}$ states with the form
$\cos k_x + \cos k_y$, $\cos {3 k_x}  +  \cos (3 k_y)$, etc,  which all change sign under  ${\bf k} \to  {\bf k} + {\bf Q}$.  An order parameter with such symmetry 
 is called an extended $s-$wave or, in shorter notations, $s^{+-}$. 

 Majority of researches  do believe that in weakly/moderately doped FeSCs the gap does have $s^{+-}$ symmetry. However, numerous studies of superconductivity in FeSCs over the last five years demonstrated that   the physics of the pairing is
 more involved than it was originally thought because of multi-orbital/multi-band nature of low-energy fermionic excitations in FeSCs. It turns out that both the symmetry and the structure of the pairing gap
 result from  rather non-trivial interplay between spin-fluctuation exchange, intraband Coulomb repulsion, and momentum structure of the interactions. In particular, an $s^\pm$ gap can be with or without nodes, depending on the orbital content of low-energy excitations.
 Besides, the structure of low-energy spin fluctuations evolves with doping, and the same spin-fluctuation
mechanism that gives rise to  $s^{+-}$ gap at small/moderate doping in a particular material can give rise  to a $d-$wave gap at strong hole or electron doping.

There is more uncertainly on the theory side.  In addition to spin fluctuations,  FeSCs also possess charge fluctuations whose strength
is the subject of debates. There are proposals~~[\onlinecite{jap,ku}] that in multi-orbital FeSCs charge fluctuations are strongly enhanced because the system is reasonably close to a transition into a state with an orbital order -- a spontaneous symmetry breaking between the occupation of different orbitals). (A counter-argument is that orbital order does not develop on its own but is induced by a magnetic order~[\onlinecite{rafael_nematic,rafael_review}]).
 If charge fluctuations are relevant, one should consider, in addition to spin-mediated pairing interaction, also the
 pairing interaction mediated  by charge fluctuations. The last interaction gives rise to a conventional, sign-preserving  $s-$wave pairing~~[\onlinecite{jap}].  A "p-wave" gap scenario (a gap belonging to $E_{2g}$ representation) has also been put forward~~[\onlinecite{p_lee}].

 From experimental side, $s$-wave gap symmetry is consistent with ARPES data on moderately doped B$_{1-x}$K$_x$ Fe$_2$As$_2$ and BaFe$_2$(As$_{1-x}$P$_x$)$_2$,
 which detected only a small variation of the gap along the FSs
centered at $(0,0)$ (Ref.~[\onlinecite{laser_arpes}]), and with the evolution of the tunneling data in a magnetic field~~[\onlinecite{hanaguri}].
However, other data on these and other FeSCs, which measure contributions from all FSs, including the FSs for which ARPES data are not available at the moment,
 were interpreted as evidence either for the full gap~~[\onlinecite{ding,chen,osborn1,carrington1}], or
 that the gap has either accidental nodes~~[\onlinecite{nodes,carrington}] or deep minima~~[\onlinecite{proz,proz_1,moler}].
  As additional level of complexity, superconductivity was also discovered in materials which only contain hole pockets, like hole-doped \KFA, or only electron pockets, like \AFS.
For these materials, the argument for $s^{+-}$ superconductivity, driven by magnetically-enhanced interaction between fermions near hole and electron pockets,
 is no longer applicable, yet both classes of materials have finite $T_c$, which is around 3K for \KFA~ and as high as 30K for \AFS~  (Refs. [\onlinecite{Guo2010}]).  For \KFA, Various experimental
probes~~[\onlinecite{KFeAs_exp_nodal}] indicate the presence of gap nodes. Laser ARPES data~[\onlinecite{shin}] were interpreted as evidence for 
 $s-$wave with nodes, while thermal conductivity data have been interpreted as evidence for both, $d-$wave and $s-$wave orders (Refs.[\onlinecite{reid}] and [\onlinecite{mats_2}], respectively).  For \AFS, ARPES results were interpreted as evidence for $s-$wave (Ref.[\onlinecite{exp:AFESE_ARPES}]), however neutron scattering  experiments~[\onlinecite{AFESE_neu}] detected a resonance peak which most naturally can be interpreted as evidence for $d-$wave~[\onlinecite{AFESE_d}] (see, however, 
 [\onlinecite{khodas_2}]).

 In this paper, I  argue that  all these seemingly very different gap structures  actually follow quite naturally from the same underlying physics idea that FeSCs can be treated as moderately interacting itinerant fermionic systems with multiple FS sheets and effective four-fermion
  intra-band and inter-band interactions in the band basis.
  I introduce the effective low-energy model with small numbers of input parameters~~[\onlinecite{maiti_11}] and use it to study the doping evolution of the pairing in hole and electron-doped FeSCs. I argue  that various approaches based on underlying microscopic models in the orbital basis reduce  to this model at low energies.

  The paper is organized as follows. In Sec. \ref{sec:2} I discuss general aspects of the band structure of FeSCs which contain hole and electron pockets.
  In Sec. \ref{sec:3} I present a generic discussion of what is needed for SDW order and superconductivity and how magnetic fluctuations help superconductivity to develop.
   In Sec. \ref{sec:pRG} I briefly review parquet renormalization group approach to FeSCs. This approach treats magnetism and superconductivity on equal footing.  I argue that, depending on input parameters and/or doping, the system  first becomes either SDW magnet or a superconductor.
   In Sec.\ref{sec:4} I review itinerant approach to magnetism. I show that for most (but not all) dopings a SDW order below $T_{sdw}$ spontaneously
    breaks $C_4$ lattice symmetry in addition to $O(3)$ symmetry of rotations in spin space.  I then review works on a pre-emptive spin-nematic instability at $T_{n} > T_{sdw}$, when the system spontaneously breaks $C_4$ symmetry down to $C_2$, but spin-rotational symmetry remains unbroken down to a smaller $T_{sdw}$.
    In Sec.\ref{sec:5}  I review an itinerant approach to superconductivity. I first present a generic symmetry consideration of a gap structure in a multi-band superconductor
   and show that a ``conventional wisdom'' that an s-wave gap
 is nodeless along the FSs, d-wave gap has four nodes, etc, has only  limited applicability in
  multi-band superconductors, and there are cases when the gap with
nodes has an $s-$wave symmetry, and the gap without nodes has a $d-$wave symmetry.  I then discuss the interplay between intra-band and inter-band interactions, for realistic multi-pocket models for FeSCs and set the conditions for an attraction in an $s-$wave or a
 $d-$wave channel. I consider 5-orbital model with local interactions, convert it into a band basis, and show the structure of the superconducting gap.
  I  use  the combination of RPA and leading angular harmonic approximation to analyze the pairing in $s-$ and $d-$wave channels at different dopings.  I show that, depending on parameters and doping,  magnetically-mediated pairing
 leads to  an $s^\pm$ superconductivity with either a near constant gap along the FSs, or gaps with deep minima, or even with the nodes.  I  briefly review  the experimental situation in Sec. \ref{sec:7} and present concluding remarks in Sec. \ref{sec:8}.

\section{The electronic structure of {\text FeSCs}.}
\label{sec:2}

\begin{figure}[tbp]
\includegraphics[angle=0,width=0.8\linewidth]{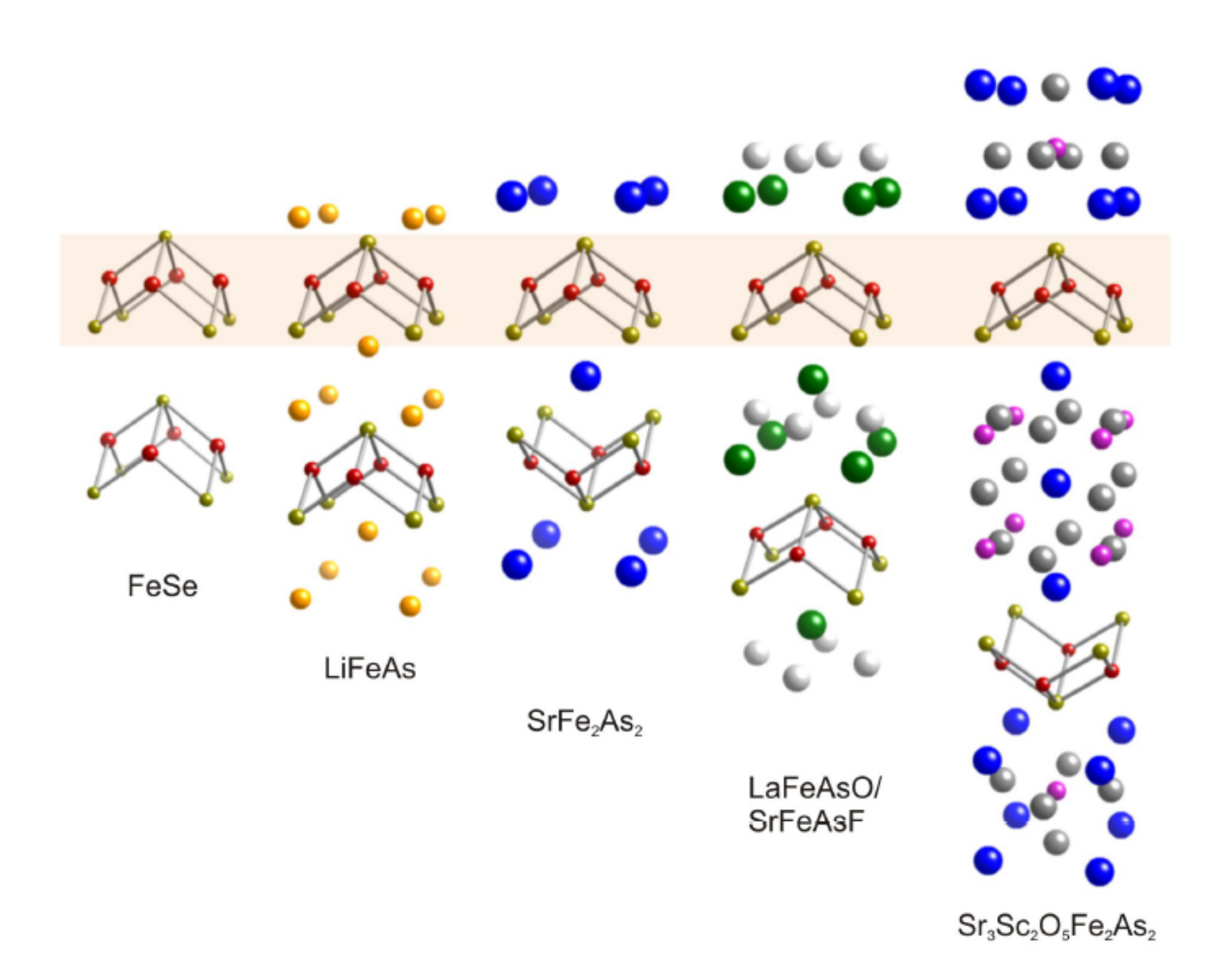}
\caption{Crystallographic structures of various families of iron-based superconductors.
  From Ref. ~[\onlinecite{review_2}]. }
\label{fig:el_structure}
\end{figure}

\begin{figure}[tbp]
\includegraphics[angle=0,width=0.8\linewidth]{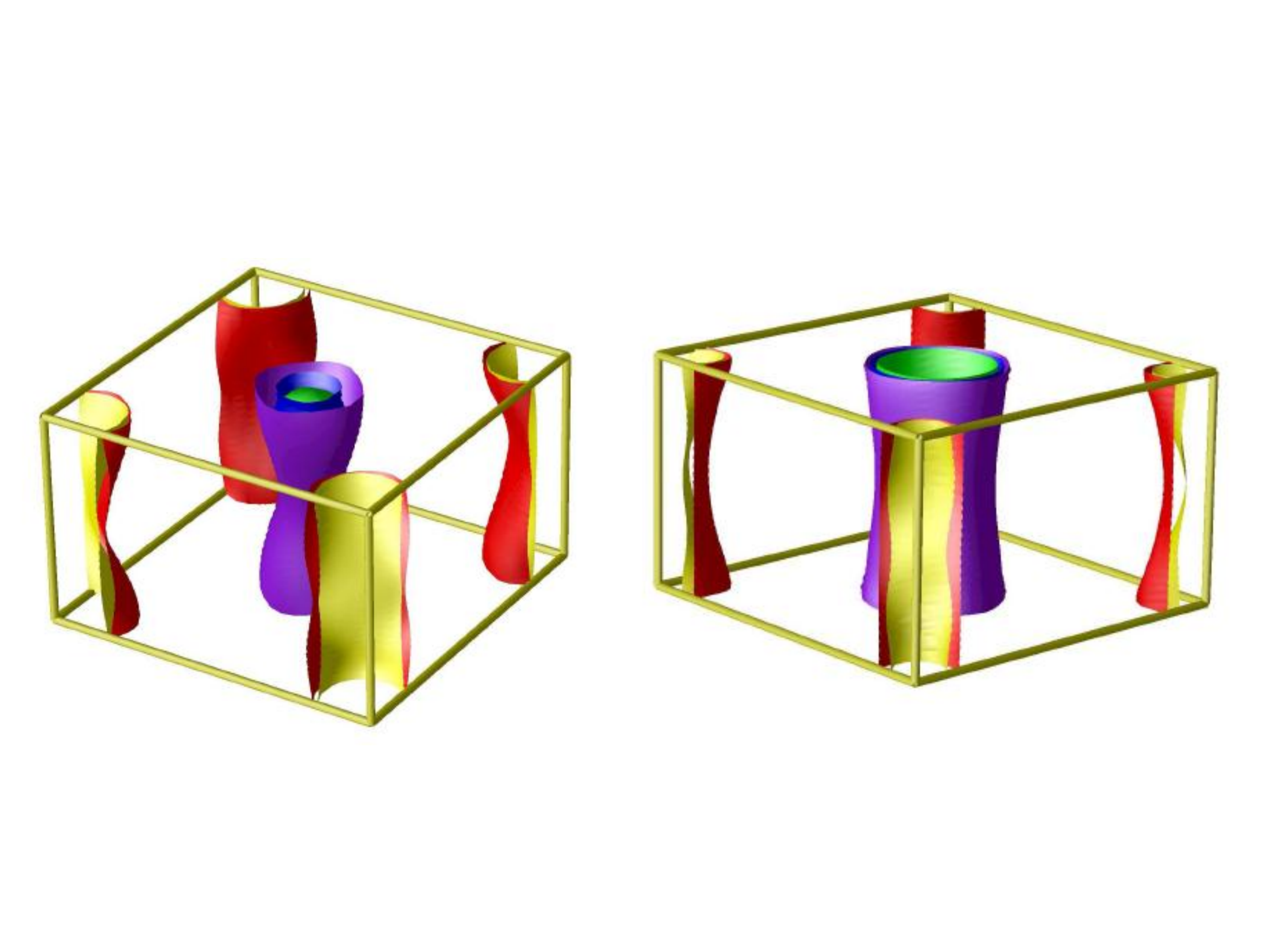}
\caption{The electronic structure of FeSCs. In  weakly and moderately electron-doped  materials (left panel)
  the FS consists of quasi-2D warped cylinders centered at $(0,0)$ and $(\pi,\pi)$ in a 2D cross-section. The ones near $(0,0)$ are hole pockets (filled states are outside cylinders), the ones near $(\pi,\pi)$ are electron pockets (filled states are inside cylinders)
  There also exists  a quasi-3D hole pocket
 near $k_z = \pi$. In hole-doped FeSCs the electronic structure is very similar, but 3D hole pocket becomes  quasi-2D warped hole cylinder.
  From Ref. ~[\onlinecite{mazin_schmalian}]. }
\label{fig:FS}
\end{figure}
The crystallographic structures of various families of iron-based superconductors is shown in Fig. \ref{fig:el_structure}.  All FeSCs contain planes made of Fe atoms, and pnictogen/chalcogene atoms are staggered in a checkerboard order above and below the iron planes. In 1111 system this order repeats itself from one Fe plane to the other, while for 122-type systems, it flips sign between neighboring planes.

The electronic structures of FeSCs at low energies are rather well established by
 ARPES~[\onlinecite{Li}] and quantum oscillation measurements~~[\onlinecite{bib:Suchitra}].
In  weakly and moderately  electron-doped  materials, like BaFe$_{1-x}$Co$_x$Fe$_2$As$_2$
  the FS contains
 several quasi-2D warped cylinders centered  at $k=(0,0)$ and $k= (\pi,\pi)$ in a 2D cross-section, and may also contain  a quasi-3D pocket
 near $k_z = \pi$ (Fig.\ref{fig:FS}).  The fermionic dispersion is electron-like
 near the FSs at $(\pi,\pi)$  (filled states are inside a FS)  and hole-like near the FSs centered at $(0,0)$
  (filled states are outside a FS).  In heavily electron-doped \Fs, like
A$_x$Fe$_{1-y}$Se$_2$ (A = K, Rb, Cs), only electron pockets remain, according to recent ARPES studies.~~[\onlinecite{exp:AFESE}]
In weakly and moderately hole-doped \Fs,
like Ba$_{1-x}$K$_x$Fe$_2$As$_2$, the electronic structure is similar to that at moderate electron doping, however the spherical FS becomes
the third quasi 2D hole FS centered  at $(2\pi,0) = (0,0)$. In addition, new low-energy hole states likely appear around $(\pi,\pi)$
 and squeeze electron pockets~~[\onlinecite{Evtush}].  At strong hole doping, electron FSs disappear and only
  only hole FSs are present~~[\onlinecite{KFeAs_ARPES_QO}] These electronic structures agree well with  first-principle calculations
~~[\onlinecite{bib:first,Boeri,review_3}], which is another argument to treat
 FeSCs as itinerant fermionic systems.
\begin{figure}[tbp]
\includegraphics[angle=0,width=\linewidth]{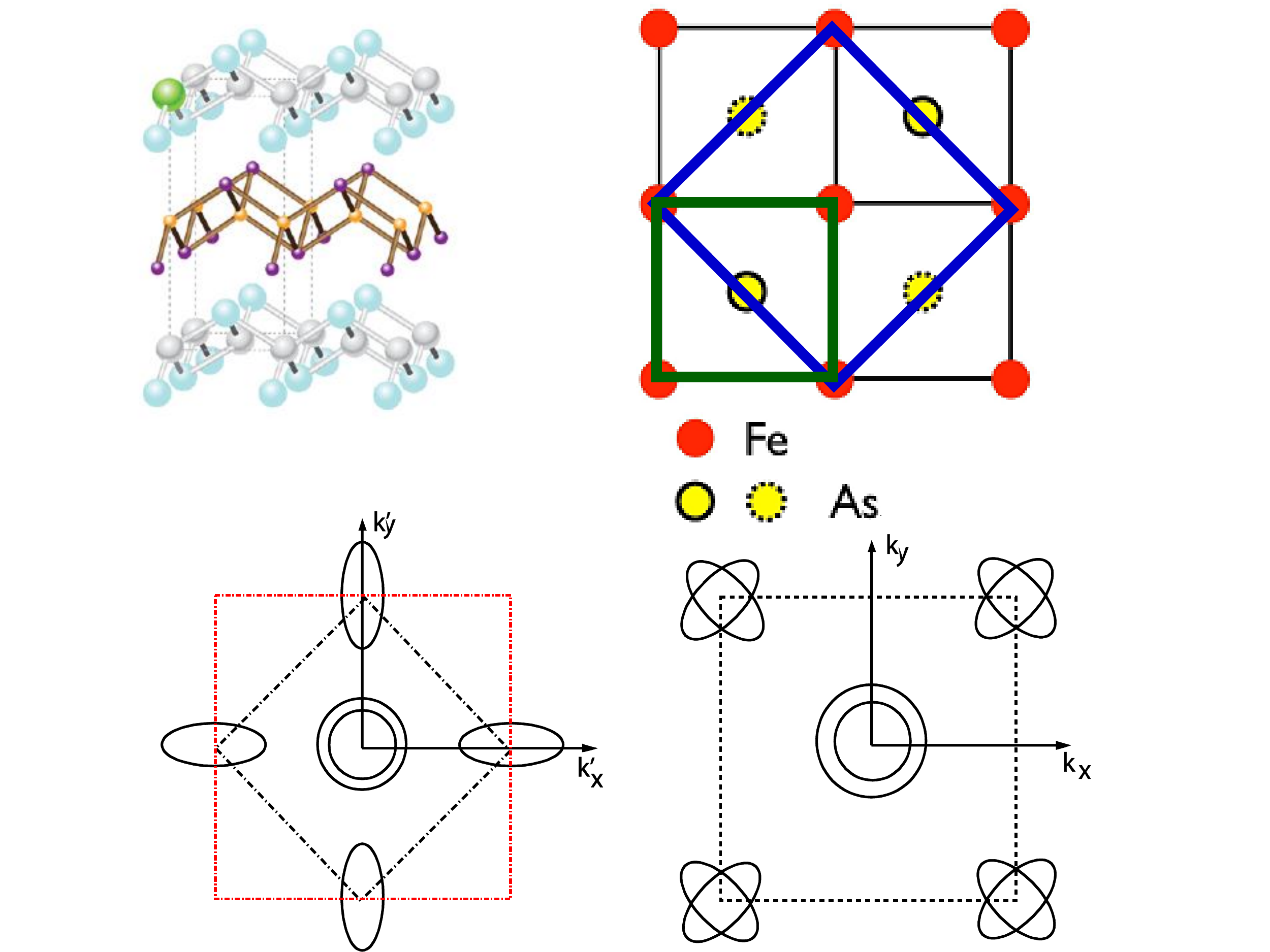}
\caption{Upper panel: 3D electronic structure of LaOFeAs (left) and its 2D cross-section (left). In only Fe states are considered, an elementary cell contains one Fe atom (green). The actual unit cell (blue) contains two Fe atoms because of two non-equivalent positions of a pnictide above and below the Fe plane.  Lower panel -- the location of hole and electron FSs in a 2D cross section  in the folded BZ (two Fe/cell, right) and
 in the unfolded BZ (one Fe/cell, left). From Refs. ~[\onlinecite{norman}], ~[\onlinecite{ising}](b) and ~[\onlinecite{vekhter_10}]b.}
\label{fig:folded}
\end{figure}

The measured FS reflects the actual crystal structure of FeSCs in which there are two non-equivalent positions of a pnictide above and below  an $Fe$ plane, and, as a result, there are two $Fe$ atoms in the unit cell (this actual 2Fe BZ is called  "folded BZ"). From theory perspective, it would be easier to work in the BZ which contains only one $Fe$ atom
in the unit cell (this theoretical 1Fe BZ is called "unfolded BZ"). I illustrate the difference between folded and unfolded BZ in Fig.\ref{fig:folded}.  In general, only folded BZ is physically meaningful. However, if by some reason a potential from
 a pnictogen (or chalcogen) can be neglected, the difference between the folded and the unfolded BZ becomes  purely geometrical:
  the  momenta ${\tilde k}_x$ and ${\tilde k}_y$ in the folded BZ are linear combinations of $k_x$ and $k_y$ in the unfolded BZ: ${\tilde k}_x = k_x + k_y$, ${\tilde k}_y = k_x -k_y$.  In this situation, the descriptions in the folded and unfolded BZ become equivalent.

  Most of the existing theory works on magnetism and on the pairing mechanism and the structure of the SC gap analyze the pairing problem in the unfolded BZ, where which two hole pockets are centered at $(0,0)$ and one at $(\pi,\pi)$, and the two electron pockets are at $(0,\pi)$ and $(\pi,0)$.
  It became increasingly clear recently that the interaction via a pnictogen/chalcogen and also 3D effects do play some role for the pairing, particularly in strongly electron-doped systems.~~[\onlinecite{3D,mazin}] However, it is still very likely that the key aspects
  of the pairing in FeSCs can be understood  by analyzing a pure 2D electronic structure with only Fe states involved. In the next three sections I assume that this is the case and consider a 2D model in the unfolded BZ with hole FSs near $(0,0)$ and $(\pi,\pi)$ and electron FSs at $(0,\pi)$ and $(\pi,0)$.

\section{The low-energy model and the interplay between magnetism and superconductivity}
\label{sec:3}

\begin{figure*}[htp]
\includegraphics[width = 2.2in]{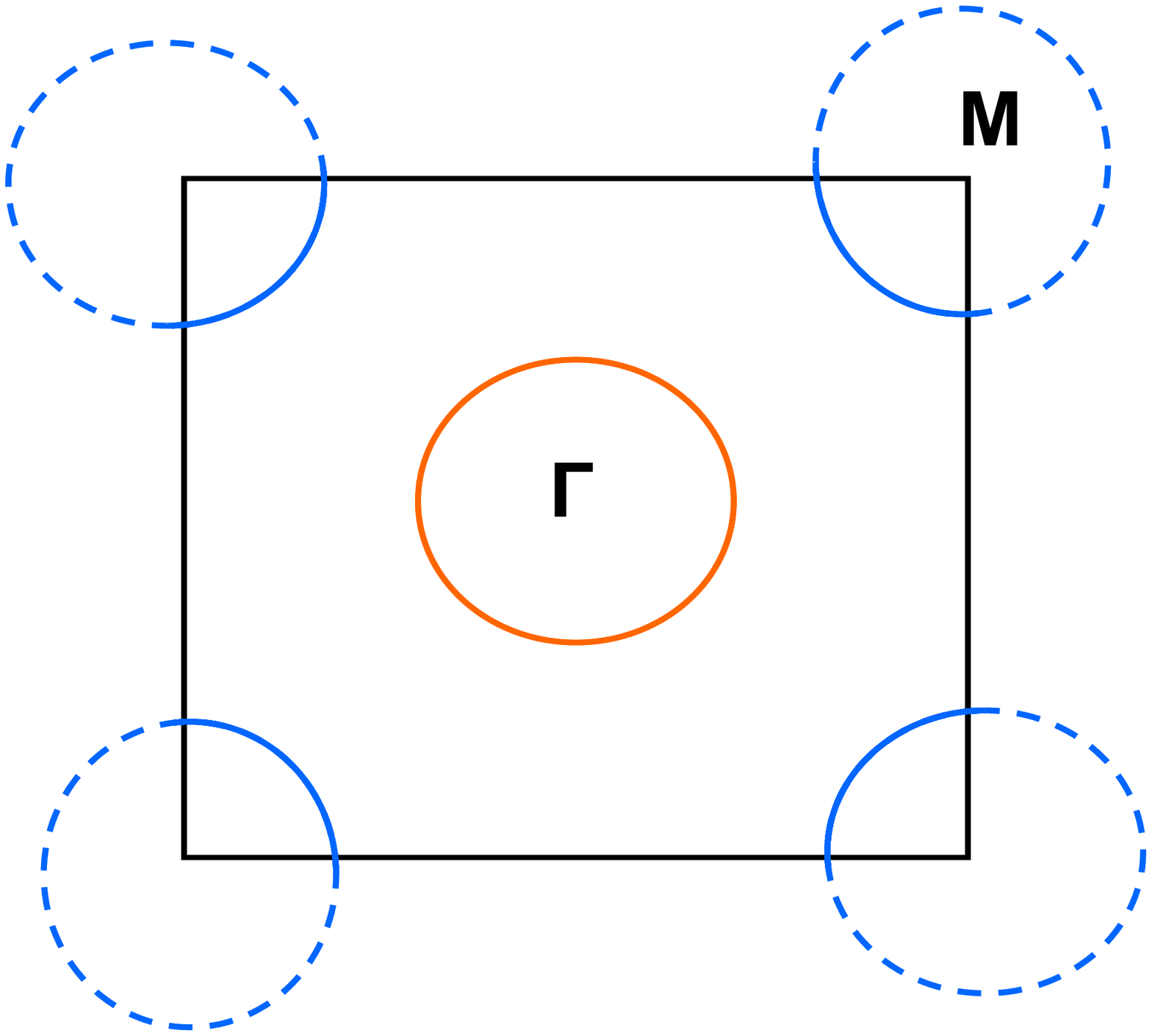}
\caption{\label{fig:FS_1}FS topology for a two-pocket model for
 FeSCs.  The two FSs are for hole-like dispersion (blue circle, filled states outside the FS) and
  electron-like dispersion (orange circle, filled states inside the FS).}
\end{figure*}

For  proof-of-concept I first consider a simple problem: a 2D
two-pocket model with one hole and one electron FS, both circular
and of equal sizes (see Fig.\ref{fig:FS_1}), and 
momentum-independent four-fermion interactions.

The free-fermion Hamiltonian is the sum of kinetic energies of
holes and electrons: \beq H_2 = \sum_{k,\sigma} \e_c
c^{\dag}_{k,\sigma}c_{k,\sigma} +\e_f
f^{\dag}_{k,\sigma}f_{k,\sigma} \eeq where $c$ stands for holes,
$f$ stands for electrons, and $\e_{c,f}$ stand for their
respective dispersions with the property that $\e_c(k)=-\e_f(k+Q)$,
where $\textbf{Q} = (\pi,\pi)$ is the momentum vector which connects the
centers of the two fermi surfaces. The  density of states $N_0$ is
the same on both pockets, and the electron pocket `nests'
perfectly within the hole pocket when shifted by $\textbf{Q}$.

\begin{figure*}[htp]
\includegraphics[width = 2.2in]{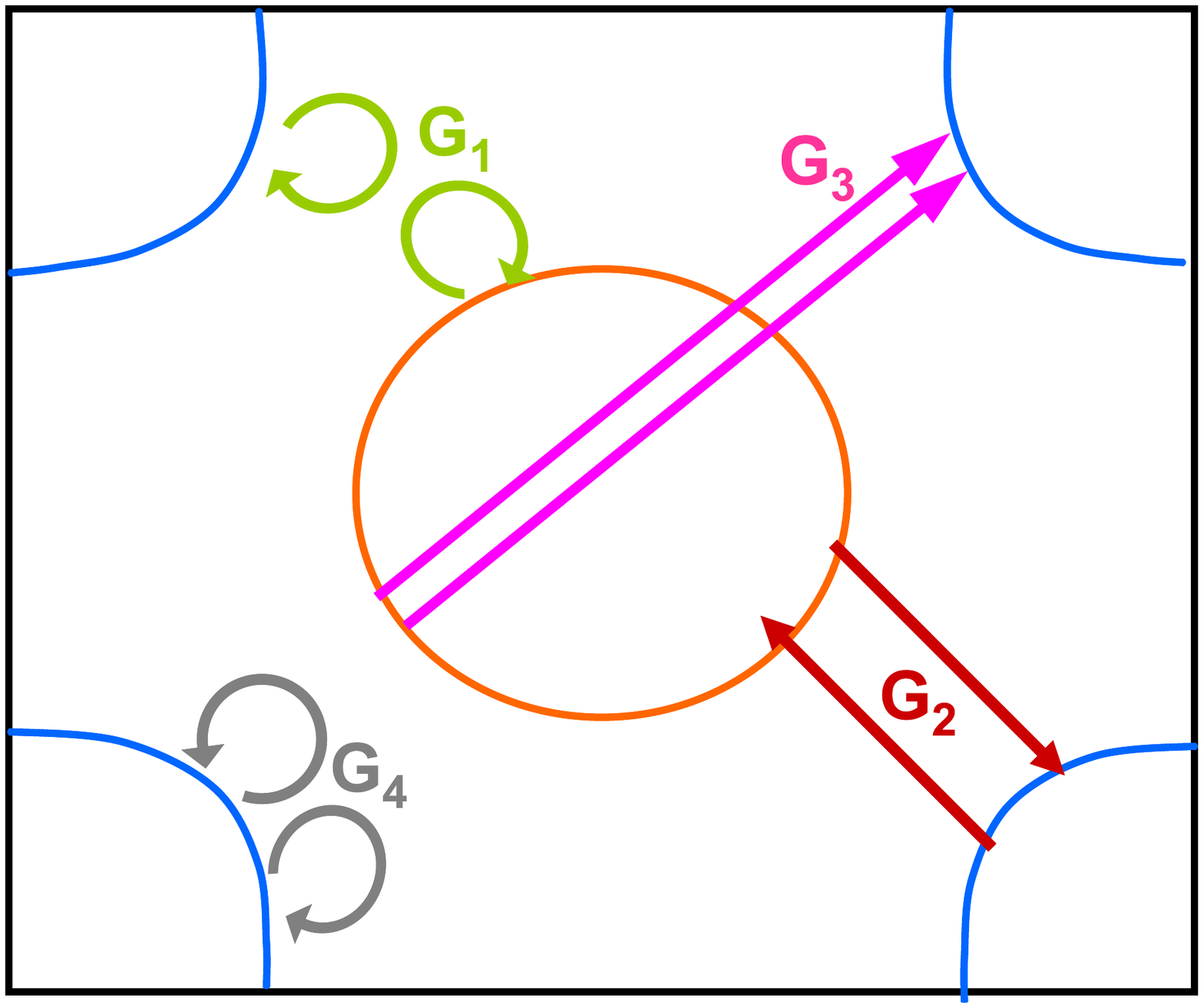}
\caption{\label{fig:int}The interactions between pockets
in the two-pocket model for Fe-pnictides.
$G_1$ is a density-density interaction between fermions from
different pockets. $G_2$ is an exchange interaction between the
pockets, $G_3$ is a pair hopping process between the
pockets, and $G_4$ is a density-density interaction within
the same pocket. All interactions are repulsive (positive). From [\onlinecite{maiti_rev,maiti_book}].}
\end{figure*}

There are five different types of interactions between low-energy
fermions: two intra-pocket density-density interactions, which I
treat as equal, interaction between densities in different
pockets, exchange interaction between pockets, and pair hopping
term, in which two fermions from one pocket transform into two
fermions from the other pocket. I show these interactions
graphically in Fig \ref{fig:int}.

In explicit form \bea\label{eq:2-poc Ham} H_{\text{int}}&=& G_1
\sum_{[k,\sigma]}  c^{\dagger}_{k_1 \sigma} f^{\dagger}_{k_2
\sigma'}  f_{k_3 \sigma'} c_{k_4 \sigma}
\nonumber\\
&&+G_2 \sum_{[k,\sigma]} f^{\dagger}_{k_1 \sigma} c^{\dagger}_{k_2
\sigma'} f_{k_3 \sigma'} c_{k_4 \sigma}
\nonumber\\
&&+ \sum_{[k,\sigma]}\frac{G_3}{2}
\left(c^{\dag}_{k_1,\sigma_1}c^{\dag}_{k_2,\sigma_2}f_{k_3,\sigma_2}f_{k_4,\sigma_1}
+
\text{h.c}\right)\nonumber\\
&&+\sum_{[k,\sigma]}\left(\frac{G_4}{2}
c^{\dag}_{k_1,\sigma_1}c^{\dag}_{k_2,\sigma_2}c_{k_3,\sigma_2}c_{k_4,\sigma_1}
+ c\leftrightarrow f\right) \nonumber\\ \eea where
$\sum{_{[k,\sigma]}}$ is short for the sum over the spins and the
sum over all the momenta constrained to $k_1+k_2=k_3+k_4$ modulo a
reciprocal lattice vector.

\begin{figure}[tbp]
\includegraphics[width=2.8in]{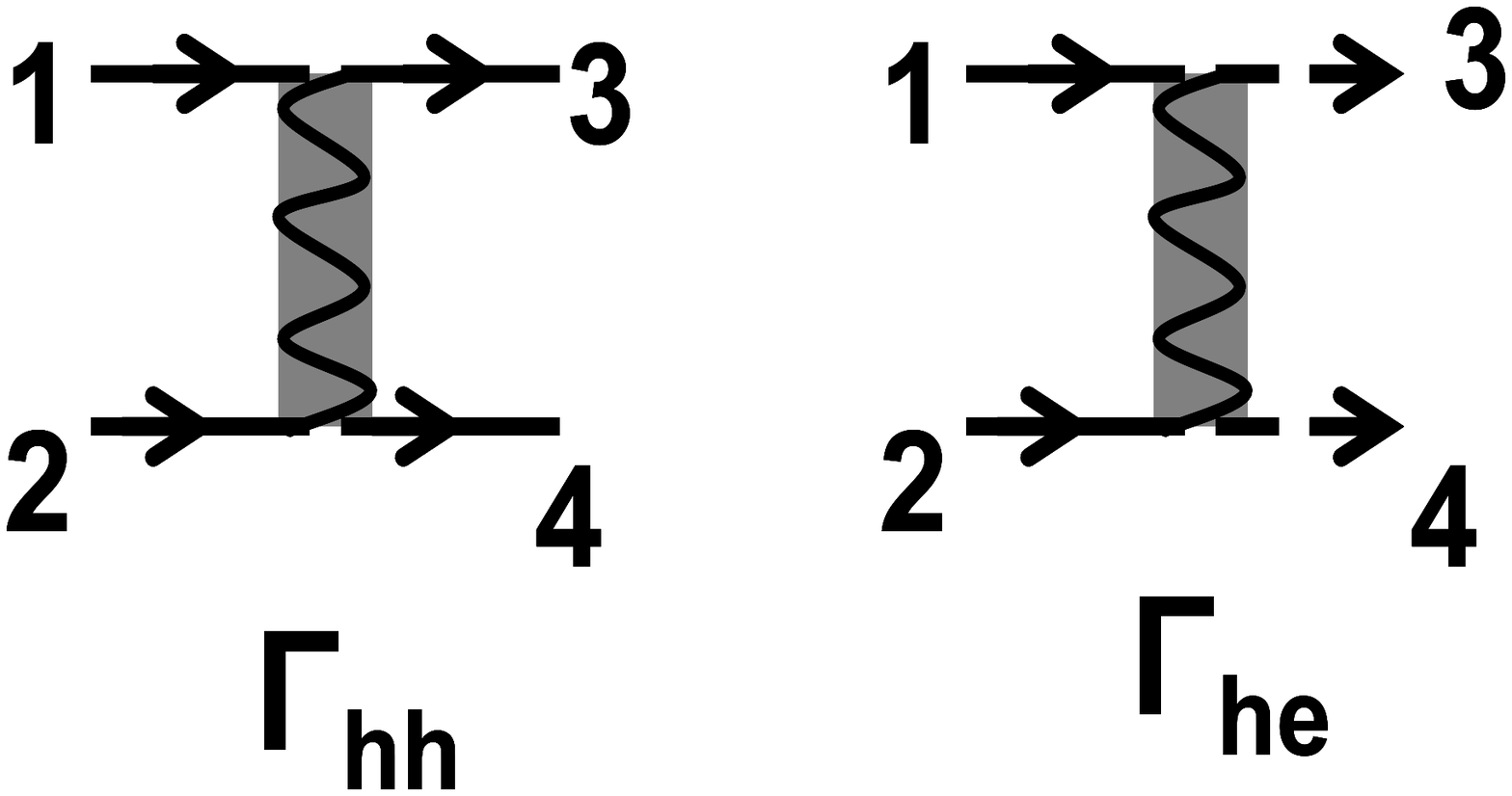}
\caption{Vertices $\Gamma_{hh}=\Gamma_{ee}$ and $\Gamma_{he}$
introduced in the 2 pocket model. Solid and dashed lines denote fermions from the two pockets.
From [\onlinecite{cee}].} \label{fig:new_vertices}
\end{figure}

\begin{figure*}[htp]
\includegraphics[width = 1\columnwidth]{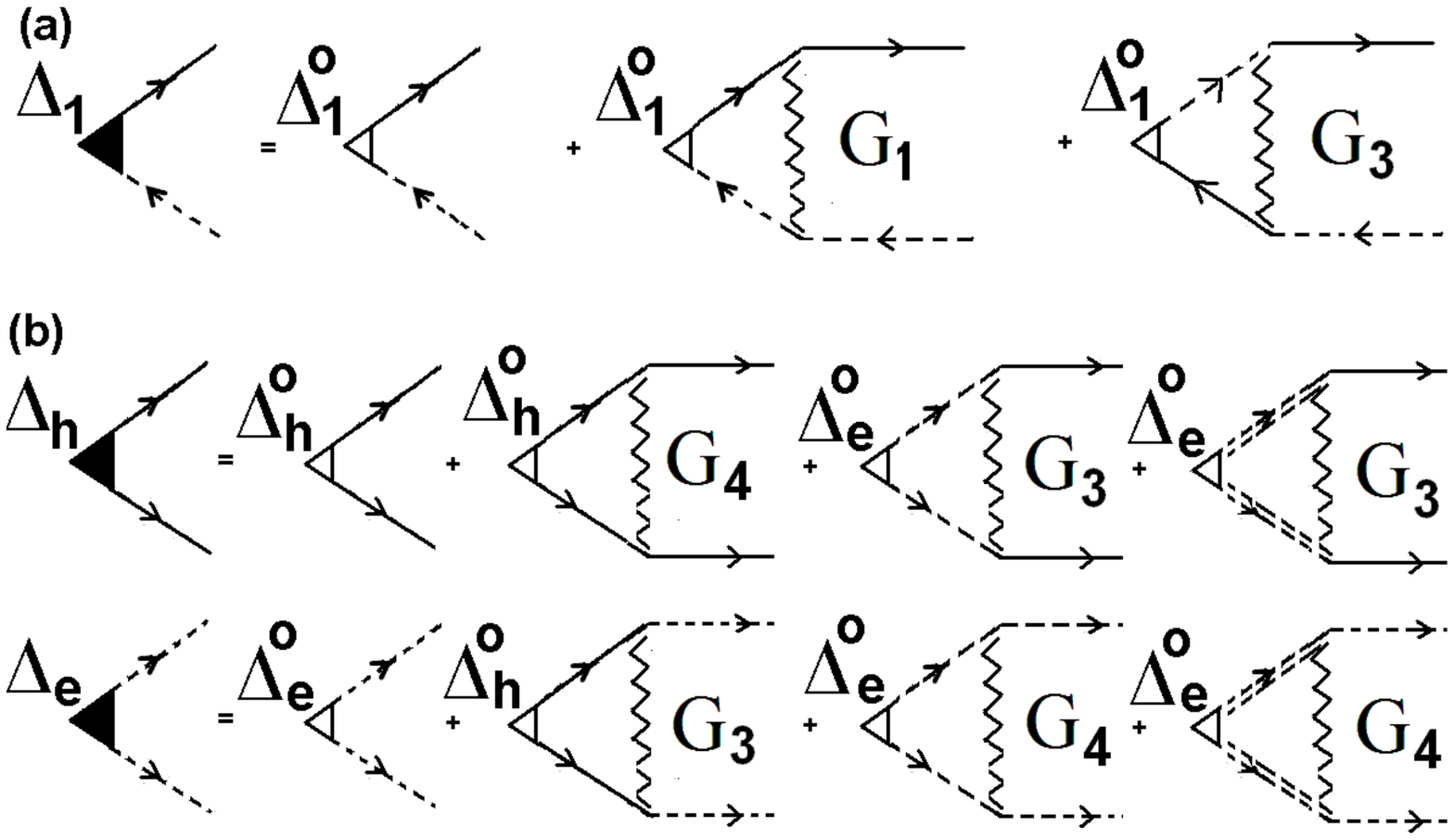}
\caption{\label{fig:mo_1} Lowest-order terms in the ladder series for the renormalizations of the SDW and superconducting vertices.
 The effective coupling in the SDW channel is $G_1 + G_3$.  The two couplings in the two SDW channels are $G_3+G_4$ and $-G_3+G_4$. From [\onlinecite{maiti_rev,maiti_book}].}
\end{figure*}

The textbook approach to analyse potential instabilities towards superconductivity and magnetism  is to consider the appearance of the poles in the
corresponding vertex functions.  For superconductivity, we need to consider vertex functions with zero total incoming momentum:
$\Gamma_{hh} ({\bf k}_F,-{\bf k}_F; {\bf p}_F,-{\bf p}_F);
\Gamma_{ee} ({\bf k}_F,-{\bf k}_F;{\bf p}_F,-{\bf p}_F)$, where
${\bf k}_F$ and ${\bf p}_F$ belong to the same pocket, and
$\Gamma_{he} ({\bf k}_F,-{\bf k}_F; {\bf p}_F,-{\bf p}_F)$, where
${\bf k}_F$ and ${\bf p}_F$ belong to different pockets (see Fig.
\ref{fig:new_vertices}). To first order in $G_i$, we have \bea
&&\Gamma^{0}_{hh}  ({\bf k}_F,-{\bf k}_F; {\bf p}_F,-{\bf p}_F) = -G_4 \nonumber \\
&&\Gamma^{0}_{ee}  ({\bf k}_F,-{\bf k}_F; {\bf p}_F,-{\bf p}_F) = -G_4 \nonumber \\
&&\Gamma^{0}_{he}  ({\bf k}_F,-{\bf k}_F; {\bf p}_F,-{\bf p}_F) =
-G_3 \eea
I follow \cite{stat_phys} and introduce the vertex function with the opposite sign compared to the interaction potential.  

For SDW order we need to consider interactions with momentum transfer ${\bf Q} = (\pi,\pi)$:  $\Gamma^{0}_{he}  ({\bf k}_F,{\bf k}'_F; {\bf p}_F,{\bf p}'_F)$,
 $\Gamma^{0}_{he}  ({\bf k}_F,{\bf p'}_F; {\bf k}'_F, {\bf p}_F)$, and $\Gamma^{0}_{he}  ({\bf k}_F,{\bf p'}_F;  {\bf p}_F,{\bf k}'_F)$, where
 ${\bf k}$ and ${\bf k}'$   belong to one pocket
 and ${\bf p}$ and ${\bf p}'$   belong to the other pocket, and ${\bf p} = {\bf k} + {\bf Q}$.
 To first order in $G$ we have
 \bea
&&\Gamma^{0}_{he}  ({\bf k}_F,{\bf k}'_F; {\bf p}_F,{\bf p}'_F) = -G_3 \nonumber \\
&&\Gamma^{0}_{he}  ({\bf k}_F,{\bf p'}_F; {\bf k}'_F, {\bf p}_F) = -G_1 \nonumber \\
&&\Gamma^{0}_{he}  ({\bf k}_F,{\bf p'}_F;  {\bf p}_F,{\bf k}'_F) = -G_2
\eea

To see which combinations of different $\Gamma$ appear in the SDW and superconducting channels, I add to the Hamiltonian
 the trial terms $\Delta_1 c^\dagger_{k,\alpha} {\bf \sigma}_{\alpha\beta} f_{k+Q,\beta}$,
 $\Delta_h c^\dagger_{k,\alpha} {i \sigma^y}_{\alpha\beta} c_{-k,\beta}$, and $\Delta_e f^\dagger_{k,\alpha} {i \sigma^y}_{\alpha\beta} f_{-k,\beta}$,
 dress them by the interactions, and express the fully renormalized $\Delta_1$, $\Delta_e$, and $\Delta_h$ via fully renormalized vertices.
 The lowest-order terms in the corresponding series are shown in Fig. \ref{fig:mo_1}.   One can easily make sure that the vertex which renormalizes $\Delta_1$ contains $G_1+G_3$, while the vertices which renormalize $\Delta_h$ and $\Delta_e$ are made out of $G_3$ and $G_4$.

 \subsection{Ladder approximation}

 To proceed further, I first assume that the two channels do not communicate with each other, i.e., the renormalization of the SDW vertex does not
 involve the interactions with zero total momentum, while the renormalization of the two superconducting vertices does not involve the interaction with momentum transfer ${\bf Q}$.  Mathematically, this approximation implies  that higher-order additions to Fig.  \ref{fig:mo_1} form ladder series.   These series can be easily summed up analytically.

 \subsubsection{The SDW vertex}

 For SDW vertex, summing up  ladder diagrams we obtain
 \beq
 \Delta_1 = \Delta_1^{(0)} \chi_{ph} (Q),  ~~ \chi_{ph} (Q) = \left(1 - \Pi_{ph} (Q) \Gamma^{full}_{sdw} \right)
 \eeq
 where
 \beq
\Gamma^{full}_{sdw} = -\frac{G_1+G_3}{1- \Pi_{ph} (Q) (G_1 + G_3)}
\label{mo_d_1}
\eeq
where $\Pi_{ph} (Q)$ is the particle-pole polarization bubble at momentum transfer $Q$.
Note that only the combination $G_1 + G_3$ appears in (\ref{mo_d_1}).  The interactions $G_2$ and $G_4$ do not participate in the renormalization of the SDW vertex.

I show the behavior of $\Pi_{ph} (q)$ at a generic $q$
 in Fig. \ref{fig:PiQ} below.  At this stage, it is just enough to observe that $\Pi_{ph} (Q)$ is positive. Eq. (\ref{mo_d_1}) then shows that
 the full vertex in the SDW channel $\Gamma^{full}_{sdw}$ and the susceptibility $\chi_{ph} (Q)$ diverge when $\Pi_{ph} (Q) (G_1 + G_3)= 1$.
 That the divergence occurs for a repulsive interaction ($G_1 + G_3 >0$) reflects the well-known fact that fermion-fermion repulsion does give rise to a Stoner-like magnetic instability.

 \subsubsection{The superconducting vertex}

 Let's now solve for
the full $\Delta_h$ and $\Delta_e$ in the ladder approximation.
 A simple analysis
shows that the two equations become
\bea
&&\Delta_h =  \Delta^{(0)}_h-\left(\Delta^{(0)}_h \Gamma^{full}_{hh} +  \Delta^{(0)}_e \Gamma^{full}_{he}\right) \Pi_{pp} \nonumber \\
&&\Delta_e = \Delta^{(0)}_e -\left(\Delta^{(0)}_e \Gamma^{full}_{ee} +  \Delta^{(0)}_h \Gamma^{full}_{he}\right) \Pi_{pp}
\label{mo_d_2}
\eea
where $\Pi_{pp} >0$ is the particle-particle polarization bubble at zero momentum transfer:
 ($\Pi_{pp} = N_0 (\log|\omega_c/\Omega| + i \pi/2)$, where $N_0$ is the density of states at the Fermi level and $\Omega$ is the total incoming frequency),
  and
\bea \Gamma^{full}_{hh}
&=& - \frac12\left(\frac{G_4 + G_3}{1 + (G_4 + G_3)\Pi_{pp}} +
\frac{G_4 - G_3}{1 + (G_4 - G_3)
\Pi_{pp}}\right)\nonumber\\
\Gamma^{full}_{ee}&=&\Gamma^{full}_{hh}\nonumber\\
\Gamma^{full}_{he} &=& - \frac12\left(\frac{G_4 + G_3}{1 + (G_4 +
G_3)\Pi_{pp}} - \frac{G_4 - G_3}{1 + (G_4 - G_3)
\Pi_{pp}}\right)\nonumber\\ \label{4_2_1} \eea
The set of equations in (\ref{mo_d_2}) decouples into
\bea
&&\Delta_h -\Delta_e= \left(\Delta^{(0)}_h - \Delta^{(0)}_e\right) \chi^{-}_{pp},~~\chi^{-}_{pp} = \frac{1}{1 + (G_4 - G_3)
\Pi_{pp}} \nonumber \\
&&\Delta_h +\Delta_e= \left(\Delta^{(0)}_h + \Delta^{(0)}_e\right) \chi^{+}_{pp},~~\chi^{+}_{pp} = \frac{1}{1 + (G_4 + G_3) \Pi_{pp}} \nonumber \\
\label{mo_d_3}
\eea
Because $\Pi_{pp} >0$,  the presence or absence of a pole in $\Gamma^{full}$ (i.e., potential divergence of $\chi_{pp}$)
depends on the signs of $G_3 + G_4$ or $G_4-G_3$. If both are
positive, there are no poles, i.e., non-superconducting state is
stable. In this situation, at small $\Omega$, $\Gamma^{full}_{hh}
\approx -1/\Pi_{pp}$, $\Gamma^{full}_{he} \approx
-(G_3/(G^2_4-G^2_3))\Pi^2_{pp}$, i.e., both vertex functions
decrease (inter-pocket vertex decreases faster). If one (or both)
combinations are negative, there are poles in the upper frequency
half-plane and fermionic system is unstable against pairing.  The
condition for the instability is $|G_3| > G_4$. $G_4$ is
inter-pocket interaction, and there are little doubts that it is
repulsive, even if to get it one has to transform from orbital to
band basis.  $G_3$ is interaction at large momentum transfer, and,
in principle, it can be either positive or negative depending on
the interplay between intra- and inter-orbital interactions.   In
most microscopic multi-orbital calculations, $G_3$ turns out to be
positive, and I set $G_3>0$ in the analysis (for the case $G_3<0$
see Ref. \cite{Kontani}).

For positive $G_3$,  the condition for the pairing instability is
$G_3 > G_4$. What kind of a pairing state do we get?  First, both
$\Gamma^{full}_{hh}$ and $\Gamma^{full}_{he}$ do not depend on the
direction along each of the two pockets, hence the pairing state
is necessary $s-$wave. On the other hand, the pole is in
$\Gamma_2$, which appears with opposite sign in
$\Gamma^{full}_{hh}$ and $\Gamma^{full}_{he}$. The pole components
of the two vertex functions then also differ in sign, which
implies that the two-fermion pair wave function changes sign
between pockets. Such an $s-$wave state  is often call $s^{+-}$ to
emphasize the sign change between the pockets.
This wave function  much resembles the second wave function from
$A_{1g}$ representation: $\cos k_x + \cos k_y$. It is still
$s-$wave, but it changes sign under ${\bf k} \to {\bf k} +
(\pi,\pi)$, which is precisely what is needed as hole and electron
FSs are separated by $(\pi,\pi)$.  I caution, however, that the
analogy should not be taken too far because the pairing wave
function is defined {\it only} on the two FSs, and any function
from $A_{1g}$ representation which changes sign under ${\bf k} \to
{\bf k} + (\pi,\pi)$ would work equally well.

\subsection{Beyond ladder approximation}

\subsubsection{How to get an attraction in the pairing channel?}

Having established the pairing symmetry, I now turn to the
central issue: how to get an attraction in the pairing channel?
 Let's start with the model with a
momentum-independent  (Hubbard) interaction in band basis.  For
such interaction, all $G_i$ are equal, i.e, $G_3 = G_4 = G_1 =G$.
The SDW vertex still diverges when $2G \Pi_{ph} (Q) =1$,  but $\chi^{-}_{pp} =1$ and
$\chi^{+}_{pp}$ vanishes at small $\Omega$.  This implies that, within ladder approximation, the only instability is a SDW.
This does not holds, however, beyond the ladder approximation, as I now demonstrate.  The consideration below follows Refs. \cite{maiti_rev,maiti_book}.

{\it Kohn-Luttinger consideration}\\

As the first step away from the ladder approximation, consider how KL physics works in our case. By this I mean that
 the intra-pocket interaction $G_4$ and pair-hopping $G_3$ are both equal to $G$ only if they are treated as bare  interactions. In reality,
  each of the two should be considered as irreducible interaction in the pairing channel. The irreducible interaction is the bare interaction plus all renormalizations except for the ones in the particle-particle channel. KL considerations includes such renormalizations to order $G^2$.  Below I label irreducible pairing vertices as ${\bar \Gamma}^{0}_{hh}$ and ${\bar \Gamma}^{0}_{he}$.

\begin{figure}[htp]
\includegraphics[width=1\columnwidth]{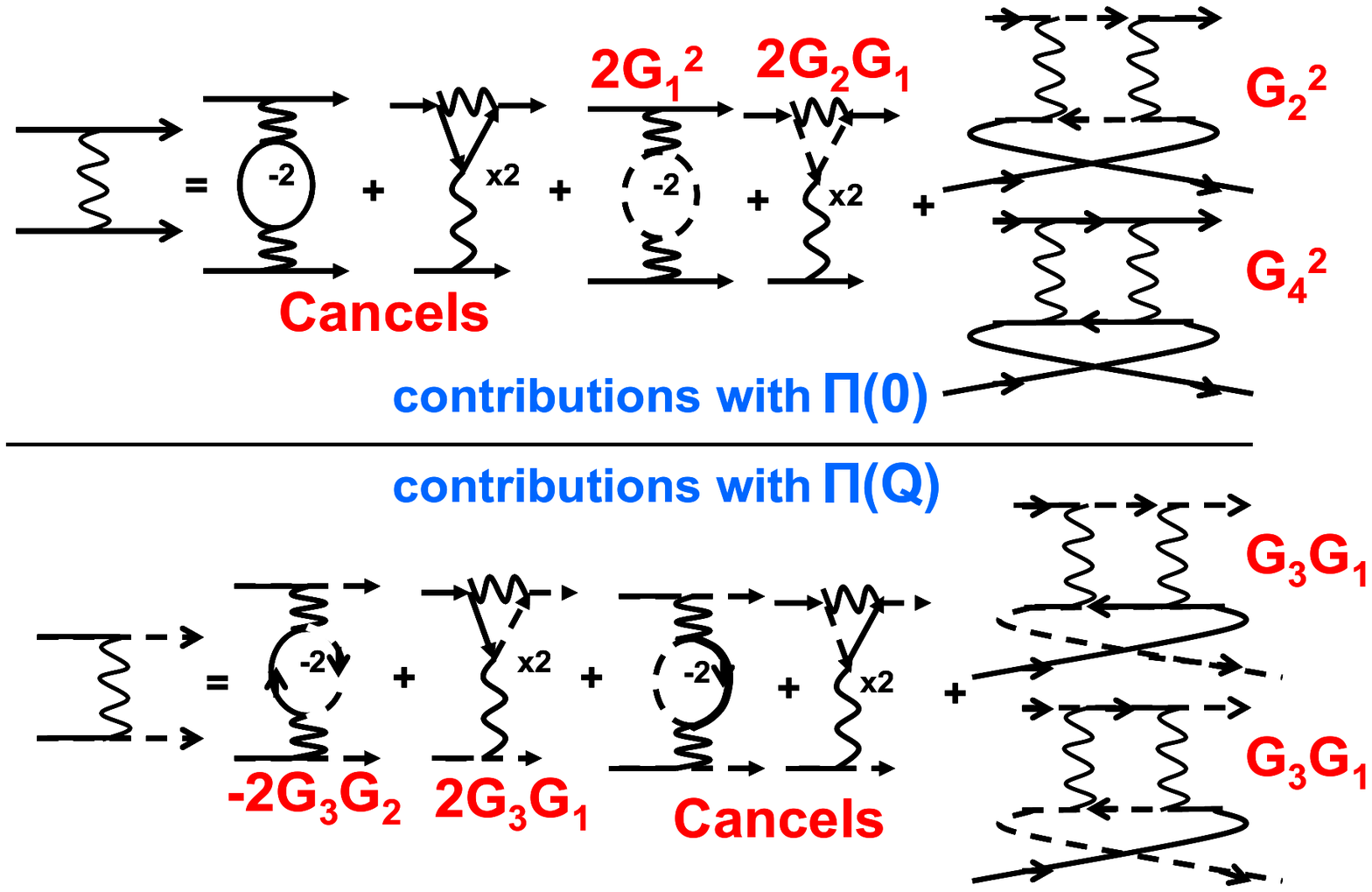}
\caption{\label{fig:Pi0} Contributions to the irreducible vertices
$\bar\Gamma_{hh}^0$(top) and $\bar\Gamma_{he}$(bottom).
$\bar\Gamma_{hh}^0$ only gets contributions form $\Pi(0)$ while
$\bar\Gamma_{he}^0$ gets contribution from $\Pi(Q)$.  From [\onlinecite{maiti_rev}].}
\end{figure}

\begin{figure}[htp]
\includegraphics[width=5in]{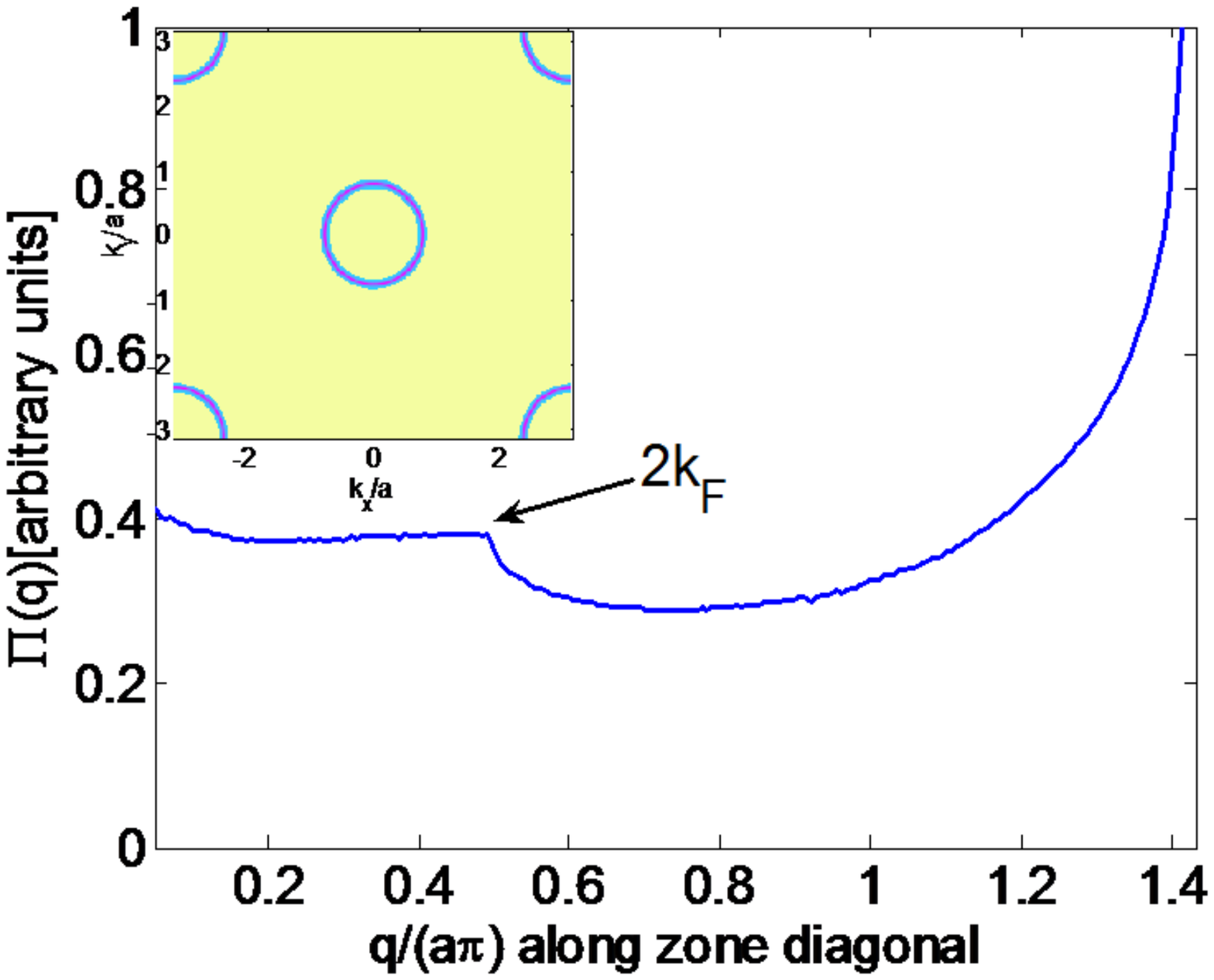}
\caption{\label{fig:PiQ}The plot of $\Pi(q)$ for a 2-pocket model
with $\vec{q}$ along the zone diagonal. When $\vec{q} <2k_F$,
$\Pi(q)$ saturates, as it is expected for a 2D system with a
circular Fermi surface. Note the $2k_F$ cusp-like feature, which
is the one-sided $2k_F$ non-analyticity of $\Pi(q)$ in 2D. At
larger $q$, $\Pi(q)$ gets larger and almost diverges at
$\vec{q}\sim \vec{Q}$ due to near-nesting. The inset shows the FS
topology for which $\Pi(q)$ has been calculated. The arcs at the
corners are parts of the electron pocket and the one in the center
is the hole pocket. From [\onlinecite{maiti_rev,maiti_book}].}
\end{figure}

The contributions to ${\bar \Gamma}^{0}_{hh}$ and
${\bar \Gamma}^{0}_{he}$ to order $G^2$ are shown in Fig \ref{fig:Pi0}. In
analytical form I have
 \bea
{\bar \Gamma}^{0}_{hh} &=& - G_4  - \left(G^2_4 + G^2_2- 2 G_1 (G_1-G_2) \right)\Pi_{ph} (0), \nonumber \\
{\bar \Gamma}^{0}_{he} &=& - G_3 - 2G_3(2 G_1 - G_2) \Pi_{ph} (Q),
\eea where, I remind,  ${\bf Q} = (\pi,\pi)$. 
For a constant $G$ this reduces
to \bea
{\bar \Gamma}^{0}_{hh} &=& - G\left(1 +  2 G\Pi_{ph} (0)\right), \nonumber \\
{\bar \Gamma}^{0}_{he} &=& - G\left(1 + 2 G \Pi_{ph} (Q) \right),
\eea 
One can show that the relation (\ref{4_2_1}) still holds if we replace  $G_3$
by $- {\bar \Gamma}^{0}_{he}$ and $G_4$ by $- {\bar
\Gamma}^{0}_{hh}$. Because
$\Gamma^{full}_{ee}=\Gamma^{full}_{hh}$, I will only deal
with $\Gamma^{full}_{hh}$ and $\Gamma^{full}_{he}$, which are given
by \bea
\Gamma^{full}_{hh} &=& \frac{1}{2} \left(\frac{ {\bar \Gamma}^{0}_{he} + {\bar \Gamma}^{0}_{hh}}{1 -({\bar \Gamma}^{0}_{he} + {\bar \Gamma}^{0}_{hh}) \Pi_{pp}} + \frac{{\bar \Gamma}^{0}_{hh} - {\bar \Gamma}^{0}_{he}}{1  - ({\bar \Gamma}^{0}_{hh}-{\bar \Gamma}^{0}_{he}) \Pi_{pp}}\right),\nonumber \\
\Gamma^{full}_{he} &=& \frac{1}{2} \left(\frac{ {\bar \Gamma}^{0}_{he} + {\bar \Gamma}^{0}_{hh}}{1 -({\bar \Gamma}^{0}_{he} + {\bar \Gamma}^{0}_{hh}) \Pi_{pp}} - \frac{{\bar \Gamma}^{0}_{hh} - {\bar \Gamma}^{0}_{he}}{1  - ({\bar \Gamma}^{0}_{hh}-{\bar \Gamma}^{0}_{he}) \Pi_{pp}}\right).\nonumber\\
&& \label{4_2_1_1} \eea
 The condition for the pairing
instability becomes $ |{\bar
\Gamma}^{0}_{he}| > -{\bar \Gamma}^{0}_{hh}$. Comparing the two irreducible vertex
functions, I find
 \bea {\bar \Gamma}^{0}_{hh} + |{\bar
\Gamma}^{0}_{he}| = 2 G^2 \left(\Pi_{ph} (Q) - \Pi_{ph} (0)\right)
\label{4_3_1} \eea i.e., the condition for the pairing is
satisfied when $\Pi_{ph} (Q) > \Pi_{ph} (0)$.  For a gas of
fermions with one circular FS, $\Pi_{ph} (q)$ either stays
constant or decreases  with $q$, and the condition $\Pi_{ph} (Q) >
\Pi_{ph} (0)$ cannot be satisfied. However, in our case, the
two FS's are separated by ${\bf Q}$, and, moreover, one FS is of hole
type, while the other is of electron type.  One can easily verify
that, in this situation, $\Pi_{ph} (Q)$ is enhanced  comparable to
$\Pi_{ph} (0)$. I present  the plot of $\Pi_{ph} (q)$ along
$q_x=q_y$ in Fig \ref{fig:PiQ}. Indeed, $\Pi_{ph} (Q)$ is much
larger than  $\Pi_{ph} (0)$.

We see  therefore that for the  renormalization of the bare interaction
into an irreducible pairing vertex  does give rise to an attraction in the $A_{1g}$ pairing channel.
 The attractive pairing interaction  is weak and at this stage is certainly smaller than the interaction in the SDW channel.
 On the other hand, the polarization bubble $\Pi_{ph} (Q)$ is in general some constant, while the polarization bubble $\Pi_{pp}$ diverges logarithmically when the total frequency $\Omega$ vanishes.

Before I proceed, a comment. Because we deal with fermions with
circular FSs located near particular $k-$points, polarization
operators at small momentum transfer and momentum transfer ${\bf
Q} =(\pi,\pi)$ can be approximated by constants.  Then the
irreducible vertex function has only an $s-$wave ($A_{1g}$)
harmonic, like the bare interaction, i.e. KL renormalization does
not generate interactions in other channels.   Treating pockets as
circular is indeed an approximation, because for square lattice
the only true requirement is that each FS is symmetric with
respect to rotations by multiples of $\pi/2$ ($C_4$ symmetry). For
small pocket sizes, deviations from circular forms are small, but
nevertheless are generally finite.  If we include this effect, we
find that the KL effect does generate interactions in other
channels ($B_{1g}, B_{2g}$, and $A_{2g}$), which may be
attractive, and also leads to more complex structure of the pair
wave function in $s^{+-}$ channel, which now acquires  angular
dependence along hole and electron pockets, consistent with $C_4$
symmetry\cite{AccNodes,Cvetkovic}

The Hubbard limit of a constant $G$ is a somewhat artificial case, however.
The actual bare interactions $G_i$ have to be extracted from the multi-orbital model
 and do depend on momentum transfer. In this situation 
$G_4-G_3$ is generally non-zero already before KL renormalization.
It is natural to expect that the bare interaction is a decreasing
function of momenta, in which case $G_4$, which is the interaction
at small momentum transfer, is larger than the interaction $G_3$
at momentum transfer near $Q$. Then the KL term has to compete
with the first-order repulsion. As long as $G \Pi_{ph} (Q)$ is
small, KL renormalization cannot overshoot bare repulsion, and the
bound state does not appear. The situation may change when we
include momentum dependence of the interaction and non-circular
nature of the pockets.  In this last case, there appears infinite
number of $A_{1g}$ harmonics, which all couple to each other, and
in some cases one or several eigenfunctions may end up being 
attractive \cite{A1g,maiti_11}. Besides, angle dependence generates
$d-$wave and $g-$wave harmonics, and some of eigenfunctions in
these channels may  also become attractive and
compete with $s-$wave~\cite{B1gA1g,maiti_11}. Still, however, in distinction to the
isotropic case, there is no guarantee that ``some" eigenfunction
from either $A_{1g}$, or $B_{1g}$, or $B_{2g}$, or $A_{2g}$, will
be attractive.  In other words, a lattice system may well remain in the normal
state down to $T=0$.\\

{\it{ RPA-type approach, spin-mediated interaction}}\\

How can we still get superconductivity in this situation?   One way to proceed is
 to apply another ladder summation scheme -- this time to series of
  renormalizations which transform a bare interaction into an irreducible particle-particle vertex.
  The leading terms in the series are KL terms, but full ladder series include infinite set of higher-order terms.
  This computational procedure is often called random-phase approximation (RPA) by  analogy with the analogous summation scheme to get a screened Coulomb interaction.
  I skip  the details of the calculations (they can be found in, e.g., ~\cite{maiti_rev,scalapino_1} and formally require $\Pi_{ph} (0) \gg \Pi_{ph} (2k_F)$ and 
  $\Pi_{ph} (Q) \gg \Pi_{ph} (Q + 2k_F)$) and present the result: ladder summation gives rise to an
   irreducible pairing vertex in the form
${\bar\Gamma}^{0}_{\alpha\beta,\gamma\delta} (k,-k;p,-p) =
\Gamma_c (k-p) \delta_{\alpha\gamma}\delta_{\beta\delta}+ \Gamma_s
(k-p)
\vec{\sigma}_{\alpha,\gamma}\cdot\vec{\sigma}_{\beta\delta}$, where for $k$ and $p$ on the same pocket
 \beq
\Gamma_c (0) = -\frac{G_4}{2}
\frac{1}{1+G_4\Pi_{ph}(0)}~~~\Gamma_s (0) = \frac{G_4}{2}
\frac{1}{1-G_4\Pi_{ph}(0)}, \eeq
 and  for $k$ and $p$ at different pockets, when $k-p \approx Q$
  \beq
\Gamma_c (Q) = -\frac{G_3}{2}
\frac{1}{1+G_3\Pi_{ph}(Q)}~~~\Gamma_s (Q) = \frac{G_3}{2}
\frac{1}{1-G_3\Pi_{ph}(Q)} \label{16_5} \eeq

Re-expressing
${\bar\Gamma}^{0}_{\alpha\beta,\gamma\delta} (k,-k;p,-p)$ in terms
of singlet and triplet components as \bea
&&{\bar\Gamma}^{0}_{\alpha\beta,\gamma\delta} (k,-k;p,-p) = \nonumber \\
&&\Gamma_{\text{s=0}} (k-p) \left(\delta_{\alpha\gamma}\delta_{\beta\delta}- \delta_{\alpha\delta}\delta_{\beta\gamma}\right) + \nonumber \\
&&
\Gamma_{\text{s=1}} (k-p) \left(\delta_{\alpha\gamma}\delta_{\beta\delta}+ \delta_{\alpha\delta}\delta_{\beta\gamma}\right). \nonumber \\
\eea
we obtain
\bea\label{eq:corr1}
\Gamma_{\text{s=0}}  &=& \frac{1}{2} \left(\Gamma_c-3\Gamma_s\right) \nonumber \\
\Gamma_{\text{s=1}}  &=& \frac{1}{2}
\left(\Gamma_c+\Gamma_s\right) \eea
i.e.
\bea
\Gamma_{\text{s=0}} (0) &=& -\frac{G_4}{4} \left(\frac{1}{1+ G_4 \Pi_{ph} (0)} + \frac{3}{1- G_4 \Pi_{ph} (0)} \right) \nonumber \\
\Gamma_{\text{s=1}} (0) &=& \frac{G_4}{4} \left(\frac{1}{1- G_4 \Pi_{ph} (0)} - \frac{1}{1+ G_4 \Pi_{ph} (0)}\right) \nonumber \\
\Gamma_{\text{s=0}} (Q) &=& -\frac{G_3}{4} \left(\frac{1}{1+ G_3 \Pi_{ph} (Q)} + \frac{3}{1- G_3 \Pi_{ph} (Q)} \right) \nonumber \\
\Gamma_{\text{s=1}} (Q) &=& \frac{G_3}{4} \left(\frac{1}{1- G_3
\Pi_{ph} (Q)} - \frac{1}{1+ G_3 \Pi_{ph} (Q)}\right) \label{16_1}
\eea Let's compare this result with what we obtained in the KL
formalism. Focus on the singlet channel and expand in (\ref{16_1})
to second order in $G_{3,4}$.
 We have
 \bea
&&\Gamma_{\text{s=0}} (0) \approx - \frac{G_4}{2} \left(1 + \frac{1}{1- G_4 \Pi_{ph} (0)}\right)\nonumber \\
&&\approx - G_4 \left(1+ 0.5 G_4\Pi_{ph} (0) \right) \nonumber \\
&&\Gamma_{\text{s=0}} (Q)\approx - \frac{G_3}{2} \left(1 + \frac{1}{1- G_3 \Pi_{ph} (Q)}\right) \nonumber \\
&&\approx - G_3 \left(1 + 0.5 G_3 \Pi_{ph} (Q) \right)
\label{16_2} \eea Apart from the factor of $1/2$ (which is the
consequence of an approximate RPA scheme) $\Gamma_{\text{s=0}}
(0)$ is the same as irreducible vertex ${\bar \Gamma}^0_{11}$,
which we obtained in KL calculation in the previous section, and
$\Gamma_{\text{s=0}} (Q)$ the same as ${\bar \Gamma}^0_{12}$  By
itself, this is not surprising, as in
 $\Gamma_{\text{s=0}}$ we included
the same particle-hole renormalization of the bare pairing
interaction as in the KL formalism.

I now look more closely at the spin-singlet components
\bea
&&\Gamma_{\text{s=0}}(0) = -\frac{1}{4} \left(\frac{G_4}{1+G_4 \Pi_{ph} (0)} + \frac{3G_4}{1-G_4 \Pi_{ph} (0)}\right) \nonumber \\
&&\Gamma_{\text{s=0}}(Q) = -\frac{1}{4} \left(\frac{G_3}{1+G_3
\Pi_{ph} (Q)} + \frac{3G_3}{1-G_4 \Pi_{ph} (Q)}\right), \eea For
repulsive interaction, the charge contribution  gets smaller
when we add higher terms in $G$ whereas spin contribution gets
larger. A conventional recipe in this situation is to neglect all renormalizations
in the charge channel and approximate ${\Gamma_{\text{s=0}}}$ with
the sum of a constant and the interaction in the spin channel. The
irreducible interaction in the $s+-$ channel  is
then
\bea &&\Gamma_{\text{s=0}}(0) - \Gamma_{\text{s=0}}(Q)=\nonumber\\
&&\frac{-G_4+G_3}{4} - \frac{3}{4} \left(\frac{G_4}{1-G_4 \Pi_{ph} (0)} -
\frac{G_3}{1-G_3 \Pi_{ph} (Q)}\right) \label{16_3} \nonumber\\
\eea

Like I said  before,  if $G_4\Pi_{ph}(0)$ and $G_3 \Pi_{ph} (Q)$ are
both small, $G_4-G_3$ term is the largest  and the pairing
interaction is repulsive for $G_4 > G_3$. However, we see that
there is a way to overcome the initial repulsion: if $G_3\Pi_{ph}
(Q)
> G_4\Pi_{ph} (0)$, one can imagine a situation when $G_3\Pi_{ph}
(Q) \approx 1$, and the correction term in (\ref{16_3}) becomes
large and positive and can overcome the negative first-order term.

What does it mean from physics perspective?  We found earlier that
 the condition
$G_3\Pi_{ph} (Q) =1$ signals an instability of a metal towards a SDW order
with momentum Q. We don't need the order to develop, but we need
SDW fluctuations to be strong and to mediate pairing interaction
between fermions. Once spin-mediated interaction  exceeds bare
repulsion, the irreducible pairing interaction in the
corresponding channel becomes attractive. Notice in this regard
that we need magnetic fluctuations to be peaked at large momentum
transfer $Q$. If they are peaked at small momenta, $\Pi_{ph} (0)$
exceeds $\Pi_{ph} (Q)$, and the interaction in the singlet channel
remains repulsive.\\

{\it{ Spin-fluctuation approach}}\\

What I  just described is the main idea of the
spin-fluctuation-mechanism of superconductivity.  The effective
pairing interaction can be obtained either within
RPA\cite{peter,Kuroki_2} or, using one of several advanced
numerical methods developed over the last decade, or just
introduced semi-phenomenologically.  The semi-phenomenological
model is called the spin-fermion model\cite{acs}.  Quite
often, interaction mediated by spin fluctuations also critically
affects single-fermion propagator (the Green's function), and this
renormalization has to be included into the pairing problem. As
another complication, the interaction mediated by soft spin
fluctuations has a strong dynamical part due to Landau damping --
the decay of a spin fluctuation into a particle-hole pair.  This
dynamics also has to be included into consideration, which makes
the solution of the pairing problem near a magnetic instability
quite involved theoretical problem.

There are two crucial aspects of the spin-fluctuation approach~\cite{acs,maiti_book}.
First, magnetic fluctuations have to develop at energies much
larger than the ones relevant for the pairing, typically at
energies comparable to the bandwidth $W$. It is crucial for
spin-fluctuation approach that SDW magnetism is the {\it only}
instability which develops at such high energies. There may be
other instabilities (e.g., charge order), but the assumption is
that they develop at small enough energies  and can be captured
within the low-energy  model with spin fluctuations already
present\cite{Efetov3,Senthil,wang}. Second, spin-fluctuation approach
is fundamentally not a weak coupling approach. In the absence of
nesting, $\Pi_{ph} (Q)$ and $\Pi_{ph} (0)$ are generally of order
$1/W$, and $\Pi_{ph} (Q)$ is only larger numerically. Then the
interaction $G_{3}$ must
be of order $W$ in order to get a strong
magnetically-mediated component of the pairing interaction,

One way to proceed in this situation is to introduce the
spin-fermion model with static magnetic fluctuations built into
it, and then assume that within this model the interaction between
{\it low-energy} fermions ${\bar g}$ is smaller than $W$ and do
controlled low-energy analysis treating $\bar{g}/W$ as a small
parameter\cite{acs,Efetov3,Senthil}. There are several ways to
make the assumptions ${\bar g} \ll W$ and $G \sim W$ consistent
with each other, e.g., if microscopic interaction has length
$\Gamma_0$ and $\Gamma_0k_F/\hbar \gg 1$, then ${\bar g}$ is small
in $1/(\frac{\Gamma_0k_F}{\hbar})$ compared to $G$
(Refs.\cite{cm,Dzero}). At the same time, the properties of the
spin-fermion model do not seem to crucially depend on ${\bar g}/W$
ratio, so the hope is that, even if the actual ${\bar g}$ is of
order $W$, the analysis based on expansion in ${\bar g}/W$
captures the essential physics of the pairing system behavior near
a SDW instability in a metal. \\

\section{Interplay between SDW magnetism and superconductivity, parquet RG approach}
\label{sec:pRG}

I now return to weak coupling, where I have control over calculations, and
 ask the question whether one can still
 get an attraction in at least one pairing channel despite that $G_4 > G_3$, i.e., the bare
pairing interaction is repulsive in all channels.  The answer is,
actually, yes, it is possible, but under a special condition that
$\Pi_{ph} (Q)$ is singular and diverges logarithmically at zero
frequency or zero temperature, in the same way as the
particle-particle bubble $\Pi_{pp} (0)$.  This condition is
satisfied exactly when there is a perfect nesting between
fermionic excitations separated by $Q$.   For Fe-pnictides, it implies that hole and
electron FSs perfectly match each other when one is shifted by
$Q$.

I  show below that $\Pi_{ph} (Q)$ and $\Pi_{pp} (0)$ do have
exactly the same logarithmic singularity at perfect nesting. At
the moment, let's take this for granted and compare the relevant
scales. First, no fluctuations develop  at energies/temperatures
of order $W$ because at such high scales the logarithmical
behavior of $\Pi_{pp}$ and $\Pi_{ph}$ is not yet developed and
both bubbles scale as $1/W$.  At weak coupling $G/W <<1$, hence
corrections to bare vertices are small at these energies. Second, we know that the
pairing vertex evolves at $(G_3-G_4)\Pi_{pp} (0) \sim O(1)$, and
that corrections to the bare irreducible pairing vertex become of
order one when $G_3\Pi_{ph} (Q) \sim O(1)$.  But we also know
from,  e.g., (\ref{16_5}) that at the same scale the SDW vertex
begins to evolve. Moreover other inter-pocket interactions, which
we didn't include so far: density-density and exchange
interactions (which here and below we label  as $G_1$ and $G_2$,
respectively) also start evolving because their renormalization
involves terms $G_1 \Pi_{ph} (Q)$ and $G_2\Pi_{ph} (Q)$,  which also
become of $O(1)$, provided that all bare interactions are of the same order.
Once $G_{1,2} \Pi_{ph} (Q)$ becomes of order one, the
renormalization of $G_3$ by $G_1$ and $G_2$ interactions also
becomes relevant. The bottom line here is that renormalization of
all interactions become relevant at the same scale where
$G_i \Pi_{ph} (Q) \sim G_i \Pi_{pp} (0) \sim 1$. 
 At this scale
we can expect superconductivity, if the corrections to $G_4-G_3$
overcome the sign of the pairing interaction, and we also 
we can expect an instability towards SDW and, possibly, towards some other
order. The issue then is whether it is possible to construct a
rigorous description of the system behavior in the situation when
all couplings are small compared to $W$, but $G_i \Pi_{ph} (Q)$
and $G_i\Pi_{pp} (0)$ are of order one. The answer is yes, and the
corresponding procedure is called a parquet renormalization group (pRG).

The pRG is a controlled weak coupling approach.  It assumes that no correlations develop at energies
comparable to the bandwidth, but that there are several competing
orders whose fluctuations develop simultaneously at smaller energies. Superconductivity is one of them, others include SDW and
potential charge-density-wave (CDW), nematic and other orders. The
pRG approach treats superconductivity, SDW, CDW and other
potential instabilities on equal  footings. Correlations in each
channel grow up with similar speed, and fluctuations in one
channel affect the fluctuations in the other channel and vise
versa. For superconductivity, once the corrections to the pairing
vertex become of order one, and there is a potential to convert
initial repulsion into an attraction. We know that second-order
contribution to the pairing vertex from SDW channel works in the
right direction, and one may expect that higher-order corrections
continue pushing the pairing interaction towards an attraction.
However even if attraction develops, there is no guarantee that
the system will actually undergo a SC transition because it is
entire possible that SDW instability comes before SC instability.

The pRG approach addresses both of these issues. It can be also
applied to a more realistic case of non-perfect nesting if
deviations from nesting are small in the sense that there exists a
wide range of energies where $\Pi_{ph} (Q)$ and $\Pi_{pp} (0)$ are
approximately equal.  Below some energy scale, $\omega_0$, the
logarithmical singularity in $\Pi_{ph} (Q)$ is cut.  If this scale
is smaller than the one at which the leading instability occurs, a
deviation from a perfect nesting is an irrelevant perturbation. If
it is larger, then pRG runs up to $\omega_0$, and at smaller
energies only SC channel continues to evolve in BCS fashion.

There also exists a well-developed numerical computational
procedure called functional RG (fRG)\cite{frg1,frg2,dhl}. Its
advantage is that it is not restricted to a small number of
patches and captures the evolution of the interactions in various
channels  even if the interactions depend on the angles along the
FS. The ``price" one has to pay is the reduction in the control
over calculations -- fRG includes both leading and subleading
logarithmical terms. If only logarithmical terms are left, the
angle dependencies of the interactions do not evolve in the process of RG flow,
 only the overall magnitude changes\cite{RG_SM} So far, the
results of fRG and pRG analysis for various systems fully agree.
Below I focus on the pRG approach. For the thorough tutorial on the
RG technique, see Ref. \cite{shankar}. In the discussion below and in Sec. 8.5 I follow Refs. [\onlinecite{maiti_rev,maiti_book}].

\subsection{Parquet Renormalization Group: The Basics}

 I recall that in Fe-pnictides a bubble with
momentum transfer $Q$ contains one hole (c) and one electron (f)
propagator, and at perfect nesting the dispersions of holes and
electrons are just opposite, $\varepsilon_c (k) =-\varepsilon_f
(k+Q)$.  The particle-hole and particle-particle bubbles are

\bea \label{eq:bubbles} \Pi_{pp}(0)  &=& -i \int \frac{d^2
k\;d\omega}{(2\pi\hbar)^3}
G^c(k,\omega)G^c(-k,-\omega) \nonumber\\
\Pi_{ph}(Q) &=&i \int \frac{d^2 k\;d\omega}{(2\pi\hbar)^3}
G^c(k,\omega)G^f(Q+k,\omega),
 \eea where

$G^{c,f} = \frac{1}{\omega - \varepsilon^{c,f}_k+i\delta
sgn(\omega)}$. Substituting into Eq. \ref{eq:bubbles} and using
$\varepsilon_c (k) =-\varepsilon_f (k+Q)$ one can easily make sure
that the two expressions in Eq. \ref{eq:bubbles} are identical.
Evaluating the integrals we obtain \beq \Pi_{pp}(0) = \Pi_{ph}(Q)
= N_0 L +... \eeq where $N_0 = m/2\pi\hbar^2$ is the 2D density of
states, \beq L = \frac{1}{2} \log \left( \frac{W}{E} \right), \eeq
$E$ is a typical energy of an external fermion, and the dots stand for
non-logarithmic terms.  The factor $1/2$ is specific to the pocket
model and accounts for the fact that for small pocket sizes, the
logarithm comes from integration  over positive energies $W > E >
E_F$. At non-perfect nesting, the particle-particle channel is
still logarithmic, but the particle-hole channel gets cut by the
energy difference ($\delta E$) associated with the nesting
mismatch, such that

\beq \Pi_{ph}(Q) = N_0 \log \frac{W}{\sqrt{E^2 + \delta E^2}} \eeq

The main idea of pRG (as of any RG procedure) is to consider $E$
as a running variable, assume that initial $E$ is comparable to
$W$ and $G_i \log \left( \frac{W}{E} \right) = G_i L$ is small,
calculate the renormalizations of all couplings by fermions  with
energies larger than $E$, and find how the couplings evolve as $E$
approaches the region where $G_i L = O(1)$.

This procedure can be carried out already in BCS theory, because
Cooper renormalizations are logarithmical. For an isotropic
system, the evolution of the interaction $U_l$ in a channel with
angular momentum $l$ due to Cooper renormalization can be
expressed in RG treatment as 
 an equation for the running coupling $U_l (L)$
\beq \frac{d U_l (L)}{dL} = - N_0
\left(U_l(L)\right)^2. \label{17_1} \eeq
  The solution of (\ref{17_1}) is
\beq U_l (L)  = \frac{U_l}{1 + U_l N_0 L}. \eeq
  Similar formulas can be
obtained in lattice systems when there are no competing
instabilities, i.e., only renormalizations in the pairing channel
are relevant. For example, in the two-pocket model for the
pnictides, the equations for the vertices $\Gamma_{hh} (L)
= -G_4 (L)$ and $\Gamma_{he} (L) = - G_3 (L)$, Eqs.
(\ref{4_2_1}),  can be reproduced by solving the two coupled RG
equations \bea
&&\frac{d G_3 (L)}{dL} = -2 N_0 G_3 (L) G_4 (L) \nonumber \\
&&\frac{d G_4 (L)}{dL} = -N_0 \left(\left(G_3 (L)\right)^2 +
\left(G_4 (L)\right)^2\right) \label{17_2} \eea with boundary
conditions $G_4 (L=0) = G_4$, $G_3 (L=0) = G_3$. The
set can be factorized by introducing $G_A (L) = G_3 (L) +
G_4 (L)$ and $G_B (L) = G_4 (L) - G_3 (L)$
 to
\beq \frac{d G_A (L)}{dL} = -N_0 \left(G_A (L)\right)^2,
~~\frac{d G_B (L)}{dL} = -N_0 \left(G_B (L)\right)^2 \eeq
The solution of the set yields \bea
&&G_A (L) = G_4 (L) + G_3 (L) = \frac{G_3 + G_4}{1 + N_0 L (G_3 + G_4)} \nonumber \\
&&G_B (L) = G_4 (L) - G_3 (L) = \frac{G_4-G_3}{1 + N_0 L
(G_4 - G_3)} \eea Solving this set and using $\Gamma_{hh} (L) =
-G_4 (L)$, $\Gamma_{he} (L) = -G_3 (L)$, we reproduce
(\ref{4_2_1}). This returns us to the same issue as we had before,
namely if $G_4 > G_3$, the fully renormalized pairing interaction
does not diverge at any $L$ and in fact decays as $L$ increases:
$G_4 (L)$ decays as $1/L$ and $G_3 (L)$ decays even faster,
as $1/L^2$.

I now consider how things change when $\Pi_{ph} (Q)$ is also
logarithmical and the renormalizations in the particle-hole
channel have to be included on equal footings with
renormalizations in the particle-particle channel.

\subsection{pRG in a 2-pocket model}

Because two types of renormalizations are relevant, we need to
include into consideration all vertices with either small total
momentum or with momentum transfer near $Q$ i.e., use the full
low-energy Hamiltonian of Eq. (\ref{eq:2-poc Ham}). There are
couplings $G_3$ and $G_4$ which are directly relevant for
superconductivity, and also the couplings $G_1$ and $G_2$ for
density-density and exchange interaction between hole and electron
pockets, respectively. These are shown in Fig. \ref{fig:int}.

The strategy to obtain one-loop pRG equations, suitable to our
case, is the following: One has to start with perturbation theory
and obtain the variation of each full vertex $\delta G_i$ to order
$G_iG_j L$. Then one has to replace $\delta G_i/L$ by $d
G_i (L)/dL$ and  also replace $G_i G_j$ in the r.h.s. by
$G_i (L) G_j (L)$. The result is the set of coupled
differential equations  for $d G_i (L)/dL$ whose right sides
are given by bilinear combinations of  $G_i (L) G_j (L)$.
The procedure may look a bit formal, but one can rigorously prove
that it is equivalent to summing up series of corrections to $G_i$
in powers of $G_i L$, neglecting corrections terms with higher
powers of $G_i$ than of $L$. One can go further  and collecting
correction terms of order $G_i G_j G_k  L$. This is called 2-loop
order, and 2-loop terms give contributions of order $(G (L))^3$
to the right side of the equations for $d G_i (L)/dL$.  2-loop
calculations are, however, quite involved\cite{brazil} and have not been re-checked.
Below I only consider 1-loop pRG equations.

\begin{figure*}[htp]
$\begin{array}{cc}
\includegraphics[width = 0.45\columnwidth]{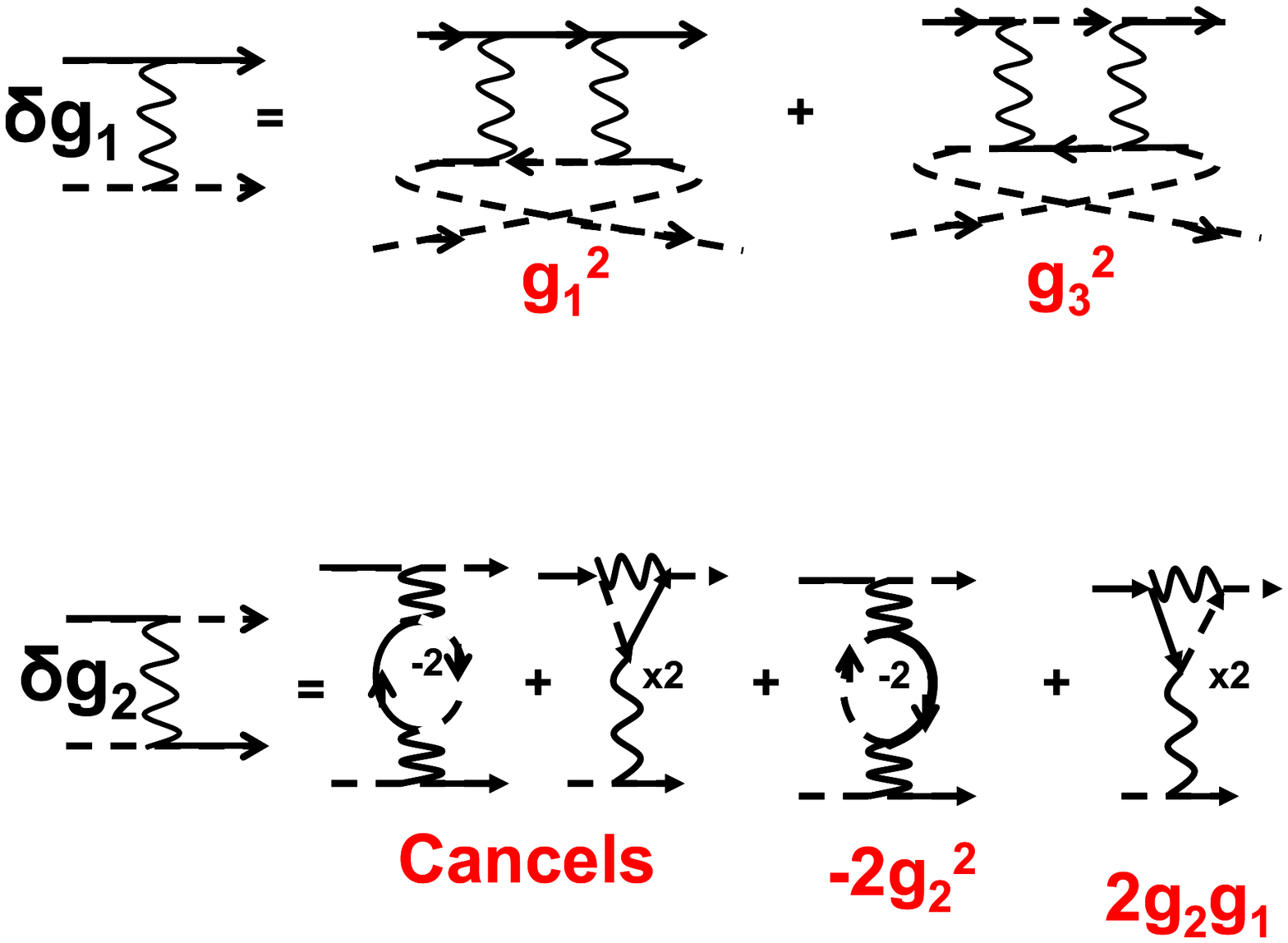}&
~~~~~~\includegraphics[width = 0.45\columnwidth]{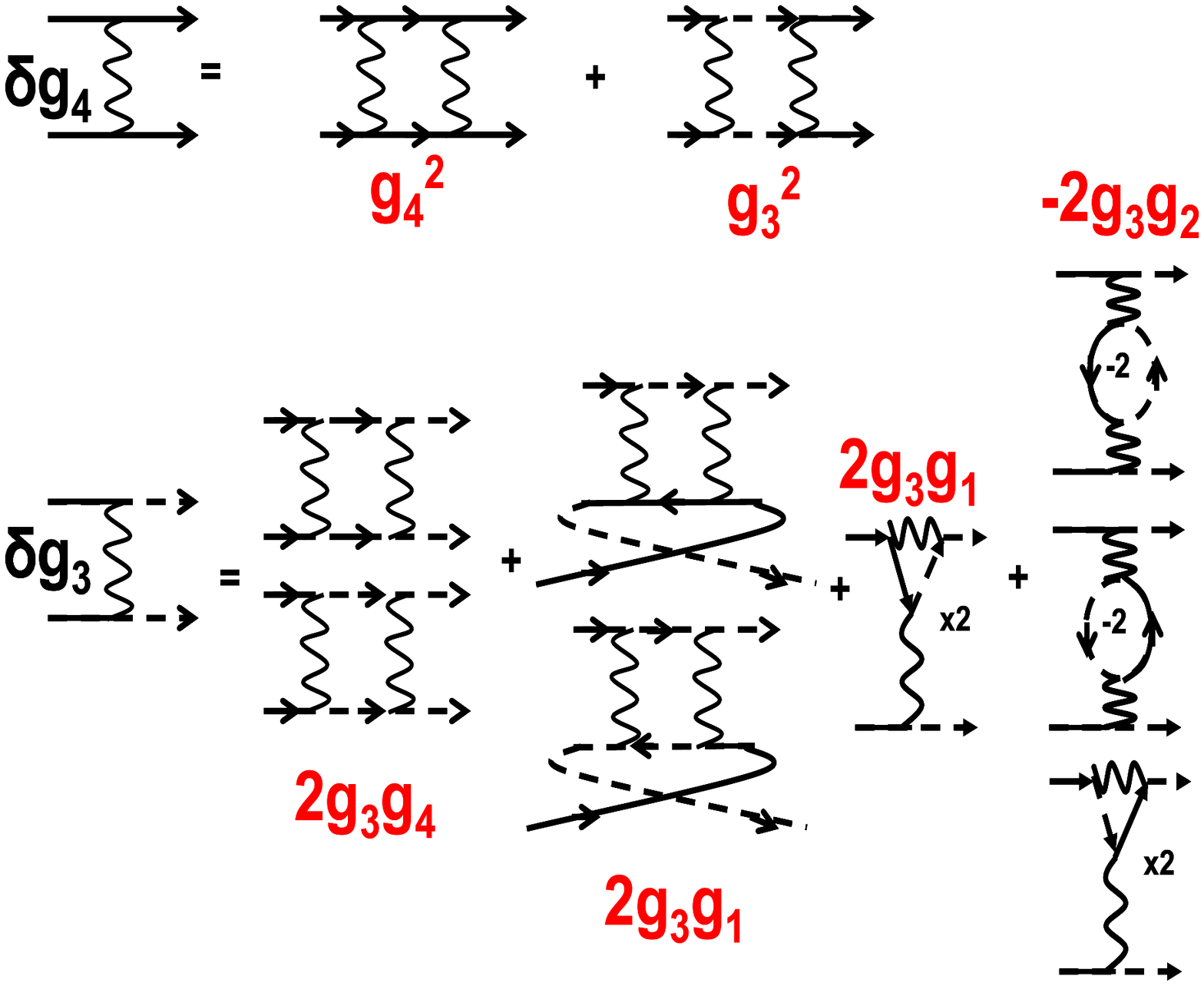}
\end{array}$
\caption{\label{fig:g_3g_4}The pRG diagrams to one loop order, which
contribute to the parquet flow of  $g_1$, $g_2$, $g_3$ and $g_4$
vertices. From [\onlinecite{maiti_rev,maiti_book}].}
\end{figure*}

The $G^2$ corrections to all four couplings are shown in
Fig.\ref{fig:g_3g_4}. Evaluating the integrals and following the
recipe we obtain
\begin{eqnarray}\label{eq:RG_2poc}
&&\dot{g}_1 = g_1^2 + g_3^2\nonumber \\
&&\dot{g}_2 = 2  g_2(g_1 - g_2 ) \nonumber \\
&&\dot{g}_3 =  2 g_3(2 g_1 - g_2 - g_4)\nonumber \\
&&\dot{g}_4 = -g_3^2 - g_4^2 \nonumber \\
\end{eqnarray}
where we introduced $g_i \equiv g_i (L) = G_i (L) N_0$ and $\dot{g}_i =
dg_i/dL$

\begin{figure}[t]
\includegraphics[width = 1\columnwidth]{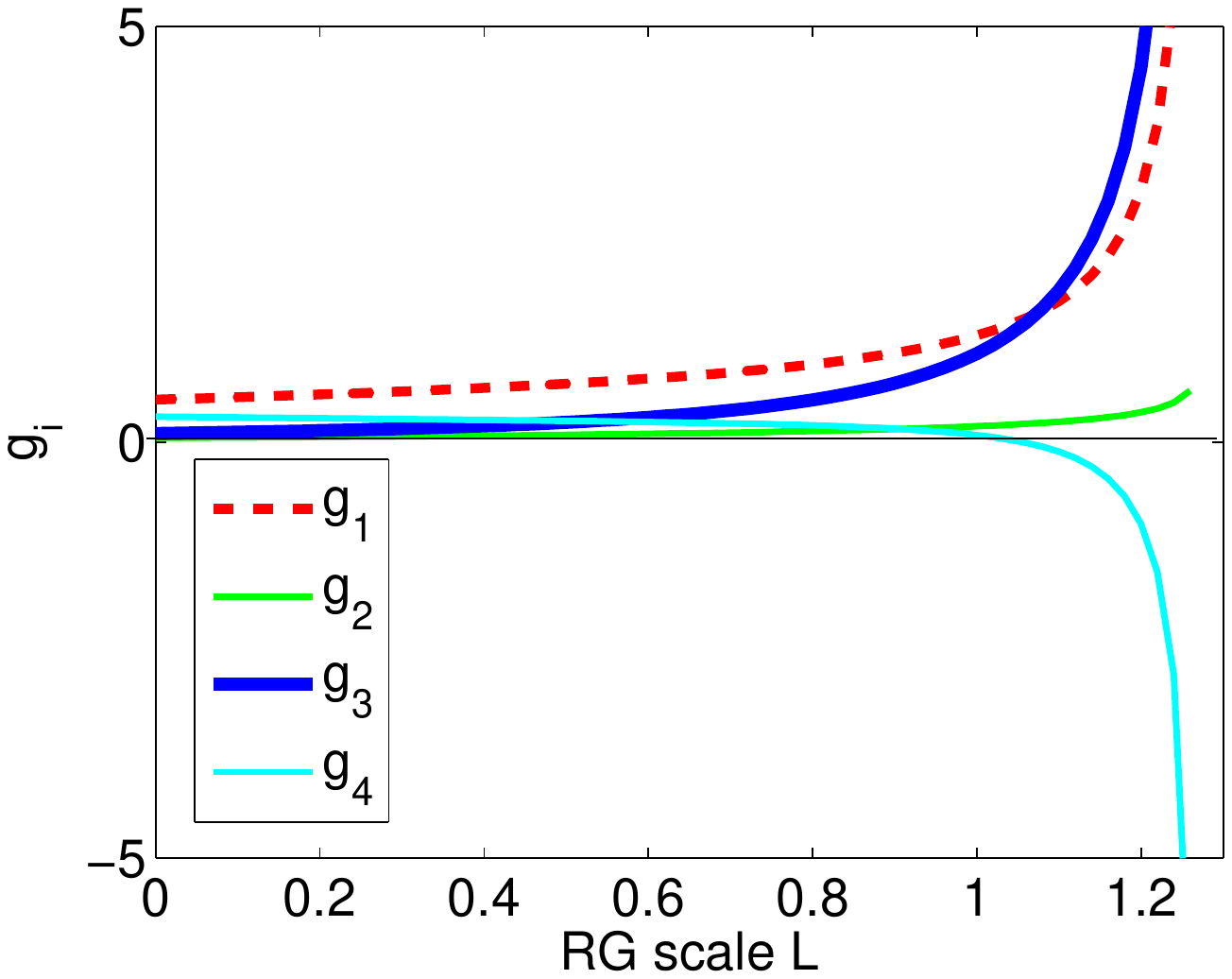}
\caption{\label{fig:Fe_flow} The flow of dimensionless couplings
$g_{1,2,3,4}$. $g_3$ grows and eventually crosses $g_4$, which
becomes negative at a large enough RG scale. From [\onlinecite{cee}].}
\end{figure}

We note that the renormalizations of $g_4$ are still only in the
Cooper channel and causes $g_4$ to reduce. But for $g_3$ we now
have a counter-term from $g_1$, which pushes $g_3$ up. And the
$g_1$ term is in turn pushed up by $g_3$. Thus already at this
stage one can qualitatively expect $g_3$ to eventually get larger.
Fig \ref{fig:Fe_flow} shows the solution of (\ref{eq:RG_2poc})--
the flow of the four couplings for this model. We see that, even
if $g_3$ is initially smaller than $g_4$, it flows up with
increasing $L$, while $g_4$ flows to smaller values. At some
$L =L_0$, $g_3$ crosses $g_4$, and at larger $L$ the pairing
interaction $g_4-g_3$ becomes negative (i.e., attractive). In
other words, in the process of pRG flow, the system self-generates
attractive pairing interaction. I remind that the attraction
appears in the $s^{+-}$ channel. The pairing interaction in
$s^{++}$ channel: $g_3 + g_4$ remains positive (repulsive) despite
that $g_4$ eventually changes sign and becomes negative. It is
essential that for $L \sim L_0$ the renormalized $g_i$ are still
of the same order as bare couplings, i.e., are still small, and
the calculations are fully under control. In other words, the sign
change of the pairing interaction is a solid result, and
higher-loop corrections may only slightly shift the value of $L_0$
when it happens.

At some larger $L= L_c$, the couplings diverge,  signaling the
instability towards an ordered state (which one I discuss later).
One-loop pRG is valid "almost" all the way to the instability, up
to $L_c - L \sim O(1)$, when the renormalized $g_i$ become of
order one.  At smaller distances from $L_c$ higher-loop
corrections become relevant. It is very unlikely, however, that
these corrections will change the physics in any significant way.

The sign change of the pairing interaction can be detected also if
the nesting is not perfect and $\Pi_{ph} (Q)$ does not behave
exactly in the same way as $\Pi_{pp} (0)$.  The full treatment of
this case is quite involved. For illustrative purposes I  follow
the approach first proposed in Ref.\cite{Furukawa} and measure the
non-equivalence between $\Pi_{pp} (0)$ and $\Pi_{ph} (Q)$ by
introducing a phenomenological parameter $d_1 = \Pi_{ph}
(Q)/\Pi_{pp} (0)$ and treat $d_1$ as an $L-$ independent constant
$0<d_1<1$, independent on $L$.  This is indeed an approximation,
but it is at least partly justified by our earlier observation
that the most relevant effect for the pairing is the sign change
of $g_4-g_3$ at some scale $L_0$, and around this scale $d_1$ is
not expected to have strong dependence on $L$.  The case $d_1 =1$
corresponds to perfect nesting, and the case $d_1 =0$ implies that
particle-hole channel is irrelevant,  in which case, I remind,
$g_4-g_3$ remains positive for all $L$.

\begin{figure*}[t]
$\begin{array}{ccc}
\includegraphics[width = 2.3in]{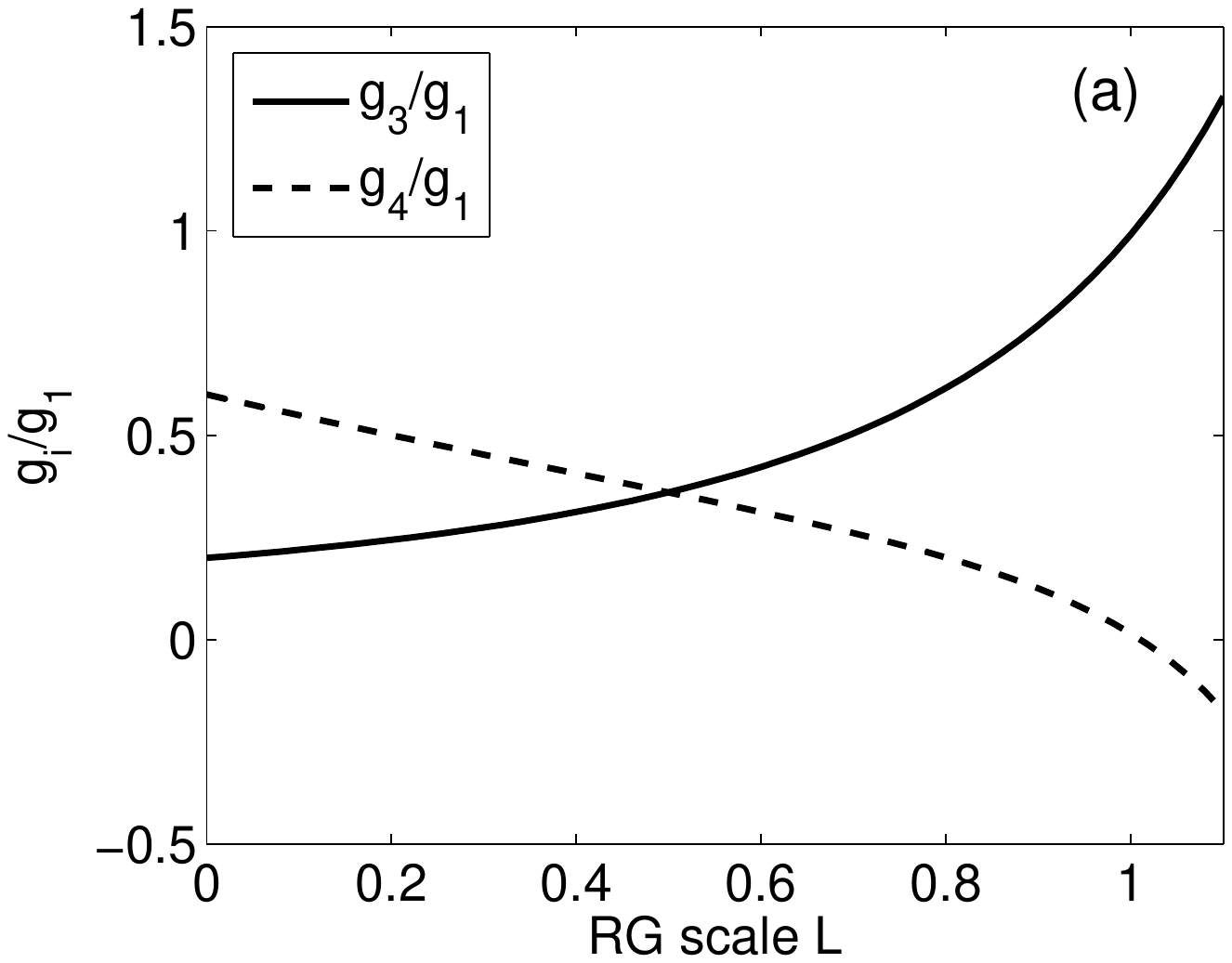}&
\includegraphics[width = 2.3in]{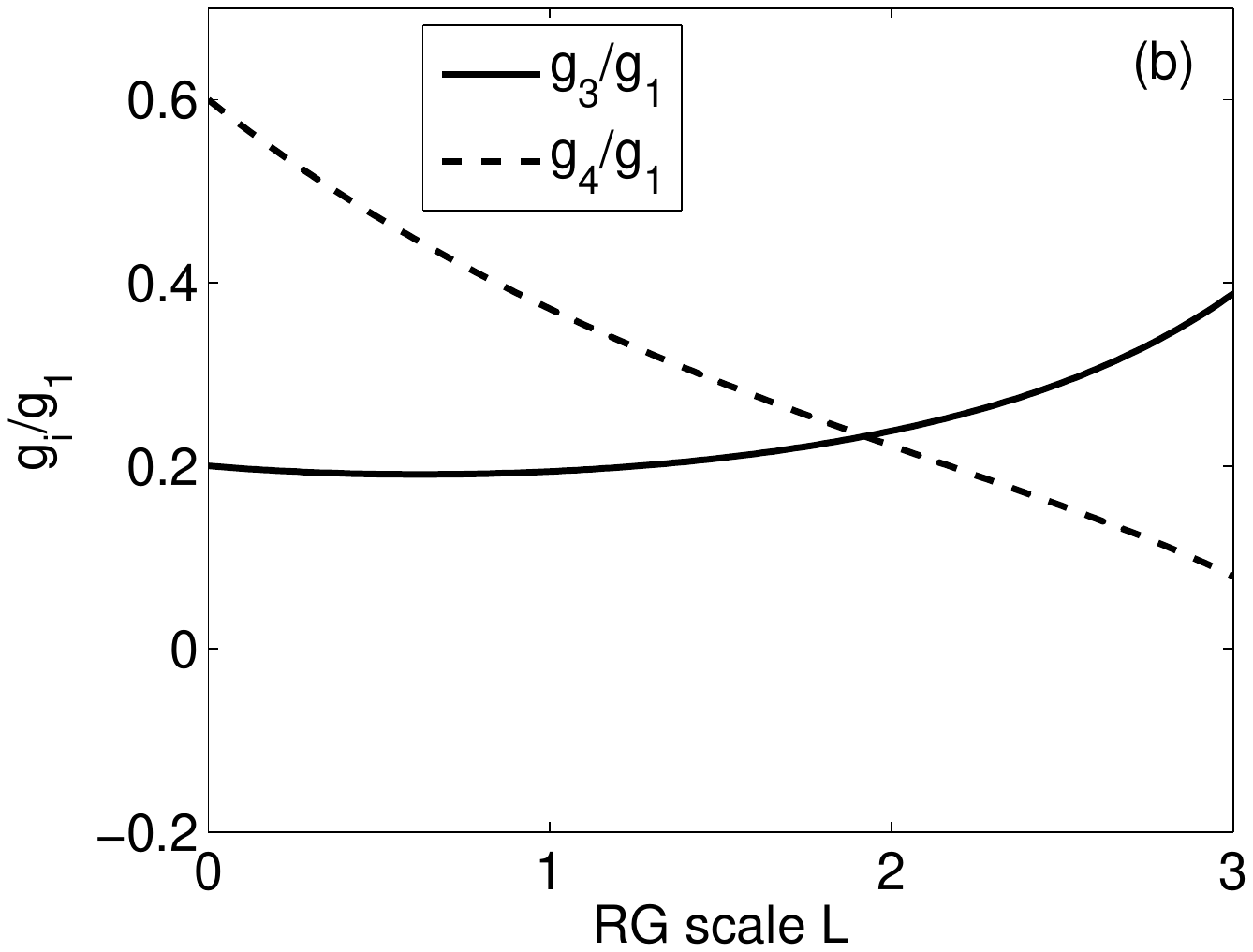}&
\includegraphics[width = 2.3in]{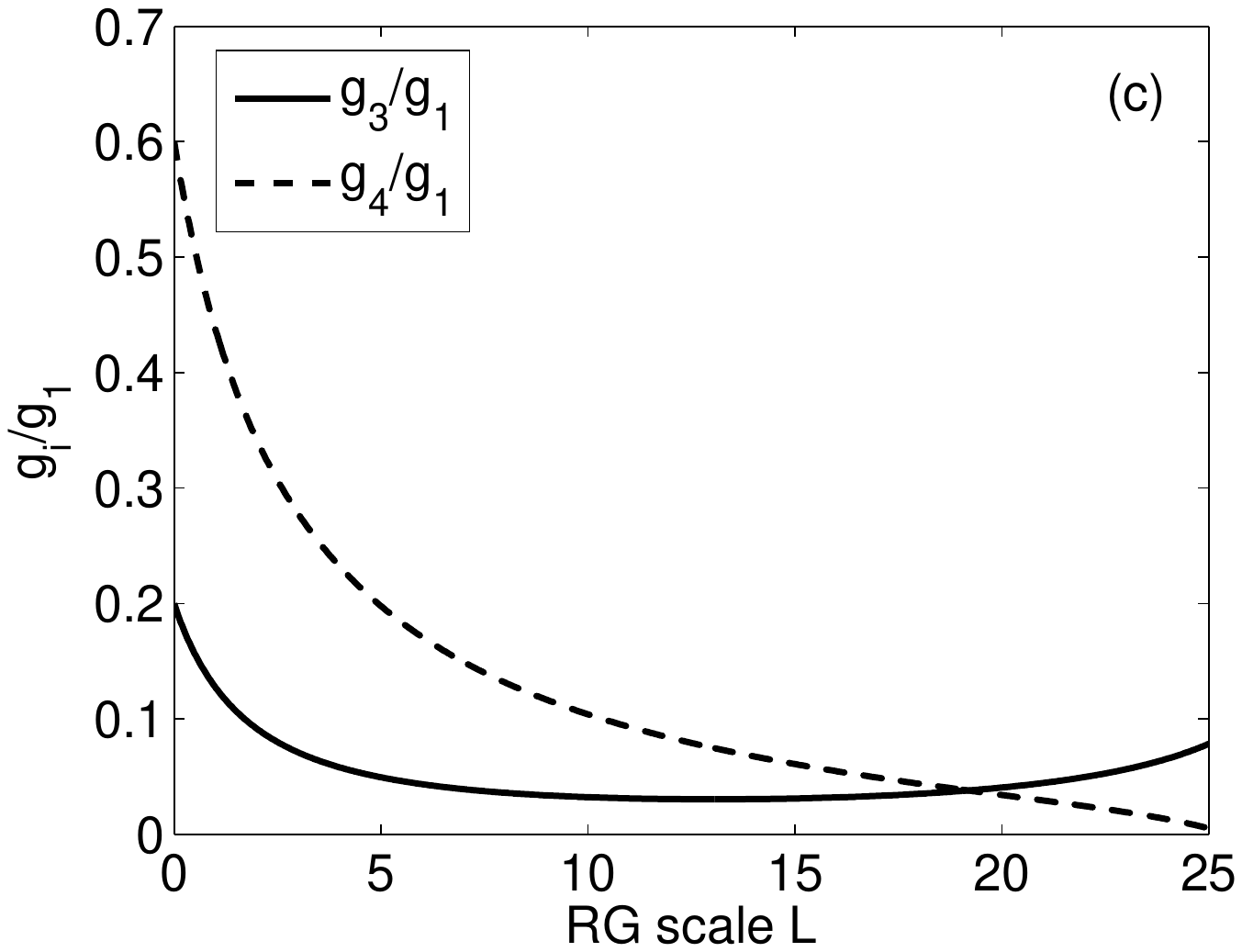}
\end{array}$
\caption{\label{fig:Fe_flows} The flow of ratio of couplings
$g_3/g_1$ and $g_4/g_1$ for different nesting parameters
$d_1=1$(a), $d_1=0.3$(b),$d_1=0.05$(c). All cases are
qualitatively similar in that $g_3/g_1$ eventually crosses
$g_4/g_1$. The smaller is the nesting parameter, the `later' is
this crossing. If $d_1=0$, this crossing will never happen and
$g_4>g_3$ for all $L$.}
\end{figure*}

The pRG equations for arbitrary $d_1$ are straightforwardly
obtained using the same strategy as in the derivation of
(\ref{eq:RG_2poc}), and the result is~\cite{cee,rahul,podolsky}
\begin{eqnarray}\label{eq:RG_2poc_1}
&&\dot{g}_1 = d_1(g_1^2 + g_3^2)\nonumber \\
&&\dot{g}_2 = 2 d_1 g_2(g_1 - g_2 ) \nonumber \\
&&\dot{g}_3 =  2 d_1g_3(2 g_1 - g_2)- 2g_3 g_4\nonumber \\
&&\dot{g}_4 = -g^2_3 - g^2_4 \nonumber \\
\end{eqnarray}
In Fig \ref{fig:Fe_flows} I show the behavior of the couplings
for representative $0<d_1<1$. Like before, I take bare value of
$g_4$ to be larger than the bare $g_3$, i.e., at high energies the
pairing interaction is repulsive. This figure and analytical
consideration shows that for {\it any} non-zero $d_1$ the behavior
is qualitatively the same as for perfect nesting, i.e., at some
$L_0 < L_c$  the running couplings $g_3$ and $g_4$ cross, and for
larger $L$ (smaller energies) pairing interaction in $s^{+-}$
channel becomes attractive. The only effect of making $d_1$
smaller is the increase in the value of $L_0$. Still, for
sufficiently small bare couplings, the range where the pairing
interaction changes sign is fully under control in one-loop pRG
theory.

A way to see analytically that $g_3-g_4$ changes sign and becomes
positive is to consider the system behavior near $L = L_c$ and
make sure that in this region $g_3 > g_4$. One can easily make
sure that all couplings diverge at $L_c$, and their ratios tend to
some constant values (see discussion around Eq. (\ref{eq:NL1})
below for more detail). Introducing $g_2 = a g_1, g_3= b g_1$, and
$g_4 = c g_1$, and substituting into (\ref{eq:RG_2poc_1}) we find
an algebraic set of equations for $a, b$, and $c$. Solving the
set, we find that
$b=\frac{\sqrt{\sqrt{16d_1^4-4d_1^2+4}+2-d_1^2}}{d_1}$and $c =
\frac{d_1}{2}(3-b^2)$. The negative sign of $c$ and positive sign
of $b$, combined with the fact that $g_1$ definitely increases
under the flow and surely remains positive,  imply that near
$L_c$, $g_4$ is negative, while $g_3$ is positive (this is also
evident from the Fig \ref{fig:Fe_flows}). Obviously then, $g_3$
and $g_4$ must cross at some $L_0 < L_c$.

The reason for the sign change of the pairing interaction is clear
from the structure of the pRG equation for $g_3$ the r.h.s. of
which contains the term $4 d_1 g_3 g_4$, which pushes $g_3$ up. We
know from second-order KL calculation that the upward
renormalization of $g_3$ comes from the magnetic channel and can
be roughly viewed as the contribution from spin-mediated part of
effective fermion-fermion interaction.  Not surprisingly, we will
see below that $g_1$ does, indeed, contribute to the SDW vertex.
From this perspective, the physics of the attraction in pRG (or in
fRG, which brings in the same conclusions as pRG) and in
spin-fermion model is the same:  magnetic fluctuations push
inter-pocket/inter-patch interaction up, and below some energy
scale the renormalized inter-pocket/inter-patch interaction
becomes larger than repulsive intra-pocket/intra-patch
interaction.

There is, however, one important difference between the RG
description and the description in terms of spin-fermion model. In
the spin-fermion model, magnetic fluctuations are strong, but the
system is {\it assumed} to be at
 some distance away from an SDW
instability. In this situation, SC instability definitely comes
ahead of SDW magnetism. There may be other instabilities produced
by strong spin fluctuations, like 
CDW\cite{Max,efetov,Physics,wang,sachdev_new,pdw}, which compete with SC and, by
construction, also occur before SDW order sets in.

In RG treatment (pRG or fRG), SDW magnetism and SC instability
(and other potential instabilities) compete with each other, and
which one develops first needs to be analyzed.  So far, we only
found that SC vertex changes sign and becomes attractive. But we
do not know whether superconductivity is the leading instability,
or some other instability comes first. This is what we will study
next. The key issue, indeed, is whether superconductivity can come
ahead of SDW magnetism, whose fluctuations helped convert
repulsion in the pairing channel into an attraction.

\section{Competition between density wave orders and superconductivity}

Thus far, we identified an instability in a particular channel
with the appearance of a pole in the upper frequency half-plane in
the corresponding vertex -- the vertex with zero total momentum in
the case of SC instability, and the vertex with the
 total momentum $Q$ in the case of SDW instability.
Since our goal is to address the competition between these states,
it is actually advantageous to use a slightly different approach:
introduce all potentially relevant fluctuating fields, use them to
decouple 4-fermion terms into a set of terms containing two
fermions and a fluctuating field, compute the renormalization of
these ``three-legged" vertices and use these renormalized vertices
to obtain the susceptibilities in various channels and check which
one is the strongest.  We will see that the renormalized vertices
in different channels (most notably, SDW and SC) do diverge near
$L_c$, but with different exponents. The leading instability will
be in the channel for which the exponent is the largest. There is
one caveat in this approach --- for a divergence of the
susceptibility the exponent for the vertex should be larger than
$1/2$ (Ref.\cite{Cvetkovic}), but we will see below that this
condition is satisfied, at
least for the leading instability.

\subsection{Two pocket model}

\begin{figure}[t]
\includegraphics[width=1\columnwidth]{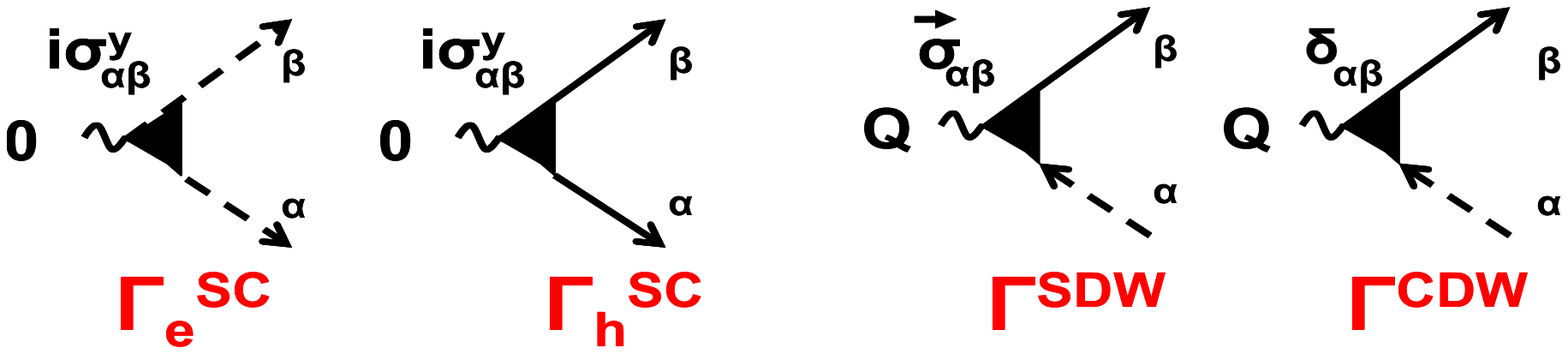}
\caption{\label{fig:new_vertex} Superconducting and density-wave
three-leg vertices. Divergence of any of these vertices
indicates that the system is likely to be unstable to the
corresponding order. $\Gamma^{SC}_{h,e}$ are superconducting
vertices, $\Gamma^{SDW}$ is SDW vertex and $\Gamma^{CDW}$ is CDW
vertex. From [\onlinecite{cee}].}
\end{figure}

\begin{figure}[t]
\includegraphics[width=1\columnwidth]{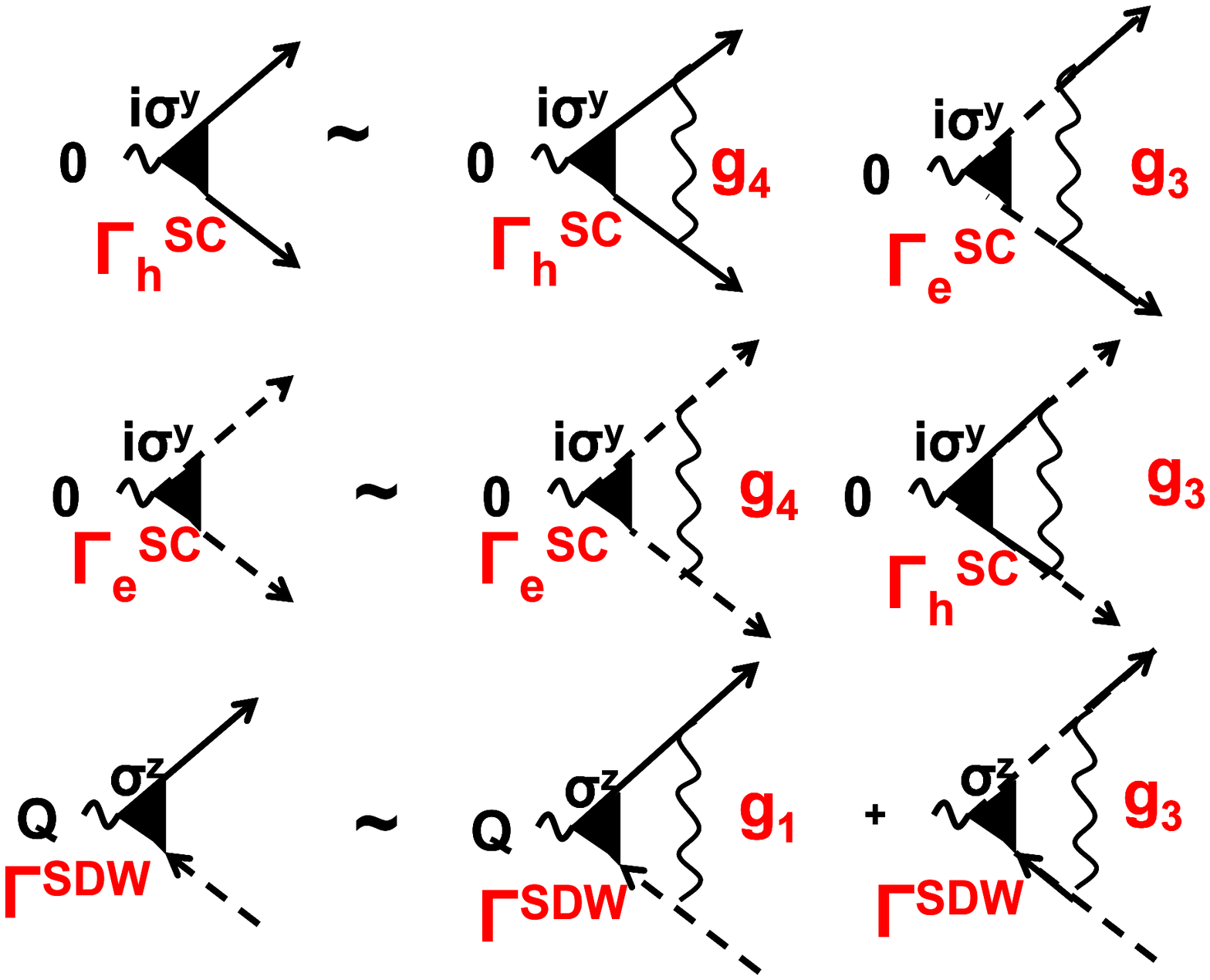}
\caption{\label{fig:new_vertex2} The flow  diagrams for
the effective vertices: SC vertex  (top two) and SDW vertex
(bottom). The couplings $g_i$'s here are running couplings in RG
sense. From [\onlinecite{maiti_rev,maiti_book}].}
\end{figure}

Let us see how it works for a two-pocket model. There are two
particle-particle three legged vertices $\Gamma_{h,e}$ as shown in
Fig \ref{fig:new_vertex}.  To obtain the flow  of these vertices,
i.e.,  $\Gamma^{SC}_{h,e} (L)$ I assume that external fermions
and a fluctuating field have energies  comparable to some  E
(i.e.,$L = \log \Lambda/E$) and collect contributions from all
fermions with energies larger than $E$. To do this with
logarithmical accuracy I  write all possible diagrams, choose a
particle-particle cross-section at the smallest internal energy
$E' \geq E$ and sum up all contributions to the left and to the
right of this cross-section, as shown in Fig
\ref{fig:new_vertex2}. The sum of all contributions to the left of
the cross-section gives the three legged vertex at energy $E'$ (or
$L' =  \log \Lambda/E'$), and the sum of all contributions to the
right of the cross-section gives the interaction $g_i$ at energy
$L'$. The integration over the remaining cross-section gives
$\int^L d L'$ (with our normalization of $g_i$), and the equation
for, e.g., $\Gamma_{h} (L)$ becomes \bea \Gamma^{SC}_h (L) =
\int^L d L' \left(\Gamma^{SC}_h (L') g_4 (L') + \Gamma^{SC}_e(L')
g_3(L')\right) \eea Differentiating over the upper limit, we
obtain differential equation for $d \Gamma^{SC}_h (L)/dL$ whose
r.h.s. contains $\Gamma^{SC}_{h,e} (L)$ and $g_{3,4} (L)$ at the
same scale $L$.

 Collecting the contributions for $\Gamma^{SC}_h (L)$ an $\Gamma^{SC}_e (L)$ we obtain
 \bea\label{eq:run1}
\frac{d\Gamma^{SC}_h}{dL}&=&\Gamma^{SC}_h g_4 + \Gamma^{SC}_e g_3\nonumber\\
\frac{d\Gamma^{SC}_e}{dL}&=&\Gamma^{SC}_e g_4 + \Gamma^{SC}_h g_3\nonumber\\
\eea

or

\bea\label{eq:run2}
\frac{d\Gamma_{++}}{dL}&=&(g_4 + g_3) \Gamma_{++}\nonumber\\
\frac{d\Gamma_{+-}}{dL}&=&(g_4 - g_3)\Gamma_{+-}\nonumber\\
\eea

where $\Gamma_{++}\equiv\Gamma^{SC}_h+\Gamma^{SC}_e$ and
$\Gamma_{+-}\equiv \Gamma^{SC}_h-\Gamma^{SC}_e$.  The first one is
for $s^{++}$ pairing, the second is for $s^{+-}$ pairing.
 We have seen in the previous section that
the running couplings $g_{3,4}$ diverge at some critical RG scale
$L_c$. The flow equation near $L_c$ is in
 the form $\dot{g}\sim g^2$, hence
\beq\label{eq:div}g_i=\frac{\alpha_i}{L_c-L}.\eeq Substituting
this into Eq. \ref{eq:run2} and solving the differential equation
for $\Gamma$ we find that the two SC three legged vertices
 behave as
\beq\label{eq:div2}\Gamma_{s^{++}} \propto
\frac{1}{(L_c-L)^{-\alpha_3-\alpha_4}},~~\Gamma_{s^{+-}}=\frac{1}{(L_c-L)^{\alpha_3-\alpha_4}},
\eeq The requirement for the divergence of $\Gamma_{s^{+-}}$ is
$\alpha_3 >\alpha_4$, which is obviously the same as $g_3 > g_4$
(see (\ref{eq:div})).

I follow the same procedure for an SDW vertex ${\vec
\Gamma}^{SDW}_{\alpha \beta} = \Gamma^{SDW} {\vec \sigma}_{\alpha \beta}$.  I introduce a particle-hole  vertex with momentum
transfer $Q$ and spin factor ${\vec \sigma}_{\alpha \beta}$, as
shown in  Fig \ref{fig:new_vertex}, and obtain the equation for $d
\Gamma^{SDW} (L)/dL$  in the same way as we did for SC
vertices. We obtain (see Fig. \ref{fig:new_vertex2})
\bea\label{eq:run3}
\frac{d\Gamma^{SDW}}{dL}&=&d_1(g_1 + g_3)\Gamma^{SDW}\nonumber\\
\eea Using Eq. \ref{eq:div} and following the same steps as above
we obtain at $L\approx L_c$

\beq\label{eq:div3}\Gamma^{SDW} \propto
\frac{1}{(L_c-L)^{d_1(\alpha_1+\alpha_3)}}\eeq

For CDW vertex (the one with the overall factor
$\delta_{\alpha\beta}$ instead ${\vec \sigma}_{\alpha\beta}$), the flow
equation is \bea\label{eq:run4}
\frac{d\Gamma^{CDW}}{dL}&=&d_1(g_1 +g_3 -2g_3 -2g_2)\Gamma^{CDW}\nonumber\\
&=& d_1(g_1 -g_3 -2g_2)\Gamma^{CDW} \eea

Using the same procedure as before we obtain
\beq\label{eq:div3_1}\Gamma^{CDW}=\frac{1}{(L_c-L)^{d_1(\alpha_1-\alpha_3-2\alpha_2)}}\eeq

The exponents $\alpha_i$ can be easily found by plugging in the
asymptotic forms in Eq. \ref{eq:div} into the RG equations. This
gives the following set of non linear algebraic equations in
$\alpha_i$ \bea\label{eq:NL1}
\alpha_1 &=& d_1(\alpha_1^2 + \alpha_3^2)\nonumber \\
\alpha_2 &=& 2 d_1 \alpha_2(\alpha_1 - \alpha_2 ) \nonumber \\
\alpha_3 &=&  2 d_1\alpha_3(2 \alpha_1 - \alpha_2)-2\alpha_3\alpha_4\nonumber \\
\alpha_4 &=& -\alpha_3^2 - \alpha_4^2 \nonumber \\
\eea

Consider first the case of perfect nesting, $d_1=1$. The solution
of the set of equations is $\alpha_1=\frac{1}{6}$, $\alpha_2=0$,
$\alpha_3=\frac{\sqrt{5}}{6}$ and $\alpha_4=-\frac{1}{6}$;
Combining $\alpha$'s, we find that the exponents  for
superconducting and spin density wave instabilities and positive
and equal:

\bea\label{eq:expo}
\alpha_{s\pm}&\equiv&\alpha_3-\alpha_4=\frac{1+\sqrt{5}}{6}\approx 0.539\nonumber\\
\alpha_{SDW}&\equiv&\alpha_1+\alpha_3=\frac{1+\sqrt{5}}{6}\approx 0.539\nonumber\\
\eea while the exponent for CDW and $s++$ vertices are negative
\bea \alpha_{CDW} &=&
\alpha_1+\alpha_3=\frac{1-\sqrt{5}}{6}\approx -
0.206 \nonumber\\
\alpha_{s++} &=& -\alpha_3-\alpha_4=\frac{1-\sqrt{5}}{6}\approx -
0.206 \eea

We see that the superconducting ($s^{+-}$) and SDW channels have
equal susceptibilities in this approximation, while CDW channel is
not a competitor.

\begin{figure}[htp]
\includegraphics[width=1\columnwidth]{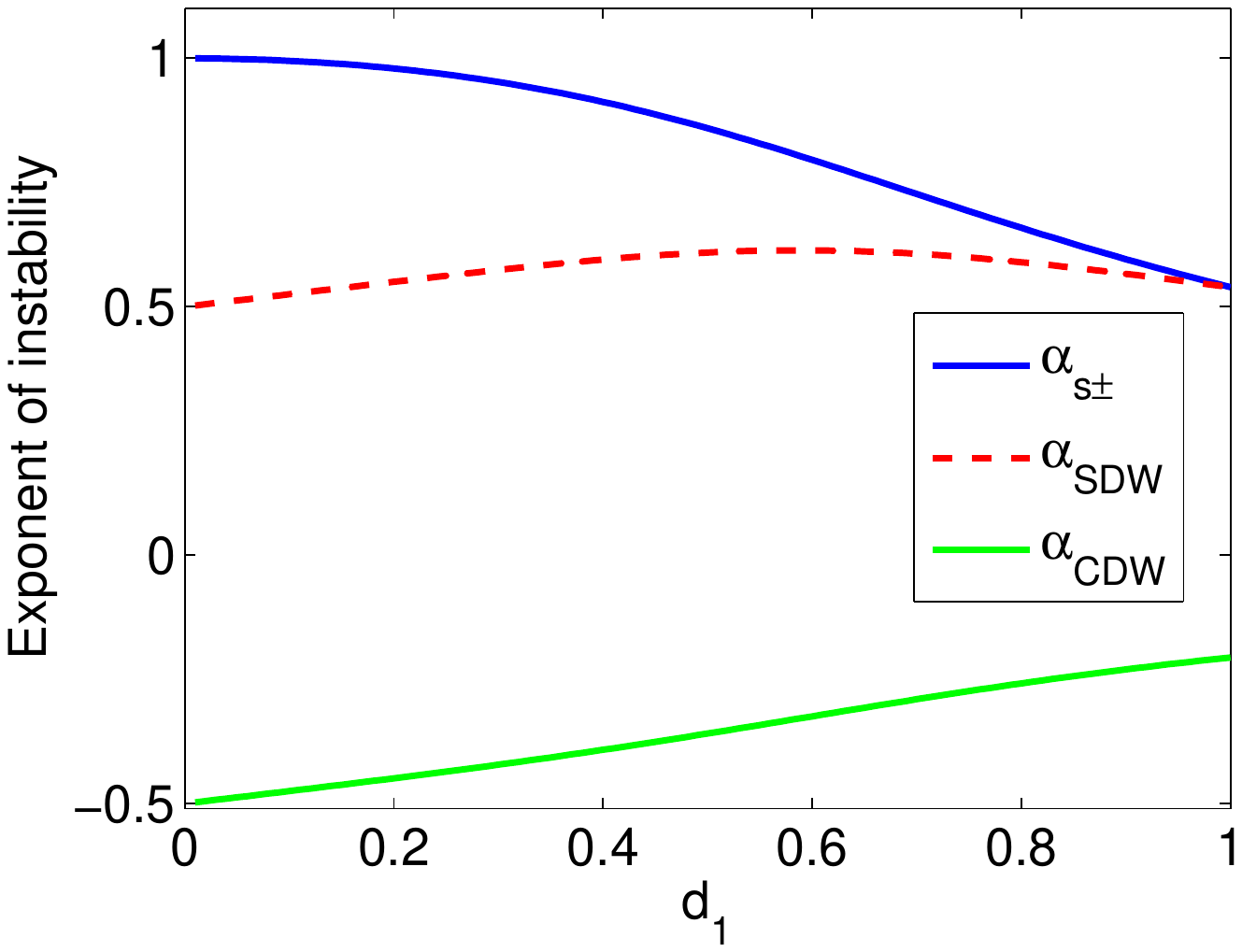}
\caption{\label{fig:alpha2} Exponents ($\alpha_{s\pm}$,
$\alpha_{SDW}$ and $\alpha_{CDW}$) for different values of the
nesting parameter $d_1$ calculated near the critical RG scale,
where the couplings diverge. The state with the largest exponent
wins. SDW and SC are degenerate when $d_1=1$ (perfect nesting) and
superconductivity wins for all other values of $d_1$. CDW is not a
competitor for all values of $d_1$.}
\end{figure}

The analysis can be extended to  $d_1<1$. I  define $\beta\equiv
\alpha_4/\alpha_1$, $\gamma\equiv \alpha_3/\alpha_1$ and obtain
\bea\label{eq:expo_sol}
\gamma^2&=&\frac{\sqrt{16d_1^4-4d_1^2+4}+2-d_1^2}{d_1^2}\nonumber\\
\beta&=&\frac{d_1}{2}\left(3-\gamma^2\right)\nonumber\\
\alpha_1&=&\frac{1}{d_1}\frac{1}{1+\gamma^2} \eea In
Fig\ref{fig:alpha2} I plot $\alpha_{s\pm}=\alpha_3-\alpha_4$,
$\alpha_{SDW}=\alpha_1+\alpha_3$, and
$\alpha_{CDW}=\alpha_1-\alpha_3$, We clearly see that (i) CDW
channel is never a competitor, and (ii) as $d_1$ decreases (the
nesting gets worse), the pairing vertex diverges with a higher
exponent that SDW channel, hence $s^{+-}$ superconductivity
becomes the leading instability, overshooting the channel which
helped SC vertex to change sign in the first place.

In real systems, pRG equations are only valid up to some distance
from the instability at $L_c$. Very near $L_c$ three-dimensional
effects, corrections from higher-loop orders and other
perturbations likely affect the flow of the couplings. Besides, in
pocket models, the pRG equations are only valid for $E$ between
the bandwidth $W$ and the Fermi energy $E_F$. At $E < E_F$,
internal momenta in the diagrams, which account for the flow of the
couplings, become smaller than external $k_F$, and  the
renormalization of $g_i$ start depending on the interplay between
all four external momenta in the vertices\cite{rev_physica,RG_SM}.
The calculation of the flow in this case is technically more
involved, but the result is physically transparent -- SDW and
$s^{+-}$ SC channels stop talking to each other, and the vertex
evolves according to Eqs. (\ref{eq:div2}) and (\ref{eq:run3}),
with $g_i$ taken at the scale $E_F$ (or $L_F = \log \Lambda/E_F$).
If $L_F > L_c$, the presence of the scale set by the Fermi energy
is irrelevant, but if $L_F < L_c$ (which is the case for the
Fe-pnictides because superconducting $T_c$ and magnetic $T_{SDW}$
are much smaller than $E_F$), then one should stop pRG flow at
$L_{E_F}$. At perfect nesting, the SDW combination $g_1 + g_3$ is
larger than $s^{+-}$ combination $g_3-g_4$ at any $L < L_c$, hence
SDW channel wins, and the leading instability upon cooling down
the system is towards a SDW order. At  non-zero doping, $\Pi_{ph}
(Q)$ is cut by a deviation from nesting, what in our language
implies that $d_1 <1$. If bare $g_3$ and $g_4$ are not to far
apart,  there exists a critical $d_1$ at which $g_3-g_4$ crosses
$d_1 (g_1 + g_3)$ at $L_{F}$, and at larger $d_1$ the crossing
occurs before $L_F$. In this situation, $s^{+-}$ SC becomes the
leading instability upon cooling off the system.

The comparison between different channels can be further extended
by considering current SDW and CDW vertices (imaginary
$\Gamma^{SDW}$ and $\Gamma^{CDW}$) and so on. I will not dwell
into this issue.

\begin{figure}[htp]
\includegraphics[width=1\columnwidth]{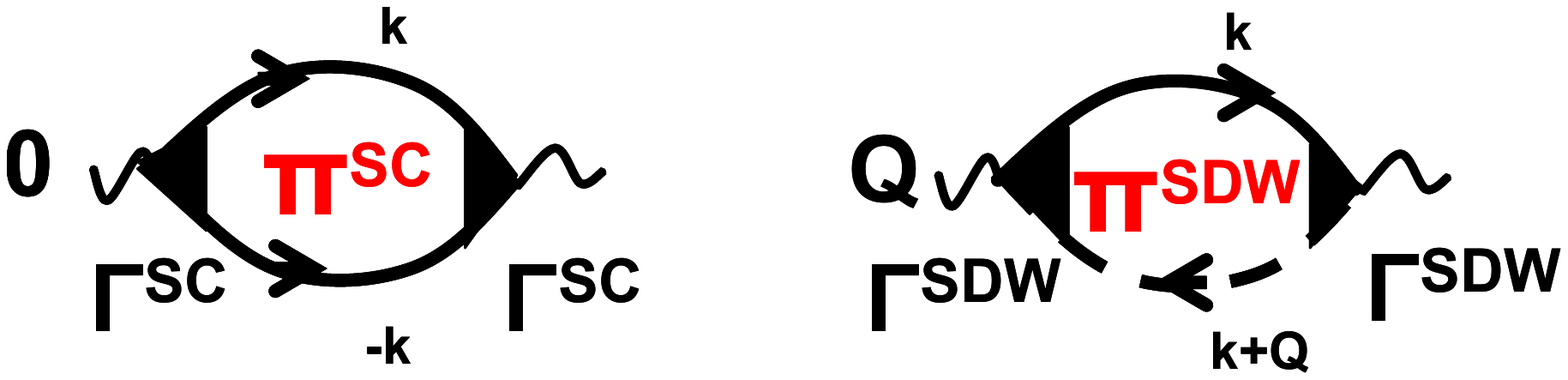}
\caption{\label{fig:fluc} (Left) The fluctuation correction to SC
pairing susceptibility. (Right) The fluctuation correction to SDW
susceptibility. 
}
\end{figure}

Before moving on, I need to clarify one more point. So far we
found that the vertices $\Gamma^{SC}$ and $\Gamma^{SDW}$ diverge
and compared the exponents. However, to actually analyze the
instability in a particular channel one has to compute fluctuation
correction to susceptibility

\beq \label{eq:susdiv}
 \chi^{i}_{fl} (L) \sim  \int d^2k (\Gamma^i)^2 \Pi_i \propto   \int^L d L' \left(\Gamma^{i} (L') \right)^2
\eeq

where $\Pi_i$ is either $\Pi^{SDW} = \Pi_{ph}$ or $\Pi^{SC} =
\Pi_{pp}$ (see Fig \ref{fig:fluc})

The fully renormalized susceptibility in a given channel is 
\beq
\left(\chi^{i} (L)\right)^{-1} =  r^i_0 - \chi^{i}_{fl} (L) \eeq where $r^i_0$ is some
bare value of order one.  The true instability occurs at $L^*$
when $ \chi^{i}_{fl} (L^*) = r^i_0$. At weak coupling, the critical
$L^*$ is close to $L_c$, and, indeed, the instability occurs first
in the channel with the largest exponent for $\Gamma^{i}$.
However, we need $\chi^{i}_{fl} (L)$ to diverge at $L_c$,
otherwise there will no instability at weak coupling~\cite{Cvetkovic}.
This requirement sets the condition that the exponent for the
corresponding $\Gamma$ must be larger than $1/2$. Fortunately,
this  condition is satisfied in the two-pocket model. For $d_1=1$,
this is evident from (\ref{eq:expo}). For $d_1 <1$, the exponent
for the SC channel only increases, while the one in SDW channel
decreases but still remains larger than $1/2$ as it is evidenced
from Fig\ref{fig:alpha2} where I plotted the exponents for SC and
SDW vertices as a function of $d_1$. In the limit $d_1\rightarrow
0$, \beq\alpha_{SDW}\approx \frac{1}{2}+\frac{d_1}{4}\eeq.

The fact that both $\alpha_{SC}$ and $\alpha_{SDW}$ are larger
than $1/2$ implies that in Landau-Ginzburg expansion in powers of
SC and SDW order parameters ($\Delta$ and $M$, respectively), not
only the prefactor for $\Delta^2$ changes sign at $T_c$, but also
the prefactor for $M^2$ term changes sign and becomes negative
below some $T_m < T_c$. This brings in the possibility that at low
T SC and SDW orders co-exist. The issue of the co-existence,
however, requires a careful analysis of the interplay of
prefactors for fourth order terms $M^4$, $ \Delta^4$, and $M^2
\Delta^2$. I do not discuss this specific issue. For details
see~\cite{VVC,FS}.

\begin{figure}[htp]
\includegraphics[width=1\columnwidth]{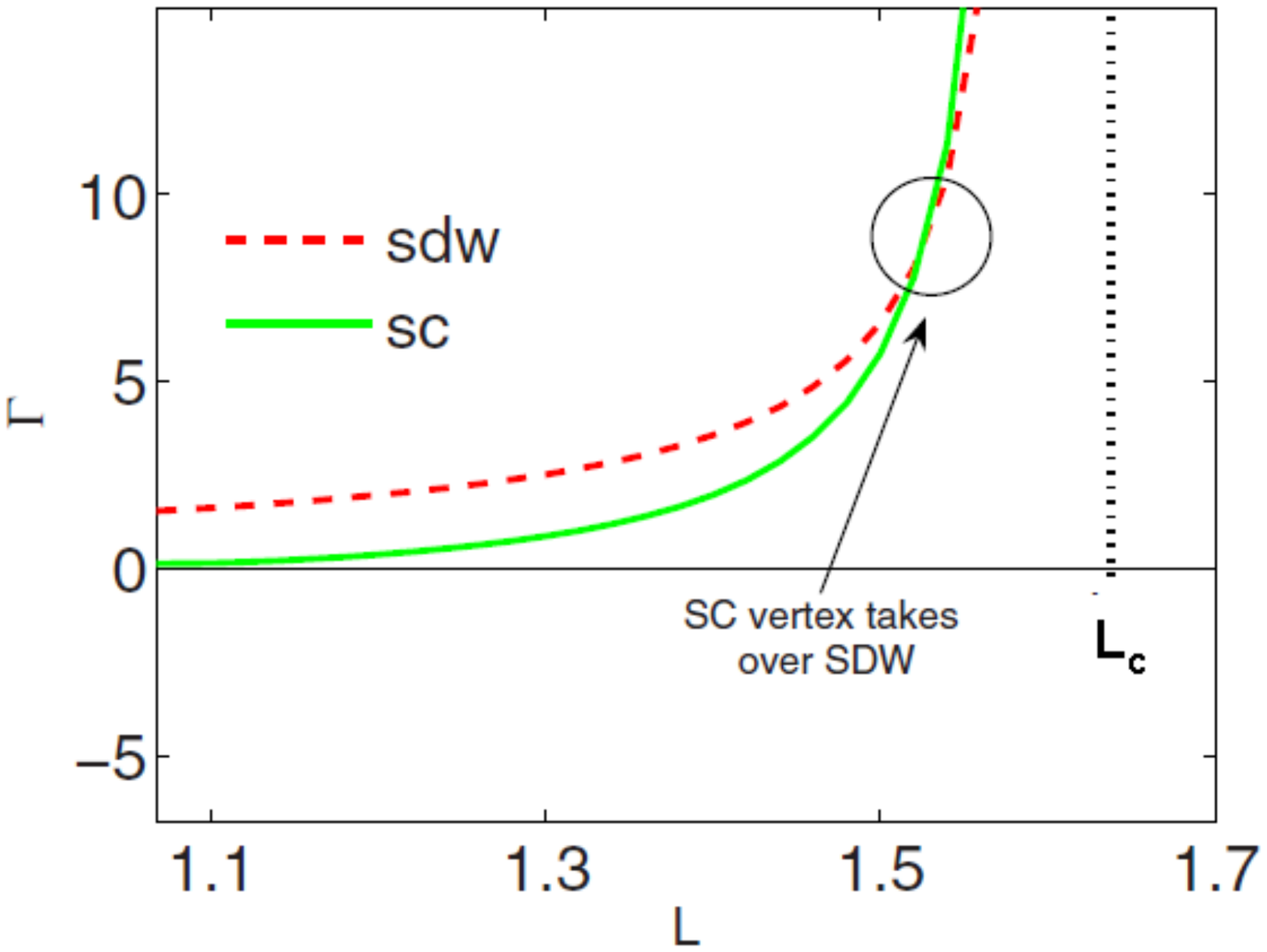}
\caption{\label{fig:RG_3poc} The flow of the SC and SDW vertices
with the RG scale. Both diverge at a critical scale, $L_c$, but
the SC vertex diverges stronger. From Ref. [\onlinecite{RG_SM}].}
\end{figure}

\subsubsection{Multi-pocket models}

The interplay between SDW and SC vertices is more involved in more
realistic multi-pocket models Fe-pnictides, with several electron
and hole pockets. I recall that weakly doped Fe-pnictides have 2
electron pockets and 2-3 hole pockets. In multi-pocket models one
needs to introduce a larger number of  intra-and inter-pocket
interactions  and analyze the flow of all couplings to decide
which instability is the leading one. This does not provide any
new physics compared to what we have discussed, but in several
cases the interplay between SC and SDW instabilities becomes such
that superconductivity wins already at perfect nesting. In
particular, in 3-pocket models (two electron pockets and one hole
pockets) the exponent for the SC vertex gets larger than the
exponent for the SDW vertex already at $d_1=1$. I show the flow
of SC an SDW couplings for 3-pocket model in
Fig.\ref{fig:RG_3poc}. Once $d_1$ becomes smaller than one, SC
channel wins even bigger compared to SDW channel.

Superconductivity right at zero doping has been detected in
several Fe-pnictides, like LaOFeAs and LiFeAs,  and it is quite
possible that this is at least partly due to the specifics of pRG
flow.

\subsection{Summary of the pRG approach}

I now summarize the key points of the pRG approach

\begin{itemize}
\item The SC vertex starts out as repulsive, but
it  eventually changes sign at some RG scale ($L_0$). This happens
due to the "push" from SDW channel, which rives rise to upward
renormalization of the inter-pocket interaction $g_3$.
\item
Both SDW and SC  vertices diverge at RG scale $L_c$ which is
larger than $L_0$. The leading instability is in the channel whose
vertex diverges with a larger exponent. At perfect nesting, SDW
instability occurs first in 2-pocket model, however in some multi-pocket models SC vertex has a
larger exponent that the SDW vertex and SC becomes the leading
instability.
\item Deviations from perfect nesting (quantified by $d_1<1$) act against SDW order by
reducing the corresponding exponent.  At sufficiently small $d_1$
SC instability becomes the leading one.
\item
The necessary condition for the instability is the diverges of the
fluctuating component of the susceptibility. This sets up a
condition $\alpha >1/2$, where $\alpha$ is the exponent for the
corresponding vertex. For the leading instability, we found
$\alpha >1/2$ in all cases. For the subleading instability,
$\alpha$ can be either larger or smaller than $1/2$.  This affects
potential co-existence of the leading and subleading orders at a
lower $T$.
\end{itemize}

\section{SDW magnetism and nematic order}
\label{sec:4}

For this section, I assume that we are in the range of parameters/dopings, where SDW instability comes first, and consider
(i) what kind of SDW order emerges and (ii) the interplay between breaking of $O(3)$ spin-rotational symmetry and breaking of a discrete
 $C_4$ symmetry of rotations on a tetragonal lattice. I consider these two issues one after the other.
In the discussions in this section I follow Refs. [\onlinecite{Eremin_10,rafael_nematic,rafael_review}].

 \subsection{Selection of SDW order}

\begin{figure*}[htp]
\includegraphics[width=0.7\columnwidth]{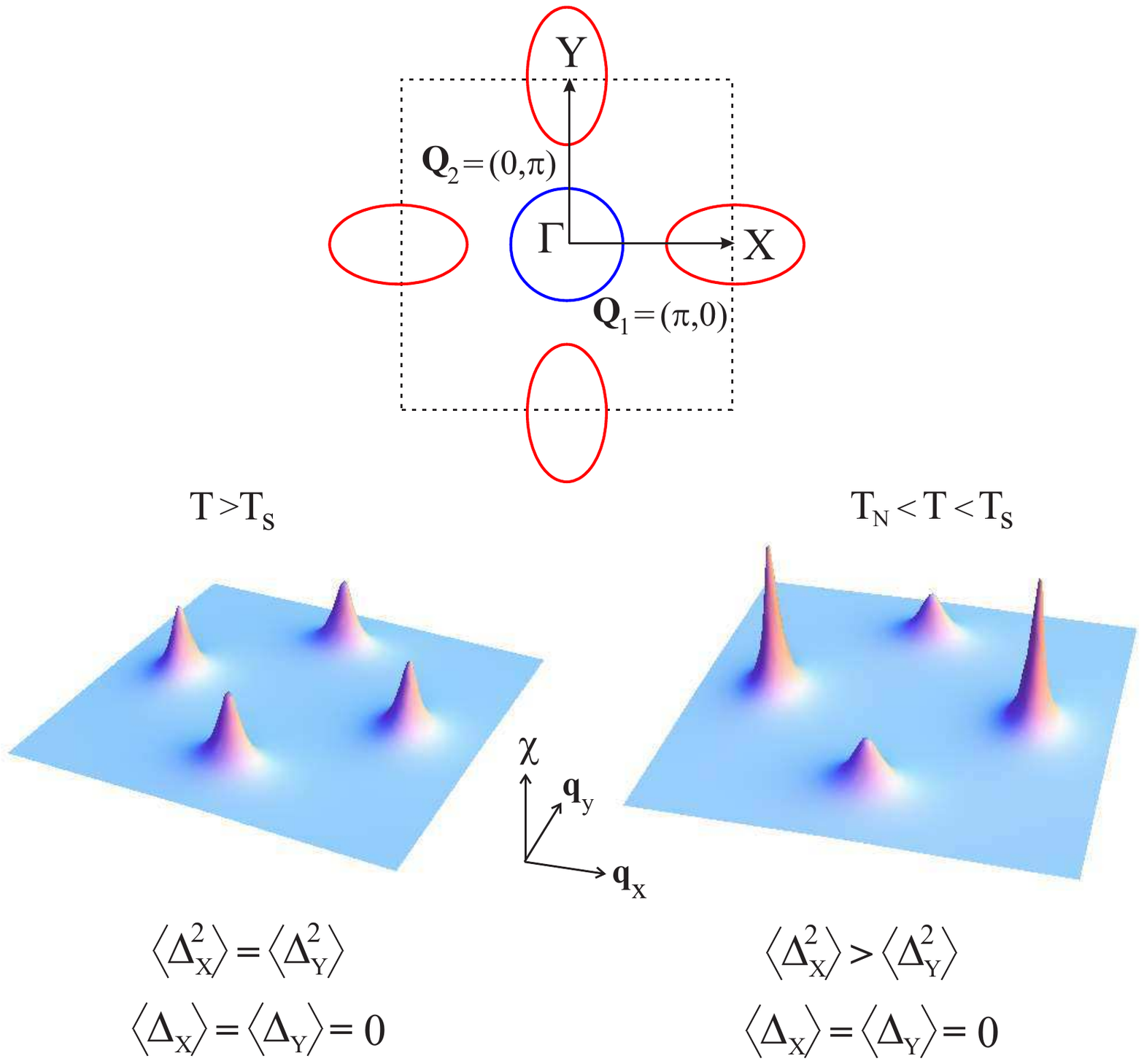}
\caption{(Color online.) (\textbf{upper panel}) The band-structure with a circular hole pocket
at $\Gamma$ and two electron pockets at $X$ and $Y$. The Brillouin
zone contains one Fe atom. (\textbf{lower panels}) Static magnetic
susceptibility $\chi_{\mathbf{q}}$ across the Brillouin zone for
different temperatures. At high temperatures, fluctuations near the
two stripe magnetic ordering vectors are equally strong, $\left\langle \Delta_{X}^{2}\right\rangle =\left\langle \Delta_{Y}^{2}\right\rangle $.
Above the magnetic ordering temperature $T_{N}$ but below the Ising-nematic
ordering temperature $T_{s}$, fluctuations associated with one of
the stripe states become stronger (in the figure, $\left\langle \Delta_{X}^{2}\right\rangle >\left\langle \Delta_{Y}^{2}\right\rangle $)
and the tetragonal symmetry is broken inside the unit cell. Stronger
fluctuations around one ordering vector yield stronger intensity and
narrower peaks. }
\label{fig:schematic_nematic_order}
\end{figure*}

I return to the model I started with, but now with interactions renormalized by pRG contributions from energies lager than $E_F$.
The only necessary extension we need to make is we need to consider  two electron pockets, one at $(0,\pi)$ and another at $(\pi,0)$
in the unfolded Brillouin zone (see Fig.\ref{fig:schematic_nematic_order}). To make presentation more simple, we consider only one hole pocket, centered at $(0,0)$.
 The extension to two (or three) hole pockets is straightforward, but requites care and in some cases leads to new states~\cite{zlatko,kang}

We need to be a bit more precise and include the ellipticity of electron pockets. Accordingly, we
 approximate dispersions of fermions near hole and electron pockets by
 $\varepsilon_{\Gamma,\mathbf{k}}=\varepsilon_{0}-\frac{k^{2}}{2m}-\mu \equiv -\varepsilon$, $\varepsilon_{X,\mathbf{k+Q_{1}}}=\varepsilon-\delta_{0}+\delta_{2}\cos2\theta$,
$\varepsilon_{Y,\mathbf{k+Q_{2}}}=\varepsilon-\delta_{0}-\delta_{2}\cos2\theta$,
where $m_{i}$ denotes the band masses, $\varepsilon_{0}$ is the
offset energy, $\mu$ is the chemical potential,
 $\delta_{0}=2\mu$, $\delta_{2}=\varepsilon_{0}m(m_{x}-m_{y})/(2m_{x}m_{y})$,
and $\theta=\tan^{-1}k_{y}/k_{x}$ \cite{Vorontsov10}.

I shift the momenta of the fermions near
the $X$ and $Y$ Fermi pockets by $\mathbf{Q}_{1}$ and $\mathbf{Q}_{2}$,
respectively, i.e. $\varepsilon_{X,\mathbf{k+Q_{1}}}\rightarrow\varepsilon_{X,\mathbf{k}}$,
$\varepsilon_{Y,\mathbf{k+Q_{2}}}\rightarrow\varepsilon_{Y,\mathbf{k}}$.

This model has eight fermionic
interactions $U_{n}$ (with the same structure as in a 2-pocket model, but now there are four different inter-and intra-pocket interactions involving the two electron pockets). These interactions can be decomposed into the spin density-wave
(SDW), the charge density-wave (CDW) and the pairing channels.  For magnetism,
 I keep only the interactions in the spin channel with momenta
near $\mathbf{Q}_{1}$ and $\mathbf{Q}_{2}$. This reduces the interacting
Hamiltonian to \begin{equation}
\mathcal{H}_{\mathrm{int}}=-\frac{1}{2}g_{\mathrm{spin}}\sum\limits _{i,\mathbf{q}}\mathbf{s}_{i,\mathbf{q}}\cdot\mathbf{s}_{i,-\mathbf{q}}\label{H_int}\end{equation}
 where $\mathbf{s}_{i,\mathbf{q}}=\sum_{k}c_{\Gamma,\mathbf{k+q}\alpha}^{\dagger}\boldsymbol{\sigma}_{\alpha\beta}c_{i,\mathbf{k}\beta}$
is the electronic spin operator, with Pauli matrices $\boldsymbol{\sigma}_{\alpha\beta}$.
The coupling $u_{\mathrm{spin}}$ is the combination of density-density
and pair-hopping interactions between hole and electron states ($g_{1}$
and $g_{3}$ terms in the same notations as in previous two Sections).

\begin{eqnarray}
g_{1}c_{\Gamma,\alpha}^{\dagger}c_{\Gamma,\alpha}c_{X,\beta}^{\dagger}c_{X,\beta} & = & -\frac{g_{1}}{2}c_{\Gamma,\alpha}^{\dagger}\boldsymbol{\sigma}_{\alpha\beta}c_{X,\beta}\cdot c_{X,\gamma}^{\dagger}\boldsymbol{\sigma}_{\gamma\delta}c_{\Gamma,\delta}  +(\cdots)\nonumber \\
g_{3}c_{\Gamma,\alpha}^{\dagger}c_{X,\alpha}c_{\Gamma,\beta}^{\dagger}c_{X,\beta} & = & -\frac{g_{3}}{2}c_{\Gamma,\alpha}^{\dagger}\boldsymbol{\sigma}_{\alpha\beta}c_{X,\beta}\cdot c_{X,\gamma}^{\dagger}\boldsymbol{\sigma}_{\gamma\delta}c_{\Gamma,\delta} +(\cdots)\label{SDW_channel}\end{eqnarray}
 where the dots stand for the terms with $\delta_{\alpha,\beta}\delta_{\gamma,\delta}$,
which only contribute to the CDW channel. Combining the two contributions
for the SDW channel, I find $g_{\mathrm{spin}}=g_{1}+g_{3}$, as in (\ref{mo_d_1}).
 Once
$g_{\mathrm{spin}}$ exceeds some critical value (which gets smaller
when $\delta_{0}$ and $\delta_{2}$ decrease), static magnetic susceptibility
diverges at $(0,\pi)$ and $(\pi,0)$, and the system develops long-range
magnetic order. An excitonic-type SDW instability in Fe-pnictides,
resulting from the interaction between hole and electron pockets,
has been considered by several authors \cite{Eremin_10,Cvetkovic09,Gorkov08,Timm09,Kuroki08,DungHaiLee08,Platt08,Vorontsov09,FS,Knolle10}.

My strategy is the following: I introduce
the two bosonic fields $\boldsymbol{\Delta}_{(X,Y)}\propto\sum_{\mathbf{k}}c_{\Gamma,\mathbf{k}\alpha}^{\dagger}\boldsymbol{\sigma}_{\alpha\beta}c_{(X,Y),\mathbf{k}\beta}$
for the collective magnetic degrees of freedom, use Hubbard-Stratonovich transformation to get rid of the terms in (\ref{H_int}) with four fermions,
 integrate out the
fermions, and obtain a Ginzburg-Landau (GL) action for $\boldsymbol{\Delta}_{X}$
and $\boldsymbol{\Delta}_{Y}$.  I then analyze this action in saddle-point approximation and show
that
one of the magnetic order parameters - either $\left\langle \boldsymbol{\Delta}_{X}\right\rangle $
or $\left\langle \boldsymbol{\Delta}_{Y}\right\rangle $ - becomes
non-zero in the magnetically ordered state. This leads to stripe-type
SDW order in which spins are ordered ferromagnetically in one direction
and antiferromagnetically in the other, i.e. the ordering momentum
is either $(\pi,0)$ or $(0,\pi)$. I then  show that another state, in which $\left\langle \boldsymbol{\Delta}_{X}\right\rangle $
or $\left\langle \boldsymbol{\Delta}_{Y}\right\rangle $ emerge simultaneously, may occur at a higher doping~\cite{osborn}.
  The same tendency
 occurs in systems like Ba(Fe$_{1-x}$Mn$_x$)$_2$As$_2$, where the
local Mn moments interact with the Fe conduction electrons~\cite{rafael_last}.

\subsubsection{The action in terms of $\boldsymbol{\Delta}_{X}$ and $\boldsymbol{\Delta}_{Y}$}

A straightforward way to obtain the action in terms of $\left\langle \boldsymbol{\Delta}_{X}\right\rangle $
and $\left\langle \boldsymbol{\Delta}_{Y}\right\rangle $ is to start
with the fermionic Hamiltonian $\mathcal{H}=\mathcal{H}_{\mathrm{0}}+\mathcal{H}_{\mathrm{int}}$ and write the partition function
as the integral over Grassmann variables:
\begin{equation}
Z\propto\int dc_{i,\mathbf{k}}dc_{i,\mathbf{k}}^{\dagger}\mathrm{e}^{-\beta\mathcal{H}}\label{ch_X}\end{equation}
 and then decouple the quartic term in fermionic operators using the
Hubbard-Stratonovich transformation:

\begin{equation}
\mathrm{e}^{\frac{ax^{2}}{2}}=\frac{1}{\sqrt{2\pi a}}\int dy\,\mathrm{e}^{\left(-\frac{y^{2}}{2a}+yx\right)}\label{Hubbard_Stratononich}\end{equation}
 where, in our case, $x=\mathbf{s}_{i,\mathbf{0}}=\sum_{k}c_{\Gamma,\mathbf{k}\alpha}^{\dagger}\boldsymbol{\sigma}_{\alpha\beta}c_{(X,Y),\mathbf{k}\beta}$
and $y=\boldsymbol{\Delta}_{(X,Y)}$.  One can then integrate Eq. (\ref{ch_X})
over fermionic variables using the fact that after the Hubbard-Stratonovich
transformation the effective action becomes quadratic with respect
to the fermionic operators. The result of the integration is recast
back into the exponent and the partition function is expressed as:

\begin{equation}
Z\propto\int d\boldsymbol{\Delta}_{X}d\boldsymbol{\Delta}_{Y}\mathrm{e}^{-S_{\mathrm{eff}}\left[\boldsymbol{\Delta}_{X},\boldsymbol{\Delta}_{Y}\right]}\label{aux_action_1}\end{equation}

If relevant $\boldsymbol{\Delta}_{X}$ and $\boldsymbol{\Delta}_{Y}$
are small, which I assume to hold even if the magnetic transition
is first-order (I present the conditions on the parameters below),
one can expand $S_{\mathrm{eff}}\left[\boldsymbol{\Delta}_{X},\boldsymbol{\Delta}_{Y}\right]$
in powers of $\boldsymbol{\Delta}_{X}$ and $\boldsymbol{\Delta}_{Y}$
and obtain the Ginzburg-Landau type of action for the order parameters
$\boldsymbol{\Delta}_{X},\boldsymbol{\Delta}_{Y}$. For uniform $\boldsymbol{\Delta}_{i}$,
the most generic form of $S_{\mathrm{eff}}\left[\boldsymbol{\Delta}_{X},\boldsymbol{\Delta}_{Y}\right]$
is \begin{eqnarray}
S_{\mathrm{eff}}\left[\boldsymbol{\Delta}_{X},\boldsymbol{\Delta}_{Y}\right] & = & r_{0}\left(\boldsymbol{\Delta}_{X}^{2}+\boldsymbol{\Delta}_{Y}^{2}\right)+\frac{u}{2}\left(\boldsymbol{\Delta}_{X}^{2}+\boldsymbol{\Delta}_{Y}^{2}\right)^{2}\nonumber \\
 &  & -\frac{g}{2}\left(\boldsymbol{\Delta}_{X}^{2}-\boldsymbol{\Delta}_{Y}^{2}\right)^{2}+v\left(\boldsymbol{\Delta}_{X}\cdot\boldsymbol{\Delta}_{Y}\right)^{2}\label{action}\end{eqnarray}

Carrying out this procedure, one obtains the coefficients $r_{0}$,
$u$, $g$, and $v$ in terms of the non-interacting fermionic propagators
convoluted with Pauli matrices.
The coefficient $v$ vanishes in our model because of the anti-commutation
property of the Pauli matrices: $\sigma^{i}\sigma^{j}+\sigma^{j}\sigma^{i}=0$
for $i\neq j$. To get a non-zero $v$, one needs to include direct
interactions between the two electron pockets \cite{Eremin_10}. The
other three prefactors are expressed via fermionic propagators $G_{j,\mathbf{k}}^{-1}=i\omega_{n}-\xi_{j,\mathbf{k}}$
as \begin{eqnarray}
r_{0} & = & \frac{2}{g_{\mathrm{spin}}}+2\int_{k}G_{\Gamma,k}G_{X,k}\nonumber \\
u & = & \frac{1}{2}\int_{k}G_{\Gamma,k}^{2}\left(G_{X,k}+G_{Y,k}\right)^{2}\nonumber \\
g & = & -\frac{1}{2}\int_{k}G_{\Gamma,k}^{2}\left(G_{X,k}-G_{Y,k}\right)^{2}\label{aux_action}\end{eqnarray}
where $\int_{k}=T\sum_{n}\int\frac{d^{d}k}{\left(2\pi\right)^{d}}$
and $k=\left(\mathbf{k},\omega_{n}\right)$, with momentum $\mathbf{k}$
and Matsubara frequency $\omega_{n}=\left(2n+1\right)\pi T$. Similar
coefficients were found in Ref. \cite{Brydon11}, which focused on
the magnetic instabilities in a two-band model.
 Near $T_{N,0}$ one can expand $r_{0}$ as $r_{0}=a(T-T_{N,0})$,
with $a>0$.
Evaluating the integrals with the products of the Green's functions, we obtain
\begin{eqnarray}
u & \approx & \frac{7\zeta\left(3\right)N_{F}}{4\pi^{2}T^{2}}\nonumber \\
g & \approx & 0.024u\left(\frac{\varepsilon_{0}\delta m}{T}\right)^{2}\label{parameters_quartic}\end{eqnarray}
 for $\delta m\ll T/\varepsilon_{0}\ll1$.  The crucial
result for our consideration is that $g$  is positive for any non-zero ellipticity.

The action $S_{\mathrm{eff}}$ is exact and includes all fluctuations
of the two bosonic fields. Fluctuations need to be included for the analysis of a potential nematic order (see below),
 but the type of SDW can be analyzed  already in the mean-field approximation (see Refs~\cite{Eremin_10,rafael_nematic} for justification.)
Solving for the minimum of $S_{\mathrm{eff}}\left[\boldsymbol{\Delta}_{X},\boldsymbol{\Delta}_{Y}\right]$
in Eq. (\ref{action}), we find that, when $g=0$, the ground state
has a huge degeneracy because any configuration $\boldsymbol{\Delta}=\left\langle \boldsymbol{\Delta}_{X}\right\rangle \mathrm{e}^{i\mathbf{Q}_{1}\cdot\mathbf{r}}+\left\langle \boldsymbol{\Delta}_{Y}\right\rangle \mathrm{e}^{i\mathbf{Q}_{1}\cdot\mathbf{r}}$
with $\left\langle \boldsymbol{\Delta}_{X}\right\rangle ^{2}+\left\langle \boldsymbol{\Delta}_{Y}\right\rangle ^{2}=-r_{0}/u$
minimizes $\tilde{S}_{\mathrm{eff}}$. A non-zero $g$ gives rise
to the additional coupling $2g\Delta_{X}^{2}\Delta_{Y}^{2}$, which
breaks this degeneracy. For a positive $g$, this term favors the
states in which only one order parameter has a nonzero value, i.e.
configurations with either $\left\langle \boldsymbol{\Delta}_{X}\right\rangle \neq0$
or $\left\langle \boldsymbol{\Delta}_{X}\right\rangle \neq0$, but
not both. These are stripe phases, in which  spins order ferromagnetically along one
direction and antiferromagnetically along the other one.

For larger dopings, recent calculations~\cite{osborn} have shown that $g$ may change sign and become negative.
 Then the SDW phase does not break $C_4$ symmetry.  The transformation from a stripe SDW state to a state which preserves $C_4$ symmetry has recently been observed in Ba$_{1-x}$Na$_x$Fe$_2$As$_2$ near the end of the SDW region~\cite{osborn}.

\subsection{pre-emptive spin-nematic order}

I now analyze a possibility that $Z_2$ symmetry between $X$ and $Y$ directions gets broken
before the system develops a stripe SDW order.  To analyze this possibility, I  include
fluctuations of the $\boldsymbol{\Delta}_{X,Y}$ fields, introduce
the collective Ising-nematic bosonic variable $\phi\propto\Delta_{X}^{2}-\Delta_{Y}^{2}$
together with $\psi\propto\boldsymbol{\Delta}_{X}^{2}+\boldsymbol{\Delta}_{Y}^{2}$,
integrate over $\boldsymbol{\Delta}_{X}$ and $\boldsymbol{\Delta}_{Y}$,
and obtain an effective action in terms of $\phi$ and $\psi$. I analyze this action and check whether the system develops an instability
towards $\left\langle \phi\right\rangle \neq0$ before $\left\langle \boldsymbol{\Delta}_{X}\right\rangle $
or $\left\langle \boldsymbol{\Delta}_{Y}\right\rangle $ becomes non-zero (see Fig. \ref{fig:schematic_nematic_order}).

That the action (\ref{action}) can potentially lead to a preemptive
Ising-nematic instability is evident from the presence of the term
$g\left(\Delta_{X}^{2}-\Delta_{Y}^{2}\right)^{2}$, which can give
rise to an ordered state with $\left\langle \Delta_{X}^{2}\right\rangle -\left\langle \Delta_{Y}^{2}\right\rangle \neq0$
in a way similar to how the $\boldsymbol{s}_{i,\mathbf{q}}\boldsymbol{s}_{i,-\mathbf{q}}$
term in the Hamiltonian (\ref{H_int}) gives rise to a state with
non-zero $\left\langle \boldsymbol{s}_{i,0}\right\rangle \neq0$.
The pre-emptive Ising-nematic instability, however, does not appear
in the mean-field approximation
simply because when magnetic fluctuations are absent,
a non-zero $\left\langle \Delta_{i}^{2}\right\rangle \neq0$ appears
simultaneously to $\left\langle \Delta_{i}\right\rangle \neq0$, once
$r_{0}$ changes sign. However, it may well happen once we go beyond
 mean-field and include magnetic fluctuations.

To study a potential preemptive $Z_{2}$ transition, I need to introduce
collective variables of the fields $\boldsymbol{\Delta}_{X}$ and
$\boldsymbol{\Delta}_{Y}$. Let me introduce auxiliary scalar fields
$\phi$ for $\boldsymbol{\Delta}_{X}^{2}-\boldsymbol{\Delta}_{Y}^{2}$
and $\psi$ for $\boldsymbol{\Delta}_{X}^{2}+\boldsymbol{\Delta}_{Y}^{2}$.
The field $\psi$ always has a non-zero expectation value $\left\langle \psi\right\rangle \neq0$,
which describes Gaussian corrections to the magnetic susceptibility
$\chi_{i,q}^{-1}$ in Eq. \ref{chi_i}. Meanwhile, the field $\phi$
may or may not have a non-zero expectation value. If it does, it generates
a non-zero value of $\left\langle \boldsymbol{\Delta}_{X}^{2}-\boldsymbol{\Delta}_{Y}^{2}\right\rangle $
and the system develops an Ising-nematic order.

The effective action in terms of $\phi$ and $\psi$ is obtained by
using again the Hubbard-Stratonovich transformation of Eq. (\ref{ch_X}),
but this time the variable $x$ is either $\psi \propto \Delta_{X}^{2}+\Delta_{Y}^{2}$
or $\phi \propto \Delta_{X}^{2}-\Delta_{Y}^{2}$. Applying this transformation
 and integrating over fluctuating fields $\Delta_X$ and $\Delta_Y$,  I obtain
  the effective action in terms
on $\phi$ and $\psi$ in the form
 \begin{equation}
S_{\mathrm{eff}}\left[\phi,\psi\right]=\int_{q}\left\{ \frac{\phi^{2}}{2g}-\frac{\psi^{2}}{2u}+\frac{3}{2}\log\left[\left(\chi_{q}^{-1}+\psi\right)^{2}-\phi^{2}\right]\right\} \label{action_phi}
\end{equation}
As it is customary for the analysis of fluctuating fields $\Delta_X$ and $\Delta_Y$, we extended the mass term $r_0$ to include spatial and time variations
 of $\Delta_{X,Y}$:
 \begin{equation}
r_{0}\rightarrow\chi_{i,q}^{-1} = r_{0}+\gamma\left|\nu_{n}\right|+q^2\label{chi_i}
\end{equation}
where $\nu_{n}=2\pi Tn$ is the bosonic Matsubara frequency.

This action can be straightforwardly analyzed in the saddle-point approximation $\partial S_{\mathrm{eff}}\left[\phi,\psi\right]/\partial\phi=\partial S_{\mathrm{eff}}\left[\phi,\psi\right]/\partial\psi=0$ (for justification see Ref. \cite{rafael_nematic}).
 Differentiating, I obtain two non-linear coupled equations for $\phi$ and $\psi$:
\begin{eqnarray}
\frac{\psi}{u} & = & \int_{q}\frac{r_{0}+\psi+q^{2}+\gamma|\nu_{n}|}{\left(r_{0}+\psi+q^{2}+\gamma|\nu_{n}|\right)^{2}-\phi^{2}}\nonumber \\
\frac{\phi}{g} & = & \int_{q}\frac{\phi}{\left(r_{0}+\psi+q^{2}+\gamma|\nu_{n}|\right)^{2}-\phi^{2}}\label{self_cons}
\end{eqnarray}
The full solution of these equations at various temperatures and in different dimensions is presented in Ref. \cite{rafael_nematic}. The key point is that, for positive $g$, $\phi$ becomes non-zero at a higher temperature $(T_n)$ than the one ($T_{sdw})$ at which SDW order sets in. In the interval $T_n > T > T_{sdw}$,
 $\left\langle \boldsymbol{\Delta}_{X}^{2}-\boldsymbol{\Delta}_{Y}^{2}\right\rangle$ becomes non-zero, while $\left\langle \boldsymbol{\Delta}_{X}\right\rangle = \left\langle \boldsymbol{\Delta}_{Y}^{2}\right\rangle  =0$. Such an order breaks $C_4$ lattice symmetry down to $C_2$ and is often called Ising-nematic order.

\begin{figure*}[htp]
\includegraphics[width=0.7\columnwidth]{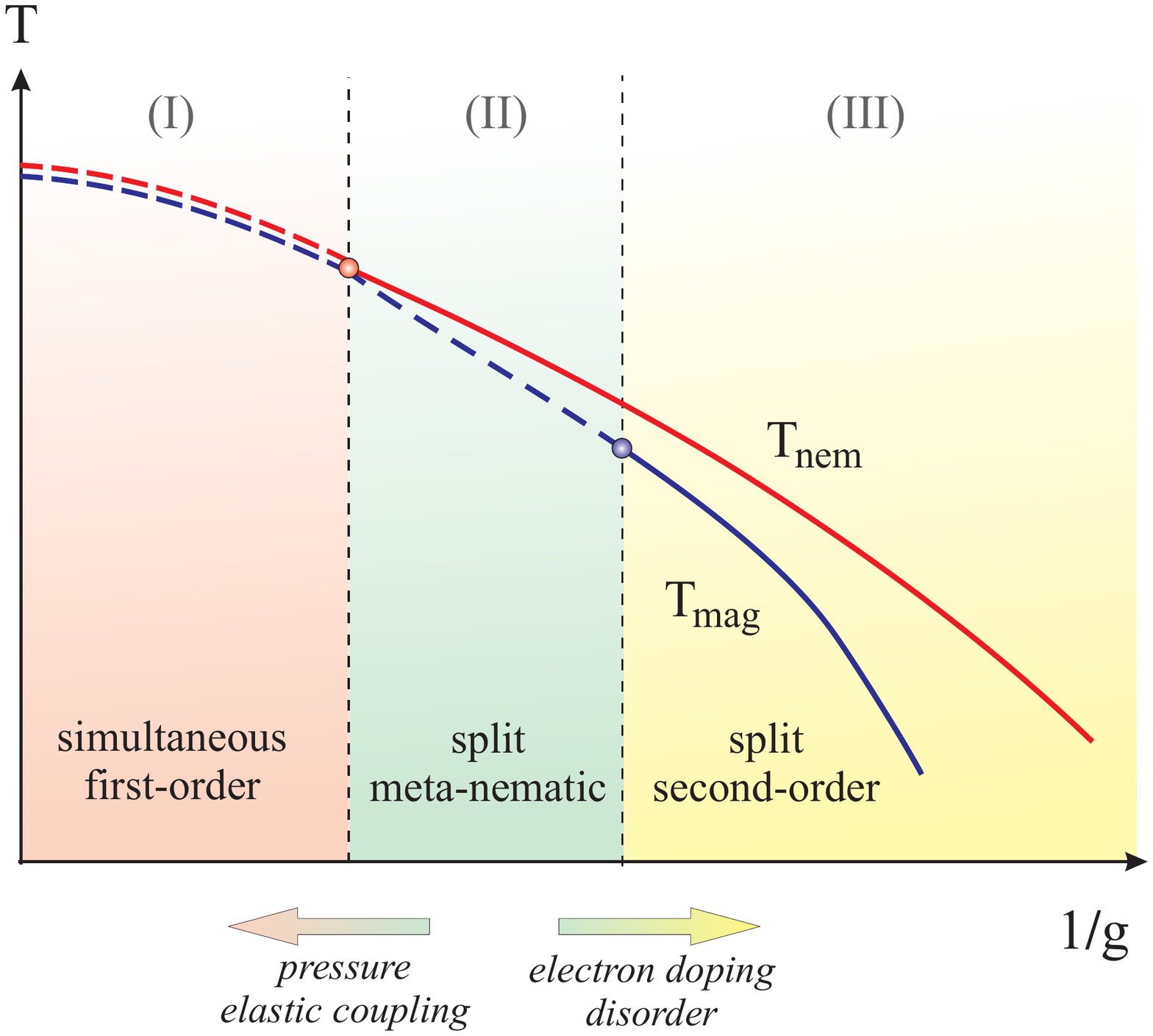}
\caption{ \textbf{Evolution of the character of the magnetic and nematic transitions
in the spin-driven nematic theory}. The control parameter is the inverse
nematic coupling $g$, which changes as function of various control
parameters within an itinerant scenario (arrows). Second-order (first-order)
lines are denotes by solid (dashed) lines. Regions (I)-(III) correspond
to those of the phase diagram in Fig. 1. The arrows show how the nematic
order parameter $g$ is expected to evolve with doping, disorder,
pressure, and elastic coupling.  From [\onlinecite{rafael_nematic}].}
\label{fig_exp_phase_diagram}
\end{figure*}

In Fig. \ref{fig_exp_phase_diagram} I present the phase diagram for anisotropic 3D system.
 The transition to an Ising-nematic state can be either second-order, or first order. A strong first-order nematic transition may instantly induce
  SDW order.

\subsection{consequences of the Ising-nematic order}

Because spin-nematic order breaks $C_4$ lattice rotational symmetry, it couples linearly to any other parameter which breaks the same symmetry,
 such as orbital and structural order parameters.  Then, once spin-nematic order becomes non-zero, it acts as "external field" to the two other parameters
  and induces non-zero values of both of them.  As a result, below $T_{n}$ the fermionic dispersion becomes
anisotropic, the occupations of  $d_{xz}$
and $d_{yz}$ orbitals become non-equal, and also the lattice constants $a$ and $b$ along
the $x$ and $y$ directions of the Fe-plane, respectively, become non-equal.  I refrain to discuss this issue in more detail here and direct a reader to a recent review~\cite{rafael_review}.  The development of the Ising-nematic order also gives rise to an increase of the magnetic correlation length, what in turn gives rise to
 a pseudo-gap-type behavior of the fermionic spectral function.

\section{The structure of the superconducting gap}
\label{sec:5}

I now turn to superconductivity. Like I did for SDW order, I assume that renormalizations captured within pRG are already included into consideration
 and consider an effective low-energy model with effective pairing interactions in the band basis.  In the discussions in this Section I follow Refs. 
 [\onlinecite{ch_annual,cee,cvv,maiti_rev,maiti_book,RG_SM,maiti_kor,maiti_11}].

\subsection{The structure of $s-$wave and $d-$wave gaps in a multi-band SC - general reasoning}

In previous sections I assumed that the interactions in the particle-particle channel (the dressed $G_3$ and $G_4$ terms)  are
 independent on the angles along the hole and electron FSs. In this situation, the only option is an $s-$wave gap, which changes sign between the FSs, but
  is a constant along each FS.  Now I consider realistic models in which the interactions in the band basis are obtained from the  underlying multi-orbital
   model. These interactions generally depend on locations of fermions along the FS.

I first display general arguments on what should be the form of the gap in different symmetries and on different FSs.
I show that an $s-$wave gap generally has angle dependence and may even have nodes, while a d-wave gap, which is normally assumed to have nodes, may in fact be nodeless on electron FSs.

A generic low-energy BCS-type model in the band basis is  described by
 \beq {\cal H} = \sum_{i,\k}
\epsilon_{i} (\k) a^\dagger_{i \k}  a_{i \k} + \sum_{i,j, \k, \p}
U_{i,j} (\k,\p) a^\dagger_{i \k} a^\dagger_{i -\k} a_{j \p}a_{j
-\p}
 \label{rev_2}
 \eeq
The quadratic  term describes low-energy
excitations near hole and electron FSs, labeled by $i$ and $j$, and the
four-fermion term describes the scattering of a pair $(k \uparrow,
-k\downarrow)$ on the FS $i$ to a pair $(p \uparrow,
-p\downarrow)$ on the FS $j$.
These interactions are either intra-pocket interactions (hole-hole $U_{h_ih_i}$ or electron-electron $U_{e_ie_i}$ ), or inter-pocket interactions
(hole-electron $U_{e_j h_i}$, hole-hole $U_{h_i \neq h_j}$, and  electron-electron $U_{e_i \neq e_j}$).

Assume for simplicity that
 the frequency dependence of $\Gamma$ can be neglected and low-energy fermions are Fermi-liquid quasiparticles with
   Fermi velocity $v_{k_F}$.  In this situation,  the gap $\Delta (k)$ also doesn't depend on frequency, and to obtain $T_c$ one 
   has to solve  the eigenfunction/eigenvalue problem:
\beq
\lambda_i\Delta_i (k) = - \int \frac{d p_\parallel}{4\pi^2  v_{{\bf p}_F}}
\Gamma({\bf k}_F,{\bf p}_F) \Delta_i (p)
\label{gap_eq}
\eeq
where $\Delta_i$ are  eigenfunctions and $\lambda_i$ are eigenvalues.  The system is unstable towards pairing if one or more $\lambda_i$ are
  {\it positive}. The corresponding $T_{c,i}$ scale as $T_{c,i} = \Lambda_i e^{-1/\lambda_i}$. Although $\Lambda_i$ are generally different for different $i$, the exponential dependence on $1/\lambda_i$ implies that, most likely, the solution with the largest positive $\lambda_i$ emerges first and establish the pairing state, at least immediately below $T_c$.

Like I discussed in the Introduction, the pairing interaction $ U(k,p)$ can be decomposed into
 representations of the tetragonal space group (one-dimensional representations are
 $A_{1g}$, $B_{1g}$, $B_{2g}$, and $A_{2g}$).
  Basis functions from different representations do not mix,
but each contains infinite number of components.  For example, $s-$wave
pairing corresponds to  fully symmetric $A_{1g}$ representation, and the
 $s-$wave ($A_{1g}$) component of $U (k,p)$
 can be quite generally expressed
as
\beq
U^{(1g)}(k,p)= U_s(k,p)=\sum_{m,n}A^{s}_{mn}\Psi^s_m(k)\Psi^s_n(p) \label{s_1} \eeq

where $\Psi^s_m (k)$ are the basis functions of the A$_{1g}$ symmetry
group: $1$, $\cos k_x \cos k_y$, $\cos k_x + \ cos k_y$, etc, and $A^{s}_{mn}$
are coefficients. Suppose that $k$ belongs to a
hole FS and is close to $k=0$. Expanding {\it any} wave function
with A$_{1g}$ symmetry near $k=0$, one obtains along $|{\bf k}| =
k_F$,
\beq
 \Psi^s_m(k) = a_m + b_m \cos 4 \phi_k + c_m \cos 8 \phi_k +...
\label{s_2}
 \eeq
where $\phi_k$ is the angle along the hole FS (which is not necessary a circle).
Similarly, for $B_{1g}$ representation the wave-functions are
$\cos k_x - \cos k_y$, $cos 2k_x - cos 2k_y$, etc, and expanding them near $k=0$ one obtains
\beq
 \Psi^d_m(k) = a^*_m \cos 2 \phi_k  + b^*_m \cos 6 \phi_k + c^*_m \cos 10 \phi_k +...
\label{d_2}
 \eeq
There are no fundamental reasons  to expect that $b_m,~c_m$ or $b^*_m, c^*_m$
are much smaller than $a_m$ or $a^*_m$, but sub-leading terms are often small
numerically. I assume that this is the case and neglect subleading terms, i.e., assume that $s-$wave interaction between fermions on the hole FSs  can
be approximated by an angle-independent $U^s_{h_ih_j} (k,p) \equiv
U_{h_ih_j}$ ($h_i$ label different hole FSs), while $d-$wave ($B_{1g}$) interaction can be approximated by $U^d_{h_ih_j} (k,p) = {\tilde U}_{h_ih_j} \cos 2 \phi_k \cos 2\phi_p$.

The situation changes, however, when I consider the pairing component
 involving fermions from electron FSs. Suppose
that $k$ are still near the center of the BZ, but $p$
are near one of the two electron FSs, say the one centered at
$(0,\pi)$.  Consider all possible $\Psi_n(p)$ with $A_{1g}$
symmetry A simple experimentation with trigonometry shows that
there are two different subsets of basis functions:
\bea
&&{\text subset}~ I: 1,~ \cos{p_x} \cos{p_y},~ \cos{2 p_x} + \cos{2p_y} ... \nonumber \\
&&{\text subset}~ II: \cos{p_x} + \cos{p_y},~\cos{3 p_x} + \cos{3 p_y}...
\label{s_3} \eea
For a circular FS centered at $(0,\pi)$, the functions
 from subset I can be again expanded in series of $\cos 4 l \phi_p$ with integer $l$.
The functions from subset II are different -- they all vanish
at $(0,\pi)$ and are expanded in series of $\cos (2\phi_p + 4 l
\phi_p)$ (the first term is $\cos 2 \phi_p$, the second is
$~\cos 6 \phi_p$, etc).  For elliptic  FS $\cos 4 l \phi_p$ and $\cos (2\phi_p + 4 l\phi_p)$ terms  appear in both subsets.  In either case, the total
\bea
 \Psi^s_m(p) &=& {\bar a}_m + {\bar b}_m \cos 4 \phi_p + {\bar c}_m
\cos 8 \phi_p + ...\nonumber\\
&&+  {\bar {\bar a}}_m \cos 2 \phi_p  + {\bar {\bar b}}_m \cos 6 \phi_p + {\bar{\bar
c}}_m \cos 10 \phi_k +...
\label{s_2p}
 \eea
For the other electron FS,  $\Psi^s_m(p)$ is the same, but   momentum components  $p_x$ and $p_y$are interchanged, hence the sign of all
$\cos (2\phi +4l\phi_p)$ components changes.

Let's make the same approximation as
before and neglect all components with $l >0$.  Then
\beq
 \Psi^s_m(p) = {\bar a}_m \pm  {\bar {\bar a}}_m \cos 2 \phi_p
\label{s_2p1}
 \eeq
where the upper sign is for one electron FS and the lower for the other.
It is essential that the angle-independent term and the  $\cos 2 \phi_p$ term
 have to be treated on equal footing because each  is the {\it leading} term in the corresponding series. Combing (\ref{s_2p1})  with the fact that $ \Psi^s_m(k)$ can be approximated by a constant, we obtain a generic form of the
 $s-$wave component of the interaction between fermions near hole and electron FSs
\bea
U^s_{e_1,h_i}(k,p) &=& U_{e,h_i} \left(1 + 2\alpha_{e,h}\cos 2
\phi_{p_{e1}} +...\right) \nonumber\\
 U^s_{e_2,h_i}(k,p) &=& U_{e,h_i} \left(1 - 2\alpha_{e,h}\cos 2
\phi_{p_{e2}} + ....\right)
\label{s_4}
\eea
where dots stand for $\cos{4 \phi_k}, \cos{4\phi_p},
\cos{6\phi_p}$, etc terms.

 By the same reasoning, $s-$wave components of  inter-pocket and intra-pocket interactions between fermions from electron FSs are
\bea
U^s_{e_1,e_1}(k,p) &=& U_{e,e} \left(1+2\alpha_{ee}
\left(\cos2\phi_{k_{e1}}+ \cos2\phi_{p_{e1}}\right)\right. \nonumber \\
&&+ 4 \beta_{ee} \cos2\phi_{k_{e1}} \cos2\phi_{p_{e1}} +... \nonumber \\
U^s_{e_2,e_2}(k,p) &=& U_{e,e} \left(1 - 2\alpha_{ee}
\left(\cos2\phi_{k_{e2}}+ \cos2\phi_{p_{e2}}\right)\right. \nonumber \\
&&+ 4 \beta_{ee} \cos2\phi_{k_{e2}} \cos2\phi_{p_{e2}} +... \nonumber \\
U^s_{e_1,e_2}(k,p) &=& U_{e,e} \left(1 + 2\alpha_{ee}
\left(\cos2\phi_{k_{e1}} - \cos2\phi_{p_{e2}}\right)\right. \nonumber \\
&&- 4 \beta_{ee} \cos2\phi_{k_{e1}} \cos2\phi_{p_{e2}} +...
\label{s_5}
\eea

Once the pairing interaction has the form of Eqs. (\ref{s_4}) and
(\ref{s_5}), the  gaps along the hole FSs are angle-independent (modulo $\cos 4 \phi$ terms), but
the gaps along the two electron FSs are of the form
\beq
\Delta^{(s)}_e (k) = \Delta_e \pm \bar{\Delta}_e \cos2\phi_k.
\eeq
 When $\bar{\Delta}_e$ is small compared
to $\Delta_e$, the angle dependence is weak, but when $|\bar{\Delta}_e| > |\Delta_e|$, $s-$wave gaps
have nodes at ``accidental'' values of $\phi$, which differ between the two electron FSs.

A similar consideration holds for $d_{x^2-y^2}$ gap.  Within the same approximation of leading angular momentum harmonics, we have
\bea
 U^d_{e_1,h_i}(k,p) &=& {\tilde U}_{e,h_i} \cos 2 \phi_{h_i}
\left(1 + {\tilde \alpha}_{e,h}\cos 2
\phi_{p_{e1}}\right) +... \nonumber\\
 U^d_{e_2,h_i}(k,p) &=&  {\tilde U}_{e,h_i} \cos 2 \phi_{h_i}
 \left(-1 + {\tilde \alpha}_{e,h}\cos 2
\phi_{p_{e2}} \right) + ...
\label{s_4_d}
\eea
and
\bea
U^d_{e_1,e_1}(k,p) &=& {\tilde U}_{e,e} \left(1+2\alpha_{ee}
\left(\cos2\phi_{k_{e1}}+ \cos2\phi_{p_{e1}}\right)\right. \nonumber \\
&&+ 4 \beta_{ee} \cos2\phi_{k_{e1}} \cos2\phi_{p_{e1}} +... \nonumber \\
U^d_{e_2,e_2}(k,p) &=& {\tilde U}_{e,e} \left(1 - 2\alpha_{ee}
\left(\cos2\phi_{k_{e2}}+ \cos2\phi_{p_{e2}}\right)\right. \nonumber \\
&&+ 4 \beta_{ee} \cos2\phi_{k_{e2}} \cos2\phi_{p_{e2}} +... \nonumber \\
U^d_{e_1,e_2}(k,p) &=& {\tilde U}_{e,e} \left(-1 - 2\alpha_{ee}
\left(\cos2\phi_{k_{e1}} - \cos2\phi_{p_{e2}}\right)\right. \nonumber \\
&&+ 4 \beta_{ee} \cos2\phi_{k_{e1}} \cos2\phi_{p_{e2}} +...
\label{s_5_d}
\eea

The solution of the gap equation then yields the gap in the
 form
\bea
&&\Delta^{(d)}_h (k) = {\tilde \Delta}_h \cos 2 \phi_k \nonumber \\
&&\Delta^{(d)}_e (k) = \pm {\tilde \Delta}_e  + \bar{{\tilde \Delta}}_e \cos2\phi_k.
\eea
Along the hole FS, the gap behaves as a conventional $d-$wave gap with 4 nodes along the
diagonals.  Along electron FSs, the two gaps differ in the sign of the angle-independent terms,
and have in-phase $\cos 2 \phi$ oscillating components. When $ \bar{{\tilde \Delta}}_e <<  {\tilde \Delta}_e$ the two electron gaps
 are simply ``plus'' and ``minus'' gaps, but when  $ \bar{{\tilde \Delta}}_e >  {\tilde \Delta}_e$, each has
accidental nodes, again along different directions on the two electron FSs.

 We see therefore that the geometry of the FSs in FeSCs affects the gap structure in quite fundamental way:
  because electron FSs are centered at the $k$ points which are not along BZ diagonals, $s-$wave gaps on these FSs have $\cos 2 \phi$
 oscillations which one normally would associate with a $d-$wave symmetry, and $d-$wave gaps have  constant (plus-minus) components
 which one would normally associate with an $s-$wave symmetry.  When these ``wrong'' components are large, the gaps
 have accidental nodes. These nodes
 may be present or absent for both $s-$wave and $d-$wave gaps.

An $s-$wave gap with nodes in one of the ``exotic'' options offered by
 the electronic structure of FeSCs. Another ``exotic''  option is  a $d-$wave state without nodes. In heavily electron-doped FeSCs, hole states are gapped, and
 only electron FSs remain. The $d-$wave gaps on these two FSs have no nodes
 if $\cos 2 \phi$ oscillation component is smaller than a constant term, hence the system will display a behavior typical for a fully gapped SC despite that the gap actually has a d-wave symmetry. There are even more exotic options offered by the actual three-dimensionality of the electronic structure and/or the hybridization of the electron FSs due to interaction via a pnictide/chalcogen, Refs. [\onlinecite{3D,mazin,khodas,khodas_3}]. 

A generic analysis of the eigenvalue/eigenfunction problem, Eq. (\ref{gap_eq}),
 reduces to the set of either four (or five) coupled equations in either s-wave or d-wave channels:  two  (or three) $\Delta$'s are the gaps on the hole FSs, and two other $\Delta$'s are angle-independent and $\cos 2 \phi$ components of the gaps
on the electron FSs.  Accordingly, there are either four or five different $\lambda_s$ and $\lambda_d$.

\subsubsection{Generic condition for a non-zero $T_c$.}

\begin{figure}[t]
$\begin{array}{cc}
\includegraphics[width =0.5\columnwidth ]{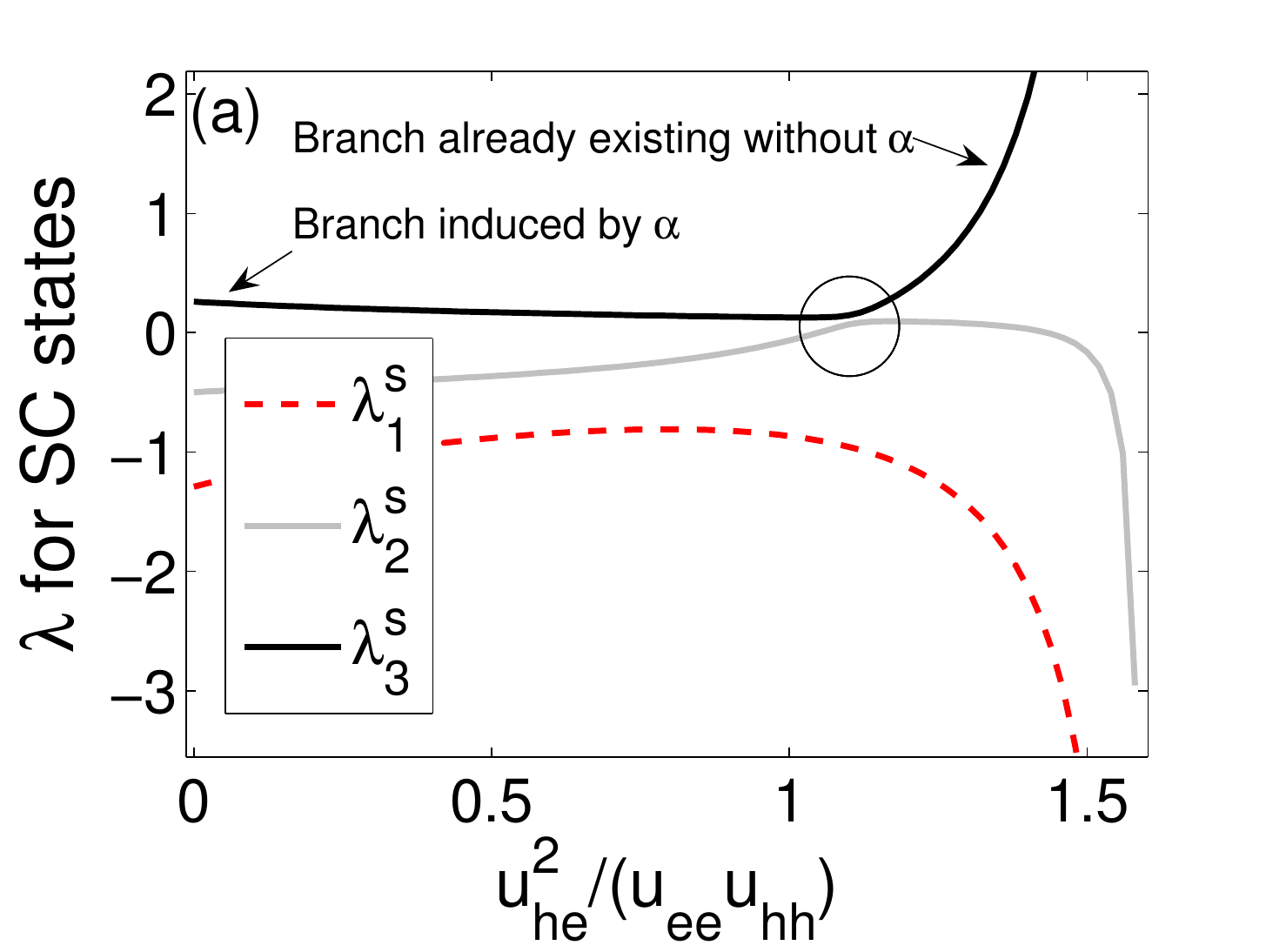}&
\includegraphics[width = 0.5\columnwidth]{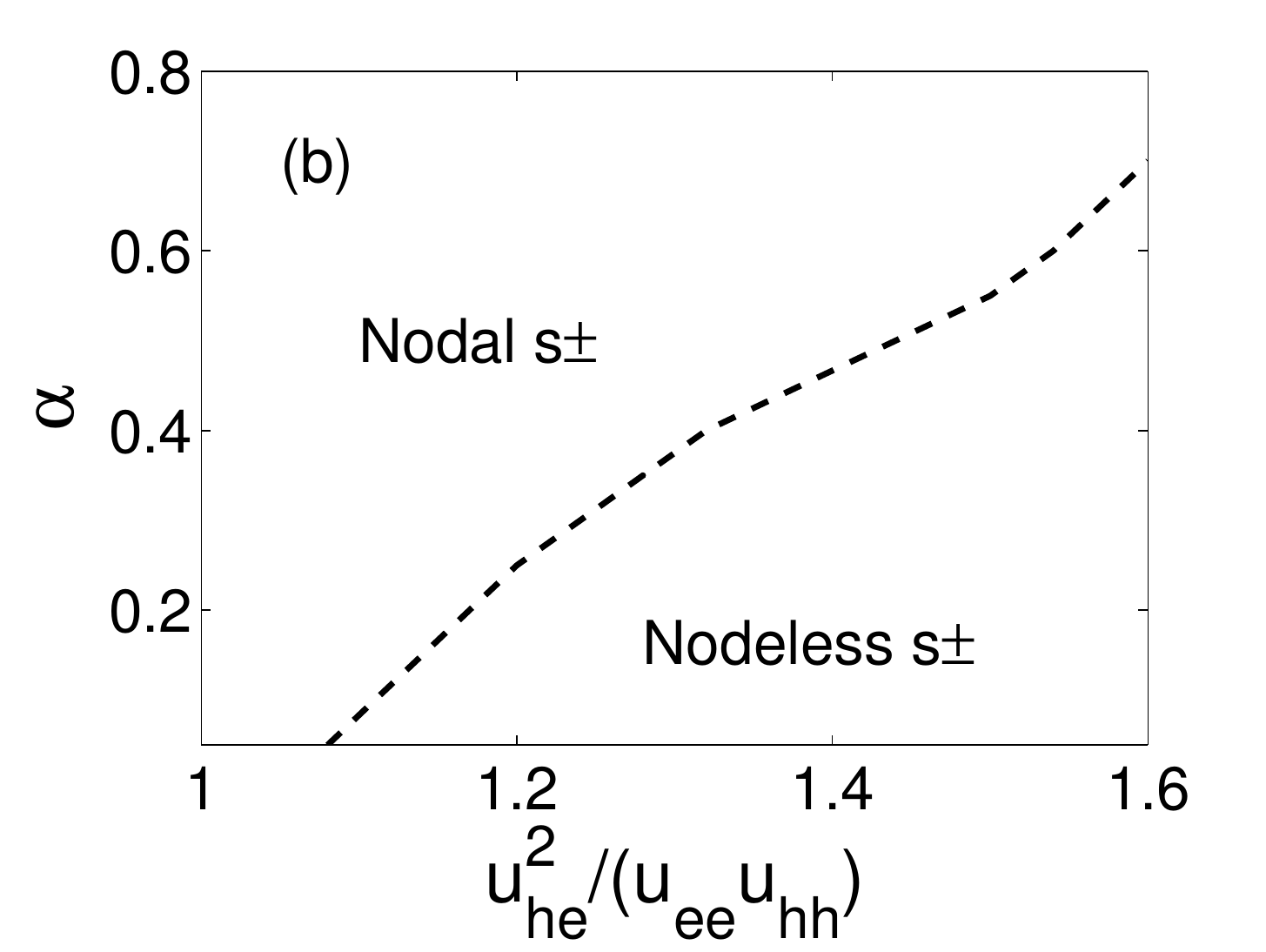}
\end{array}$
\caption{\label{fig:SC_vrtcs} (a) The three eigenvalues in the $s-$wave channel $\lambda^s_i$ as functions of $u^2_{he}/(u_{ee} u_{hh})$ for $\alpha_{ee} = \beta_{ee} =0$ and  $\alpha_{he} =0.4$. For any  $u^2_{he}/(u_{ee} u_{hh})$, one $\lambda^s_i$ is positive (attractive),  other two are negative.
 Positive $\lambda^s_i$ corresponds
to $s\pm$ pairing.
 At small  $u^2_{he}/(u_{ee} u_{hh})$ pairing is induced
by $\alpha_{he}$ and the gap has nodes on electron FSs.
At large  $u^2_{he}/(u_{ee} u_{hh})$ positive $\lambda^s_i$ exists
 already at $\alpha_{he} =0$, and the gap along electron FS has nodes only if $\alpha_{he}$ is above the threshold. The
circle marks the area where positive and negative solutions come
close to each other. The splitting between the two increases with
$\alpha_{he}$. (b) The regions of nodeless and nodal $s^\pm$ gap, depending on
$\alpha_{he}$ and $u^2_{he}/u_{ee} u_{hh}$. From Ref. ~[\onlinecite{RG_SM}].}
\end{figure}

Before I analyze specific cases of $4 \times 4$ and $5 \times 5$ gap equations, I consider the issue whether 
 in the presence of angular dependence of the interactions its is still required 
  for superconductivity that the inter-pocket interaction $u_{he}$ must exceed  the threshold  set by intra-pocket hole-hole and electron-electron interactions. Interestingly enough, this may no longer be necessary. To illustrate this, consider the case of an $s-$wave pairing in a four-pocket model and assume for simplification
 that only one hole pocket is relevant to the pairing. Then the eigenvalue
 problem reduces to the set of three equations for $\Delta_h$, $\Delta_e$, and
 ${\bar \Delta}_e$  ($\Delta_e (k) = \Delta_e + {\bar \Delta}_e \cos {2\phi_k}$).
Solving the set, we find three solutions $\lambda^s_i$ ($i=1,2,3$).  In the absence of $\cos 2\phi$ terms in $\Gamma_{ij} (k,p)$, $\lambda^s_3 =0$, and
$\lambda^s_{1,2}$ are given by
\beq \label{eq:lambda_1s}
\lambda^s_{1,2}  = \frac{-(u_{hh}+2u_{ee}) \pm
\sqrt{(u_{hh}-2u_{ee})^2+ 8u^2_{he}}}{2}
 \eeq
 I remind that $u_{ij} = U_{ij} N_0$, where $N_0$ is the density of states.  Obviously, $u_{he}$ has to exceed a threshold, otherwise $\lambda^s_{1,2} <0$.
 Once the angle dependent terms in (\ref{s_4}-\ref{s_5}) become non-zero, $\lambda^s_3$ also becomes non-zero, and its sign depends on the interplay between
 $\alpha_{he}$, $\alpha_{ee}$, and $\beta_{ee}$.
 In particular, when $u^2_{he} < u_{ee}u_{hh}$ (and, hence, $\lambda_{1,2} <0$),
 $\lambda^s_3$
is positive or negative depending on whether or not $A >0$, where
\beq
A = 4 u_{ee}u_{hh} \left(\alpha^2_{ee} -\beta_{ee}\right)+u^2_{he} \left(\alpha^2_{he}+ 2 \beta_{ee} -3 \alpha_{he} \alpha_{ee}\right)
\label{inequality}
\eeq
When the angle-dependence of the electron-electron interaction
 can be neglected, i.e., $\alpha_{ee} = \beta_{ee} =0$, $\lambda^s_3 >0$ no matter what is the ratio of  $u^2_{he}$ and  $u_{ee}u_{hh}$. In particular, for $u_{hh} u_{ee} > u^2_{he}$ and $\alpha_{he} <<1$,
 \beq
\lambda^s_3 = \alpha^2_{he} \frac{2 u^2_{he} u_{hh}}{u_{hh} u_{ee} - u^2_{he}} >0
\label{rev_1}
\eeq
 In other words, for one of
 $s-$wave solutions, $\lambda^s >0$  even if intra-pocket
 repulsions are the largest.
The full solution of the $3\times3$ set with  $\alpha_{ee} = \beta_{ee} =0$ shows that two $\lambda$'s are repulsive and one is attractive
for arbitrary  $u^2_{he}/u_{ee}u_{hh}$. When the ratio is small,
the attractive solution is  close to (\ref{rev_1}), when the ratio is large, the
 attractive solution is close to $\lambda^s_1$ in (\ref{eq:lambda_1s}).
  I illustrate this in Fig.~\ref{fig:SC_vrtcs}

There is, however, one essential difference between the cases $u^2_{he}/u_{ee}u_{hh} >1$ and   $u^2_{he}/u_{ee}u_{hh} <1$. In the first case, momentum-dependence of the interaction just modifies the ``plus-minus'' solution which already existed
 for momentum-independent interaction. In this situation, the gap along electron FS gradually acquires some $\cos 2 \phi$ variation and remains nodeless for small $\alpha_{he}$. In the second case, the solution with $\lambda >0$ is induced by the momentum dependence of the interaction, and the eigenvalue corresponding to $\lambda^s_3$ necessary has ${\bar \Delta}_e > \Delta_e$, i.e., $s-$wave gap has nodes along the electron FS~~[\onlinecite{cvv}].  In other words, the pairing occurs for all parameters but whether the gap is nodal or not at small $\alpha_{he}$
 depends on the relative strength of intra-pocket and inter-pocket interactions.  When intra-pocket interaction dominates, the gap ``adjusts'' and develops strong $\cos 2 \phi$ component which does not couple to a momentum-independent
 $u_{ee}$ term and by this effectively reduces the strength of electron-electron repulsion.

The same reasoning holds  for the case of two non-equivalent hole FSs, and for 5-pocket  models, and also for the $d-$wave channel, For all cases, the solution
 with $\lambda_i>0$ may exist even when intra-pocket interactions are the largest, but in this situation the  gaps must have accidental nodes. The
 existence or non-existence of the solution at strong intra-pocket repulsion then
 depends on the complex interplay between the prefactors of $\cos 2 \theta$  terms in electron-hole and electron-electron pairing vertices, see Eq. (\ref{inequality}).

\begin{figure}[t]
$\begin{array}{cc}
\includegraphics[width=0.45\columnwidth]{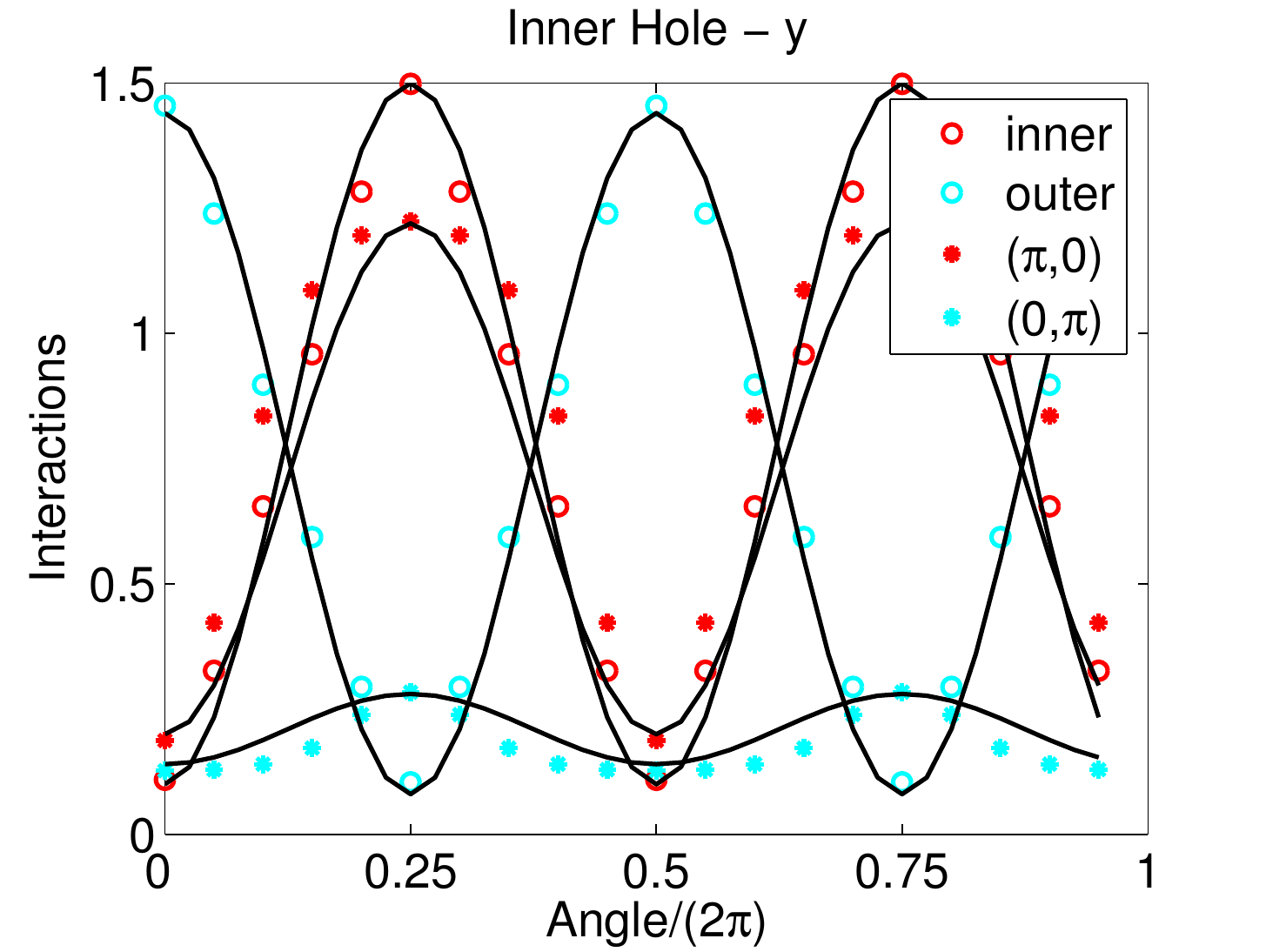}&
\includegraphics[width=0.45\columnwidth]{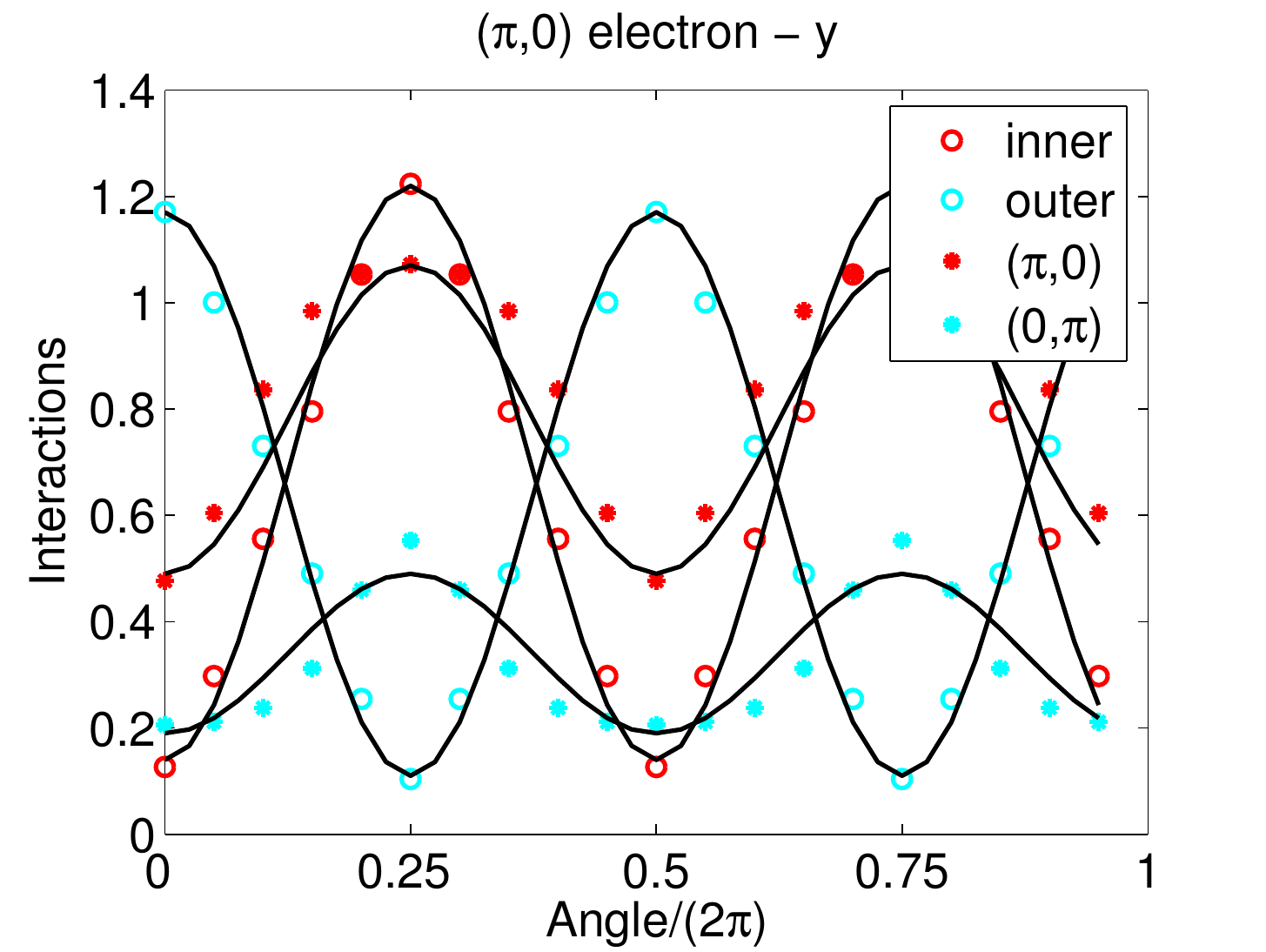}
\end{array}$
\caption{\label{fig:plots_wo_RPA} Representative fits
 of the interactions
$\Gamma_{ij} ({\bf k}_F, {\bf p}_F$) by LAHA for the 4-pocket model.
  $\Gamma_{ij}$ are obtained by
converting the Hamiltonian, Eqs. (\ref{eq:multiorb_Hubbard}), (\ref{eq:multiorb_Hubbard_1}) from the orbital to the band basis.
The symbols
represent interactions computed numerically for the 5-band orbital model using
LDA band structure, the  black lines are the fits using Eqs.
 (\ref{s_4})-(\ref{s_5_d}).  The fit is for the set $U=1.67$, $J=J'=0.21$,
$V =1.46$, and $\mu= 0.08$ (all in eV). A positive $\mu$ corresponds to electron doping.   ${\bf k}_F$ in
$\Gamma_{ij} ({\bf k}_F, {\bf p}_F)$ is selected
along $y$ direction on either an electron or a hole FS
(its location is specified on top of each figure), and ${\bf p}_F$ is varied
 along each of FSs. The angle $\phi$ is measured relative to $k_x$.}
\end{figure}

\subsection{How to extract $U_{ij} (\k,\p)$ from the orbital model?}

So far, in our discussion  $u_{ij}$,  $\alpha_{ij}$, etc, are
 treated as some phenomenological inputs.  To obtain the actual values of these parameters, one needs a microscopic model.
The most commonly considered model for FeSCs is an effective  5-orbital model
 for $Fe$ atoms with local intra-orbital and inter-orbital hopping
integrals and  intra-orbital
and inter-orbital density-density (Coulomb) repulsions, Hund-rule exchange, and the pair hopping term.
\begin{widetext}
\begin{equation}
H_{int} =  \sum_{is}U_{ii} n_{i,s\uparrow} n_{is\downarrow} +
\sum_{i,s,t\neq s} \frac{V_{st}}{2} n_{is} n_{it} - \sum_{i,s,t\neq s} J_{st}\vec{S}_{is}\cdot \vec{S}_{it}  +  \frac{1}{2}  \sum_{i,s,t\neq s} J'_{st} \sum_\sigma c_{is\sigma}^\dagger
c_{is\bar{\sigma}}^\dagger c_{it\bar{\sigma}} c_{it\sigma}\label{eq:multiorb_Hubbard}
\end{equation}
where $n_{is} =  n_{i,s\uparrow} + n_{is\downarrow}$.

The  Hamiltonian $H_{int}$ can be equivalently re-expressed via spin-independent interactions, as
\begin{equation}
H_{int} =  \sum_{is}U n_{i,s\uparrow} n_{is\downarrow} +
\sum_{i,s,t\neq s} \frac{{\bar U}}{2} n_{is} n_{it} + \sum_{i,s,t\neq s}
\frac{J}{2} c_{is\sigma}^\dagger
c_{it\sigma} c^\dagger_{it\bar{\sigma}} c_{it{\bar\sigma}}
+  \frac{1}{2}  \sum_{i,s,t\neq s} J' \sum_\sigma c_{is\sigma}^\dagger
c_{is\bar{\sigma}}^\dagger c_{it\bar{\sigma}} c_{it\sigma}\label{eq:multiorb_Hubbard_1}
\end{equation}
where ${\bar U} = V_{st} + J_{st}/2$.
\end{widetext}

The hopping integrals (36 total) are obtained
from the fit to DFT band structure.~[\onlinecite{ref:Cao}]
For the interaction parameters, the most common approximation is to assume
 that ${\bar U}$, $J$ and $J$  are independent of the orbital
indices $s$ and $t$, as long as $s\neq t$. The model can be also extended to include non-local Fe-Fe interactions via a pnictide~~[\onlinecite{orbital_J}].

The bare parameters in  (\ref{eq:multiorb_Hubbard}) and (\ref{eq:multiorb_Hubbard_1}) are inter-related due to local spin-rotation invariance~~[\onlinecite{Graser,Kuroki}], but that invariance
 is broken if we view (\ref{eq:multiorb_Hubbard}) and (\ref{eq:multiorb_Hubbard_1}) as an effective low-energy model in which the interactions are
dressed by the renormalizations coming from fermions with energies of order bandwidth. By this reason, in most studies $U$, ${\bar U}$, $J$, and $J'$ are treated as independent parameters.

\begin{table*}[htp]
\centering
\begin{tabular}{|c|c|c|c|c|c|c|c|c|c|c|c|c|c|}
\hline
&$u_{h_1h_1}$&$u_{h_2h_2}$&$u_{h_1h_2}$&$u_{h_1e}$&$\alpha_{h_1 e}$&$u_{h_2e}$&$\alpha_{h_2 e}$&$u_{ee}$&$\alpha_{ee}$&$\beta_{ee}$\\
\hline
NSF&$0.8$&$0.76$&$0.78$&$0.46$&$-0.24$&$0.4$&$-0.30$&0.77&0.14&0.09\\
\hline
SF&$2.27$&$2.13$&$2.22$&$4.65$&$-0.34$&$2.29$&$-0.22$&3.67&0.15&0.04\\
\hline
&$\tilde{u}_{h_1 h_1}$&$\tilde{u}_{h_2 h_2}$&$\tilde{u}_{h_1 h_2}$&$\tilde{u}_{h_1 e}$&$\tilde{\alpha}_{h_1 e}$&$\tilde{u}_{h_2 e}$&$\tilde{\alpha}_{h_2 e}$&$\tilde{u}_{ee}$&$\tilde{\alpha}_{ee}$&$\tilde{\beta}_{ee}$\\
\hline
NSF&$0.7$&$0.66$&$-0.68$&$-0.25$&$-0.58$&$0.24$&$-0.42$&0.11&-0.5&0.25\\
\hline
SF&$1.50$&$1.40$&$-1.50$&$-3.73$&$-0.44$&$1.44$&$-0.32$&1.03&-0.49&-0.02\\
\hline
\end{tabular}
\caption{Table for $s$-wave and $d-$wave parameters for
 the same set as in Fig.~\ref{fig:plots_wo_RPA}.
  NSF and SF mean the bare interaction without the
spin-fluctuation component and the full interaction, respectively.} \label{tab:s-set6}
\end{table*}

We now need to convert  (\ref{eq:multiorb_Hubbard}), (\ref{eq:multiorb_Hubbard_1}) into the
band basis and re-express it in the form of Eq. (\ref{rev_2}). This is done
by transforming into the momentum space, introducing new, hybridized operators, which diagonalize the hopping Hamiltonian, and re-expressing the interaction terms in
 (\ref{eq:multiorb_Hubbard}) or (\ref{eq:multiorb_Hubbard_1}) in terms of these new operators.  The end result of this procedure is the effective Hamiltonian in the band basis which has the form of Eq. (\ref{rev_2}) with $U_{ij} (\k,\p)$
 given by
\begin{eqnarray}
{U}_{ij} (\k,\k') & = & \sum_{stpq} \alpha_{i}^{t,*}(-\k)  \alpha_{i}^{s,*}(\k)
\mathrm{Re}\left[{\Gamma}_{st}^{pq} (\k,\k') \right] \nonumber \\
&& \times \alpha_{j}^{p}(\k')  \alpha_{j}^{q}(-\k'),
\label{eq:fullGamma}
\end{eqnarray}
where
$[{U}_{st}^{pq} (\k,\k')$
 are linear combinations of $U, {\bar U}$, $J$ and ${\bar J}$,
 and  $\alpha_{i}^p$ is the matrix element connecting
 the original fermionic operator $c_p$ in the orbital basis with the
new  fermionic operator $a_i$  on FS $i$ in the band basis.
  The matrix elements $\alpha_i^p$
 contain information which orbitals mostly contribute to a particular segment
 of a  particular FS~~[\onlinecite{Graser,peter}].  Because of this,
 the interaction $U_{ij} (\k,\p)$ in the band basis
generally depends on the angles
 along different FSs and  contains components in
all representations of the tetragonal $D_{4h}$ group.

The angle dependence of $s-$wave and
$d_{x^2-y^2}$ vertices agrees by symmetry with Eqs (\ref{s_4})-(\ref{s_5_d}).
What s a'priori unknown is how well the interactions can be approximated by the
 leading angle harmonics, i.e., whether the terms labeled as dots in  (\ref{s_4})-(\ref{s_5_d}) can actually be neglected.
This issue was analyzed in detail in Ref.~[\onlinecite{maiti_11}], and the answer is affirmative -- the leading anhular harmonic approximation (LAHA) works rather well. In Fig.\ref{fig:plots_wo_RPA} I show representative fits for a particular set of parameters and in Table 1, in the lines marked NSF, 
I show $u_{eh}$ and other parameters, extracted from the fit (NSF stangs for "no spin fluctuations", meaning that this is for the bare interaction, without extra spin-fluctuation component (see below)).  The results somewhat
vary depending on the values of $U$, $V$, $J$, $J'$, but in general
 intra-band  interactions in the $s-$wave channel,  $u_{ee}$ and $u_{hh}$,
 exceed interband $u_{he}$.  This is not surprising because $u_{ee}$ and $u_{hh}$ are essentially Coulomb interactions at small momentum transfers, while $u_{eh}$
 is the interaction at large momentum transfer, and it should be smaller
 on general grounds.   Only when $V=J=J'=0$,
the interaction in the band basis becomes independent on the momentum~~[\onlinecite{cee}], i.e., $u_{ee} = u_{hh} = u_{he}$ (this was termed ``Coulomb avoidance'' in Ref.~[\onlinecite{mazin_schmalian}]).  According to Table \ref{tab:s-set6},
 intra-band interactions are also larger in the d-wave channel: ${\tilde u}_{h_ih_i} {\tilde u}_{ee} > {\tilde u}^2_{h_ie}$, although the reasons why this is
 the case are not transparent.

\begin{figure}[htp]
$\begin{array}{cc}
\includegraphics[width=0.4\columnwidth]{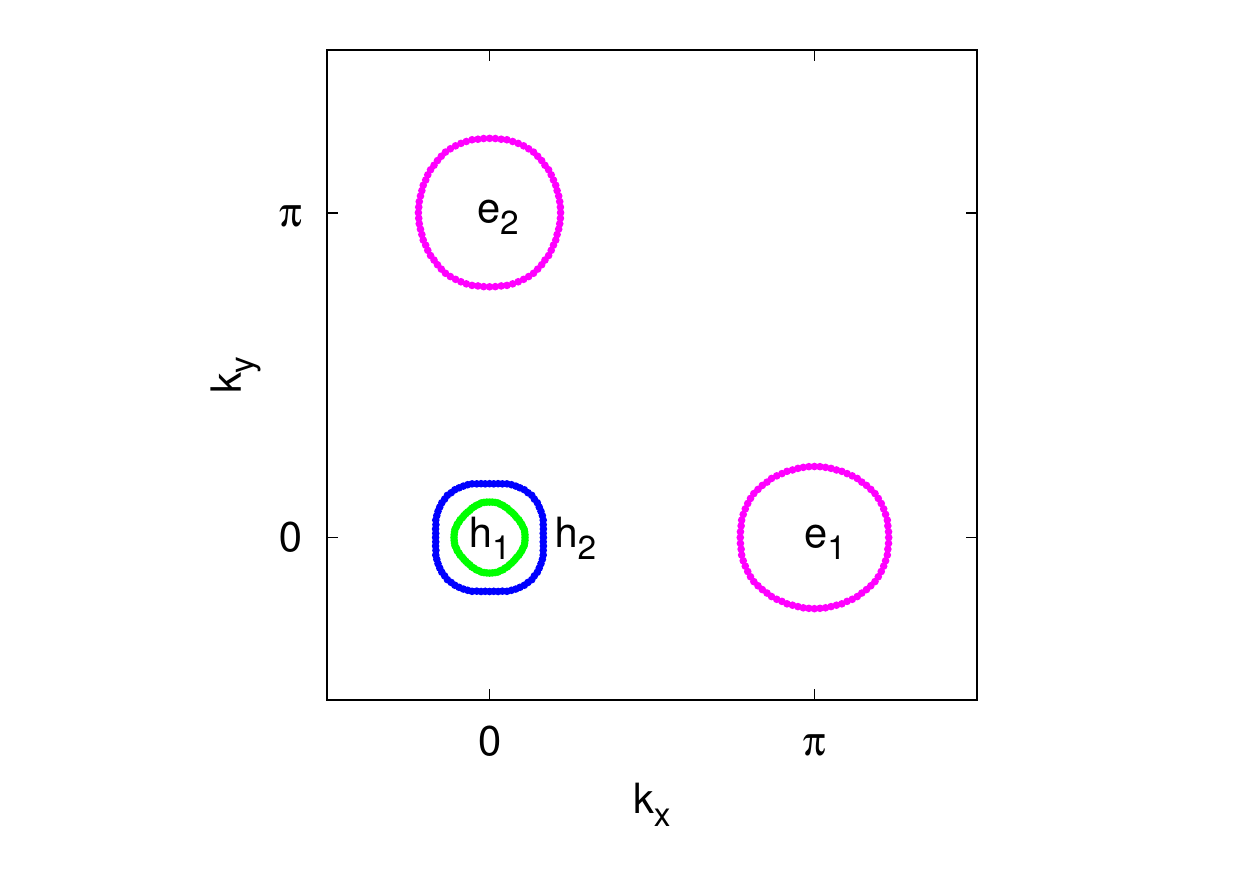}&
\includegraphics[width=0.4\columnwidth]{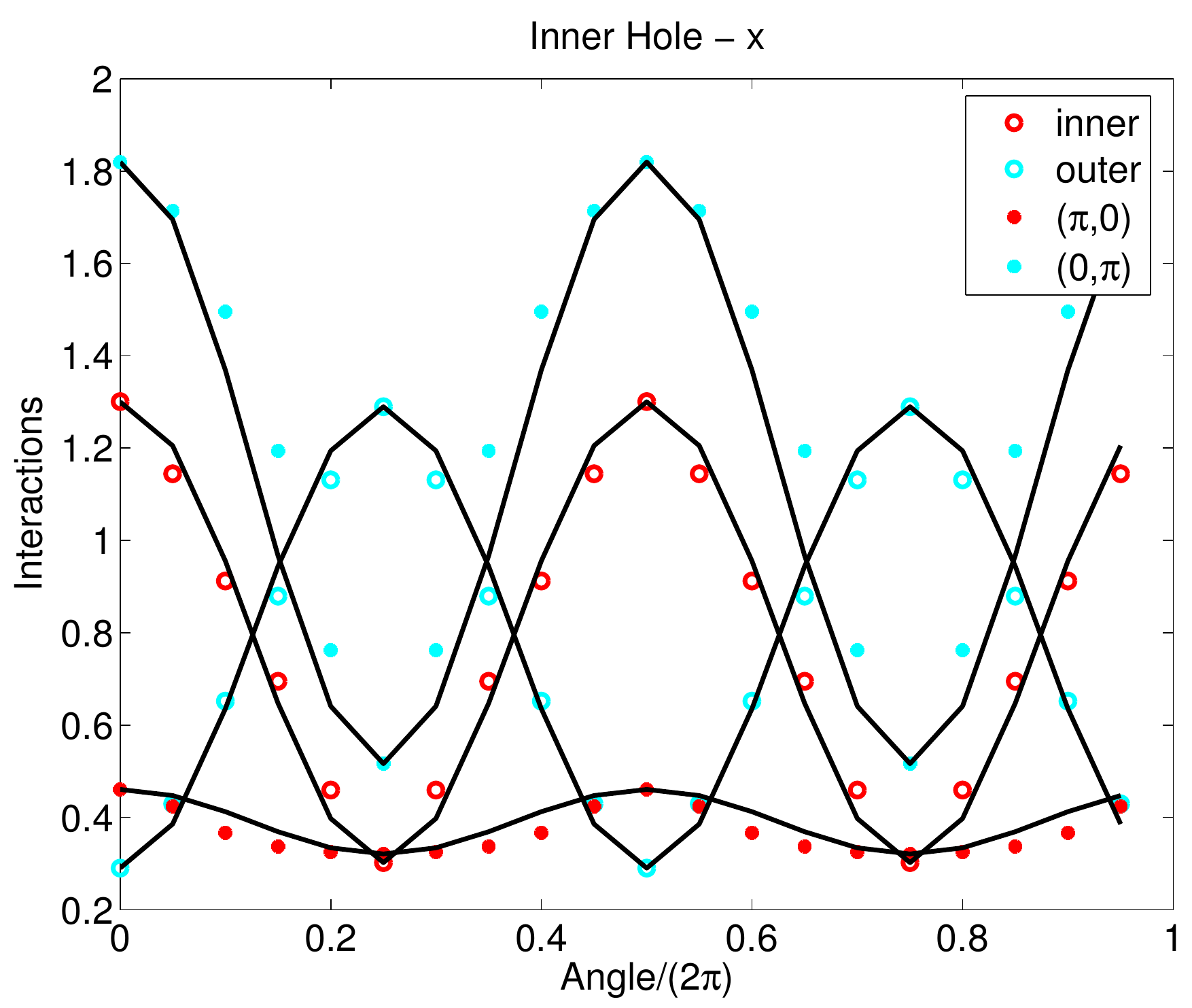}\\
\includegraphics[width=0.4\columnwidth]{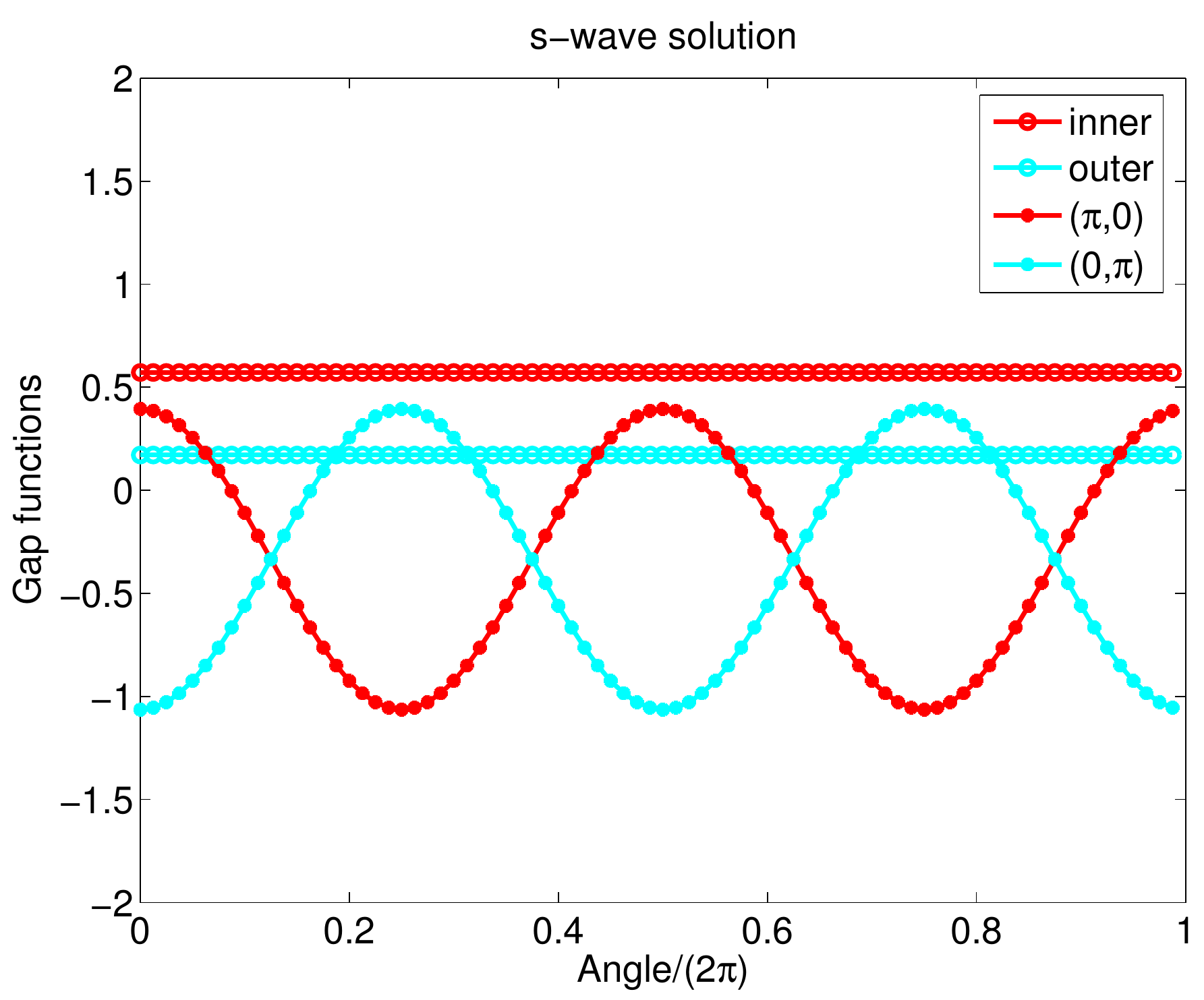}&
\includegraphics[width=0.4\columnwidth]{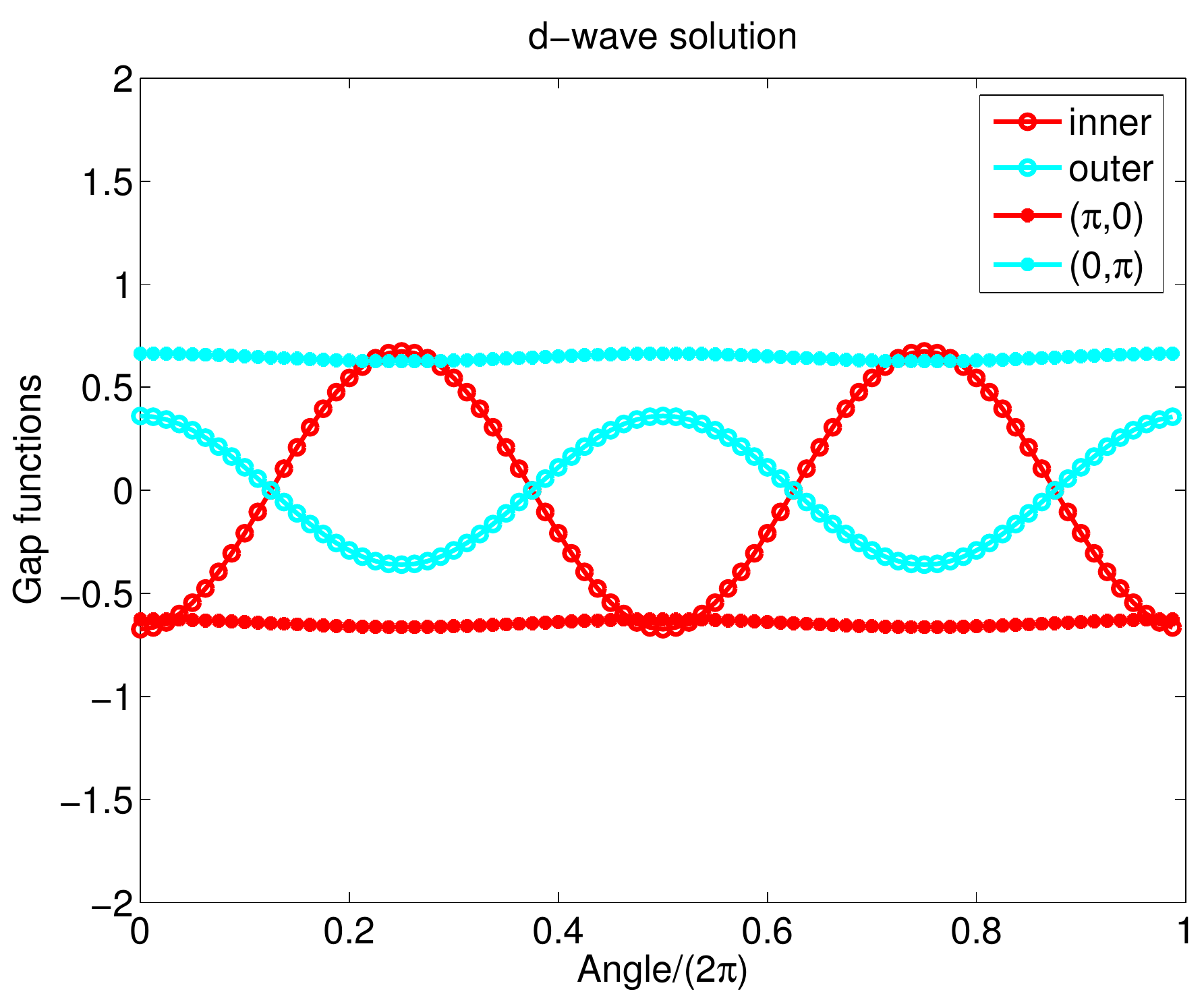}\\
\end{array}$
\caption{\label{fig:rev_2}
 Representative case of small/moderate electron doping, when both hole and electron pockets are present.
 Panel a -- the FS, panel b --  representative fits
 of the interactions by LAHA (the dots are RPA results, the lines are LAHA expressions,  Eqs (\ref{s_4})-(\ref{s_5_d})). Panels c and d --
  the eigenfunctions in $s-$wave and $d-$wave channels for the largest $\lambda^s$ and $\lambda^d$.  From Ref.~[\onlinecite{maiti_11}].}
\end{figure}

\begin{figure}[htp]
$\begin{array}{cc}
\includegraphics[width=0.45\columnwidth]{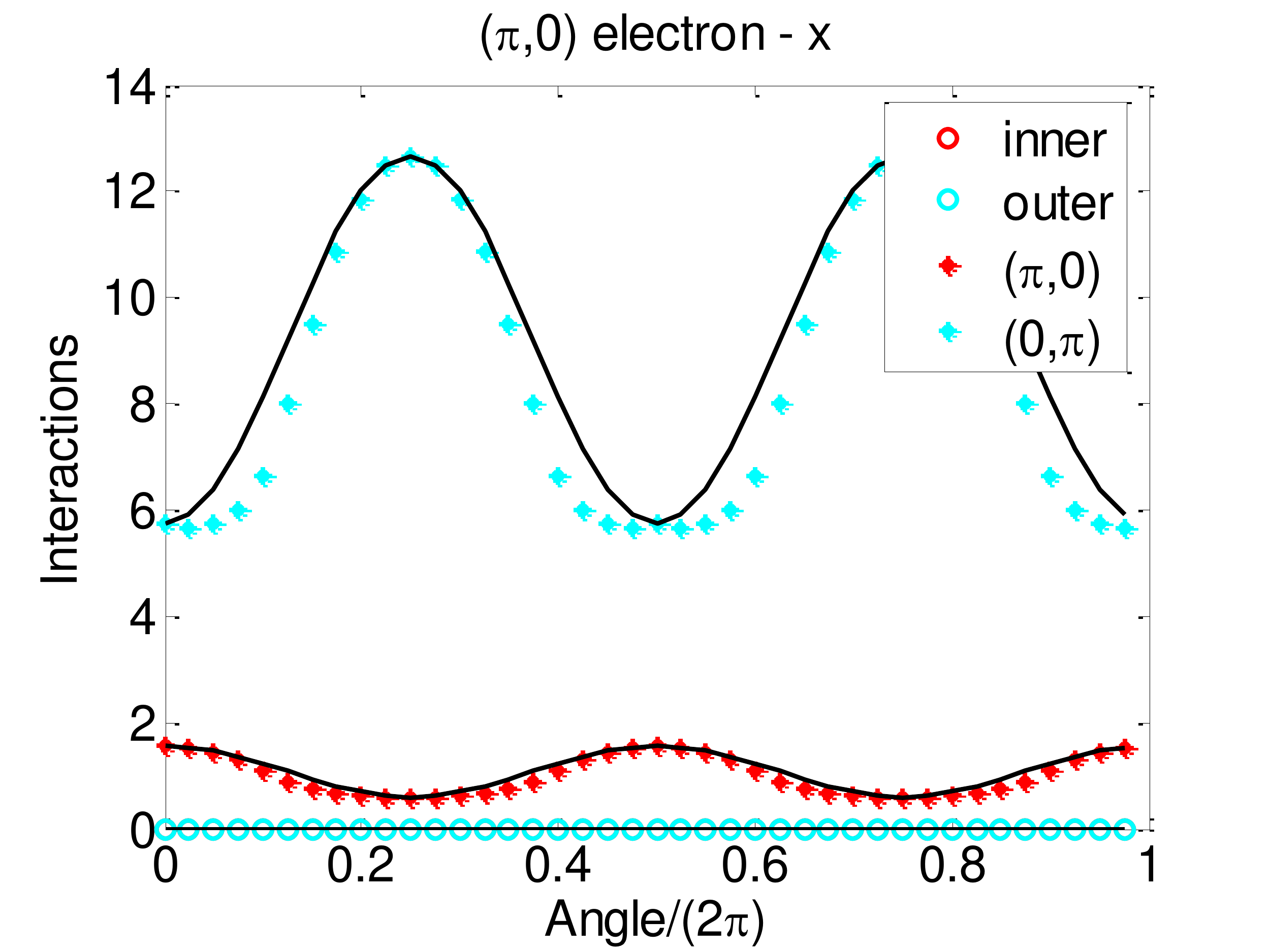}&
\includegraphics[width=0.45\columnwidth]{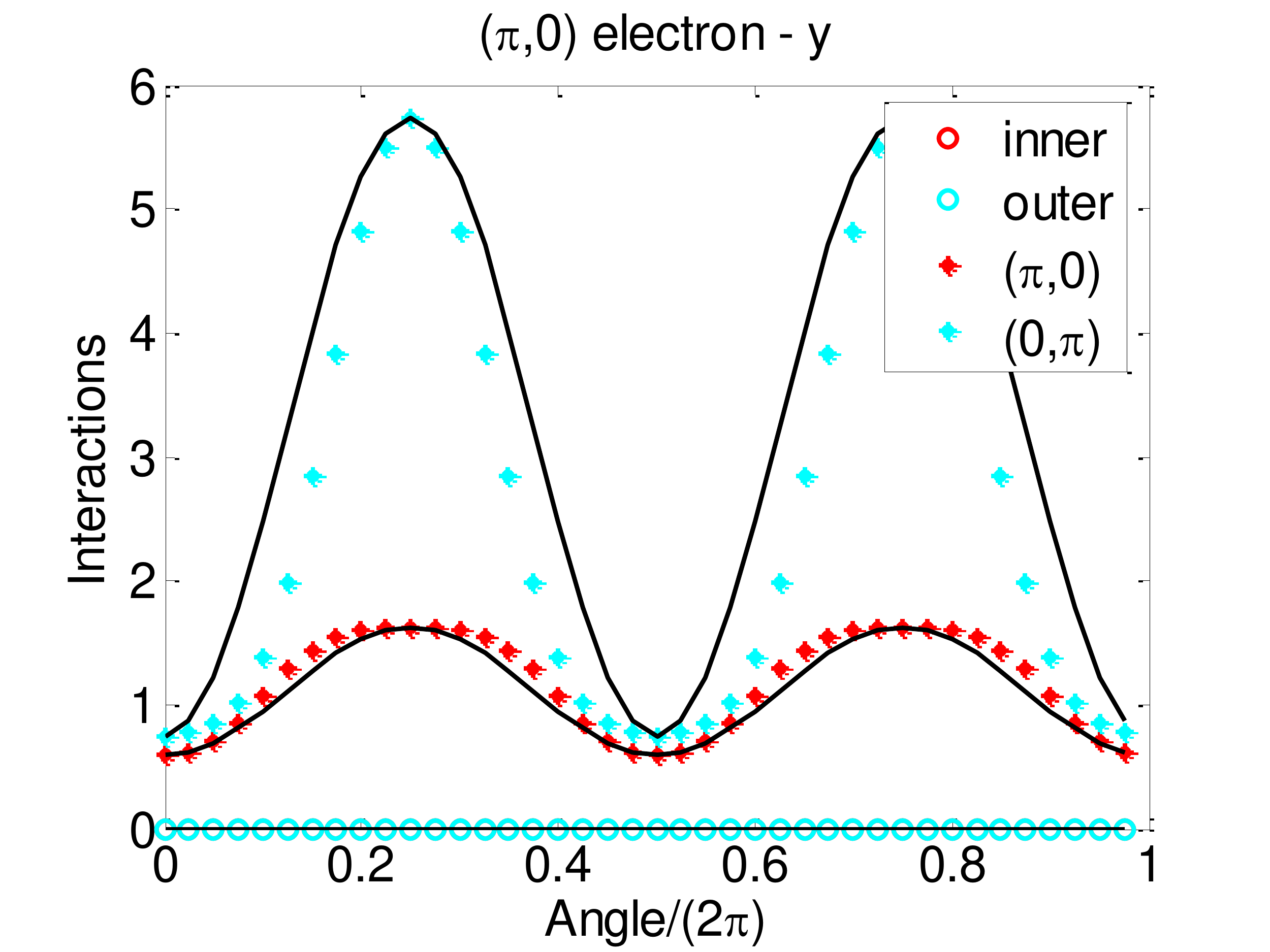}\\
\includegraphics[width=0.45\columnwidth]{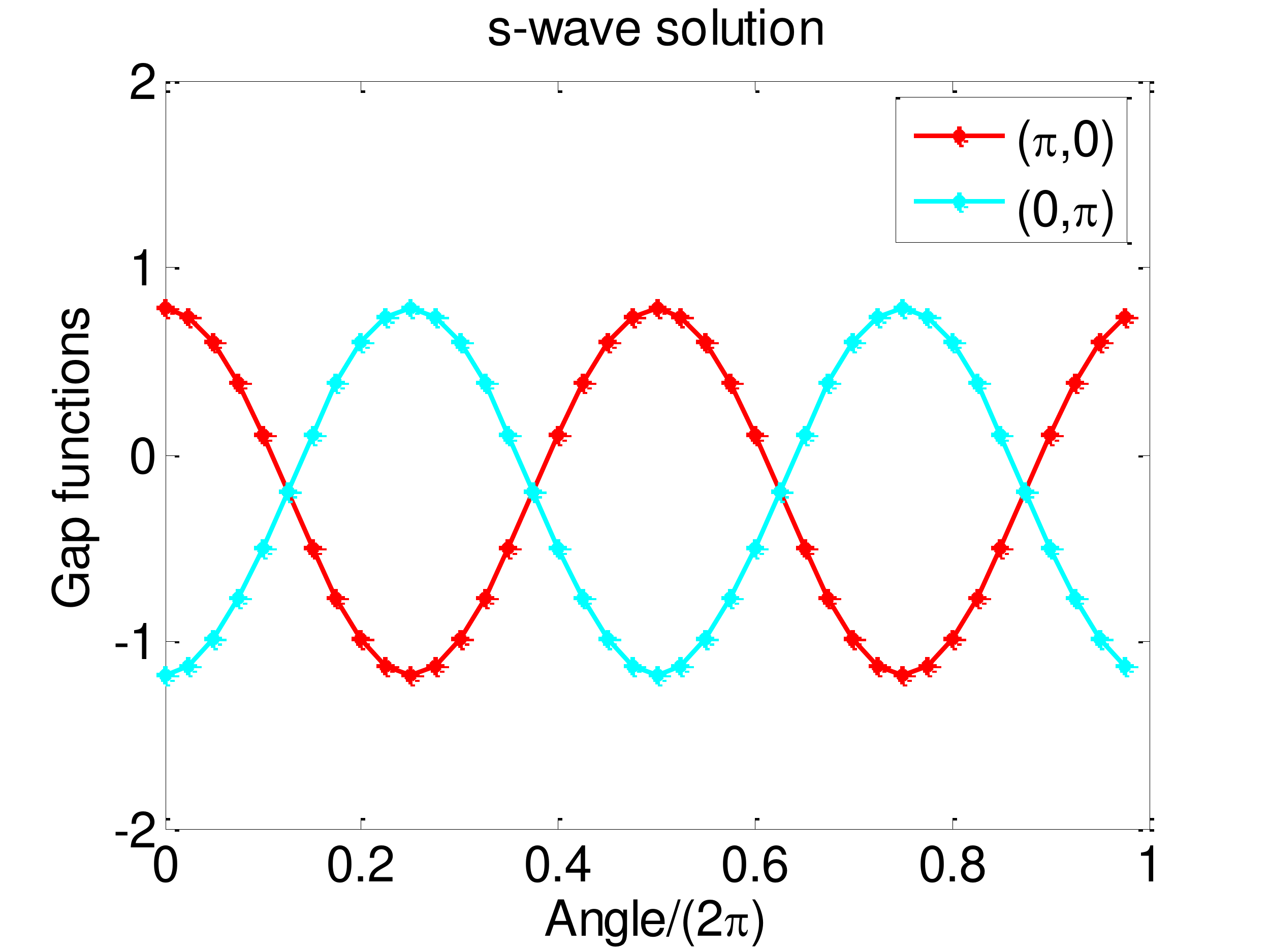}&
\includegraphics[width=0.45\columnwidth]{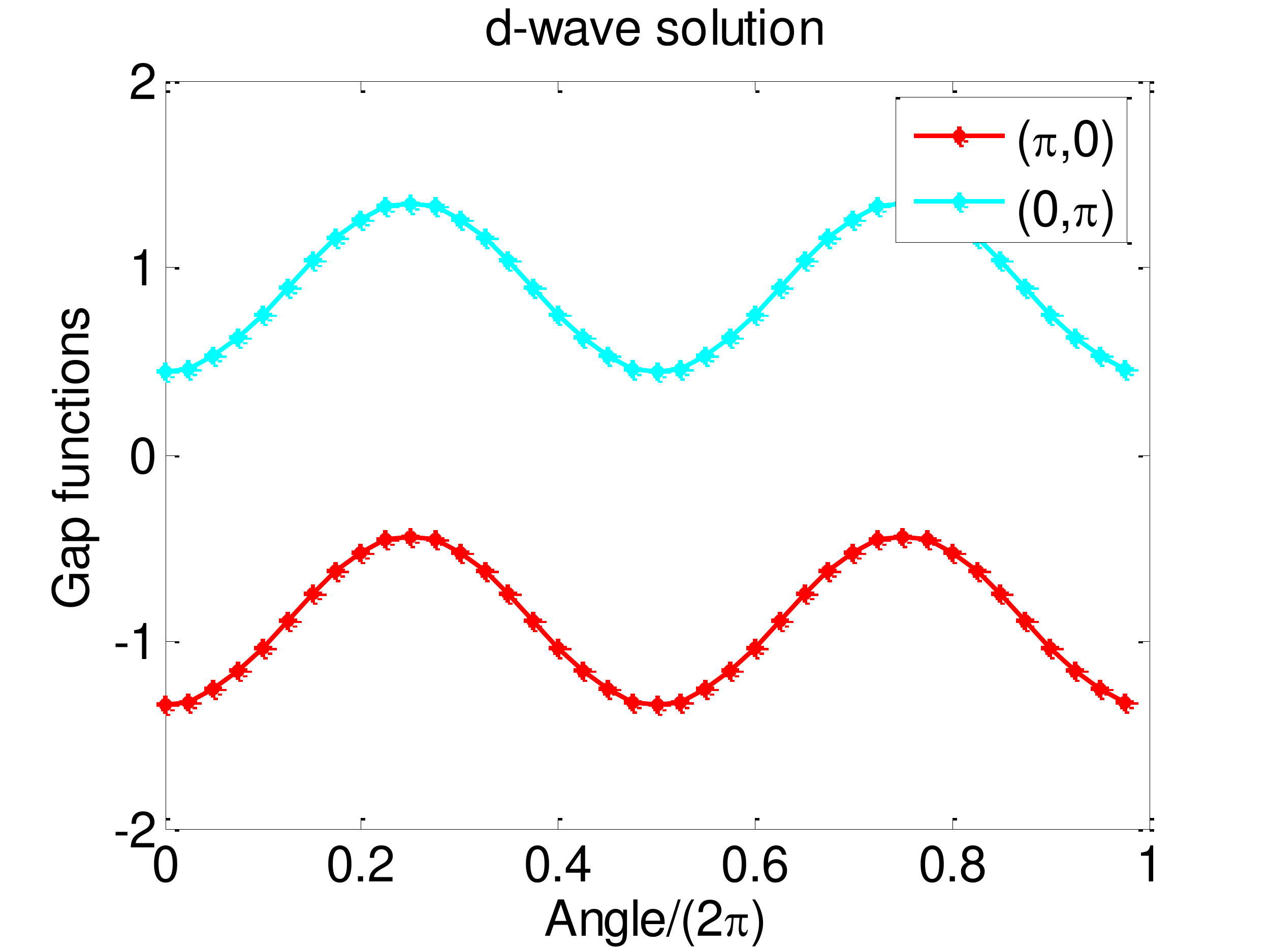}
\end{array}$
\caption{\label{fig:rev_3} The fits of the RPA
interactions by LAHA and the structure of $s-$wave and $d-$wave gaps
 for the case of heavy electron doping, when only electron
FSs are present.   From Ref.~[\onlinecite{maiti_11}].}
\end{figure}

\subsection{Doping dependence of the couplings, examples}
\label{sec:5_1}

I now present the results for the gap structure at various doping, obtained within LAHA, but including dressing of interactions by spin-fluctuations in RPA
 (lines marked "SF" in the Tables). 
The results for hole and electron doping differ, and I present them separately. I will follow  Refs. [\onlinecite{ch_annual,maiti_11}].

\subsubsection{Electron doping}

For small and moderate electron dopings, the FS consists of 4 pockets -- two hole FS at $(0,0)$ and two electron FSs at $(0,\pi)$ and $(\pi,0)$.
 Typical fits by LAHA, the parameters extracted from the fits,
 and the solutions in s-wave and d-wave channels are shown in Fig. \ref{fig:rev_2}  and in Table \ref{tab:1}.
It turns out~~[\onlinecite{maiti_11}]
 that some system properties are sensitive to the choice of
the parameters, but some are quite universal.
 The parameter-sensitive properties are the presence or absence of accidental nodes in the $s$-wave gap (although for most of parameters
the gap does have nodes, as in Fig.~\ref{fig:rev_2}) and
 the gap symmetry itself, because for most of input parameters and dopings $\lambda^s$ and $\lambda^d$ remain comparable as long as both hole and electron FSs are present (see Table  \ref{tab:1}).
That $d-$wave state is a strong competitor in 4-pocket systems
 has been first emphasized in Refs.~~[\onlinecite{Graser,Kuroki}]. The authors of~~[\onlinecite{Graser}] hinted that different FeSCs may have different symmetry even
 for the same topology of the FS.

 The universal
observation is that the driving force for attraction in both
$s$-wave {\it and} $d$-wave channels is strong inter-pocket
electron-hole interaction ($u_{h_i e}$ and ${\tilde u}_{h_i e}$
terms)  {\it no matter how small the hole or electron pockets are}.
 The gap structure actually changes only little with doping as long as both hole and electron pockets are present. \\

\begin{table*}[htp]
\caption{\label{tab:1} Some of the LAHA parameters  extracted from the LAHA fit
 in Figs. (\ref{fig:rev_2}) and (\ref{fig:rev_3})
 for electron doping. Blocks (i)  corresponds
to  Fig. (\ref{fig:rev_2}), block (ii) corresponds to  Fig. (\ref{fig:rev_3}) (no hole pockets). From Ref~~[\onlinecite{maiti_11}].}
\begin{ruledtabular}
\begin{tabular}{lcccccclccccccrccc}
& \multicolumn{6}{c}{(i)} & & \multicolumn{3}{c}{(ii)}\\
 \cline{2-7} \cline{9-11}
$s$-wave&$u_{h_1h_1}$&$u_{h_1e}$&$\alpha_{h_1 e}$&$u_{ee}$&$\alpha_{e e}$&$\lambda_s$& & $u_{e e}$&$\alpha_{e e}$&$\lambda_s$\\
&0.8&0.79&-0.19&0.91&0.05&0.25& &3.65&0.20&0.1\\
 \cline{2-7} \cline{9-11}
$d$-wave&$\tilde{u}_{h_1h_1}$&$\tilde{u}_{h_1e}$&$\tilde{\alpha}_{h_1 e}$&$\tilde{u}_{e e}$&$\tilde{\alpha}_{e e}$&$\lambda_d$& & $\tilde{u}_{e e}$&$\tilde{\alpha}_{e e}$&$\lambda_d$\\
&0.50&-0.39&-0.46&-0.04&1.5&0.37& &-2.57&0.29&5.9\\
\end{tabular}
\end{ruledtabular}
\end{table*}

{\it Extreme electron doping}\\

The situation changes qualitatively once the hole pockets disappear
(Fig.~\ref{fig:rev_3}). It is clear from  Table~\ref{tab:1} that now the
$d$-wave channel becomes the dominant one. Comparing the LAHA parameters for the two
dopings, we see the reason: once the hole pockets disappear, a
direct $d$-wave electron-electron interaction ${\tilde u}_{ee}$
becomes strong and attractive. The argument why this happens
 is as follows:~~[\onlinecite{maiti_11}] ${\tilde u}_{ee}$ is an
antisymmetric combination of intra-pocket and inter-pocket
electron-electron interactions ${\tilde u}_{ee} = u_\mathrm{intra}^{ee} - u_\mathrm{inter}^{ee}$. Both $u_\mathrm{inter}^{ee}$ and
$u_\mathrm{intra}^{ee}$ are positive (repulsive), but the
sign of ${\tilde u}_{ee}$ depends on the interplay between
$u_\mathrm{inter}^{ee}$ and $u_\mathrm{intra}^{ee}$.
 As long as hole FSs are  present, SF are peaked near $\mathbf{q}=(0,\pi)$ and
$(\pi,0)$, which are an equal distance from the relevant momenta
$\mathbf{q}=0$ for $u_\mathrm{intra}^{ee}$ and $\mathbf{q}=(\pi,\pi)$ for $u_\mathrm{inter}^{ee}$.
In this situation, $u_\mathrm{intra}^{ee}$ and $u_\mathrm{inter}^{ee}$
 remain close in magnitude,
and ${\tilde u}_{ee}$ is small.
 Once the hole pockets disappear, the peak in the RPA spin
susceptibility shifts towards $(\pi,\pi)$ ~~[\onlinecite{graser_11}] and
$u_\mathrm{inter}^{ee}$ increases more due to the SF component than
$u_\mathrm{intra}^{ee}$. A negative $u_\mathrm{intra}^{ee} -
u_\mathrm{inter}^{ee}$ then gives rise to a ``plus-minus'' gap on
the two electron FSs. The gap changes sign under $k_x \to k_y$ and
therefore has $d_{x^2-y^2}$ symmetry. This pairing mechanism is
 essentially identical to spin-fluctuation scenario for d-wave pairing in the cuprates~~[\onlinecite{scalapino_1}].

There are other proposals for the gap structure at extreme electron doping. The authors of 
Refs.~~[\onlinecite{yu_11,bernevig}]  argued that the gap symmetry may be nodeless $s-$wave (equal sign of the gap on the pockets at $(0,\pi)$ and $\pi,0)$,
 if one uses for
 electron-electron interaction the weak coupling version of the
$J_1-J_2$ model.  Another proposal for strongly electron-doped FeSCs is $s^{++}$ pairing driven by orbital fluctuations~~[\onlinecite{kontani_se}].
And yet another proposal~\cite{khodas,mazin_a,hu} is that the pairing state in FeSCs with only electron pocket present is $s^{+-}$, with the sign
  change between the hybridized electron pockets.  Such a state emerges if one includes into consideration the hybridization of the two electron pockets. 
   In this novel $s^{+-}$ state, all electron states are gapped, yet because of sign change, there is a spin resonance at momenta which is roughly a distance between the electron pockets in the unfolded BZ~\cite{khodas_2}.
\begin{figure}[htp]
$\begin{array}{cc}
\includegraphics[width=0.45\columnwidth]{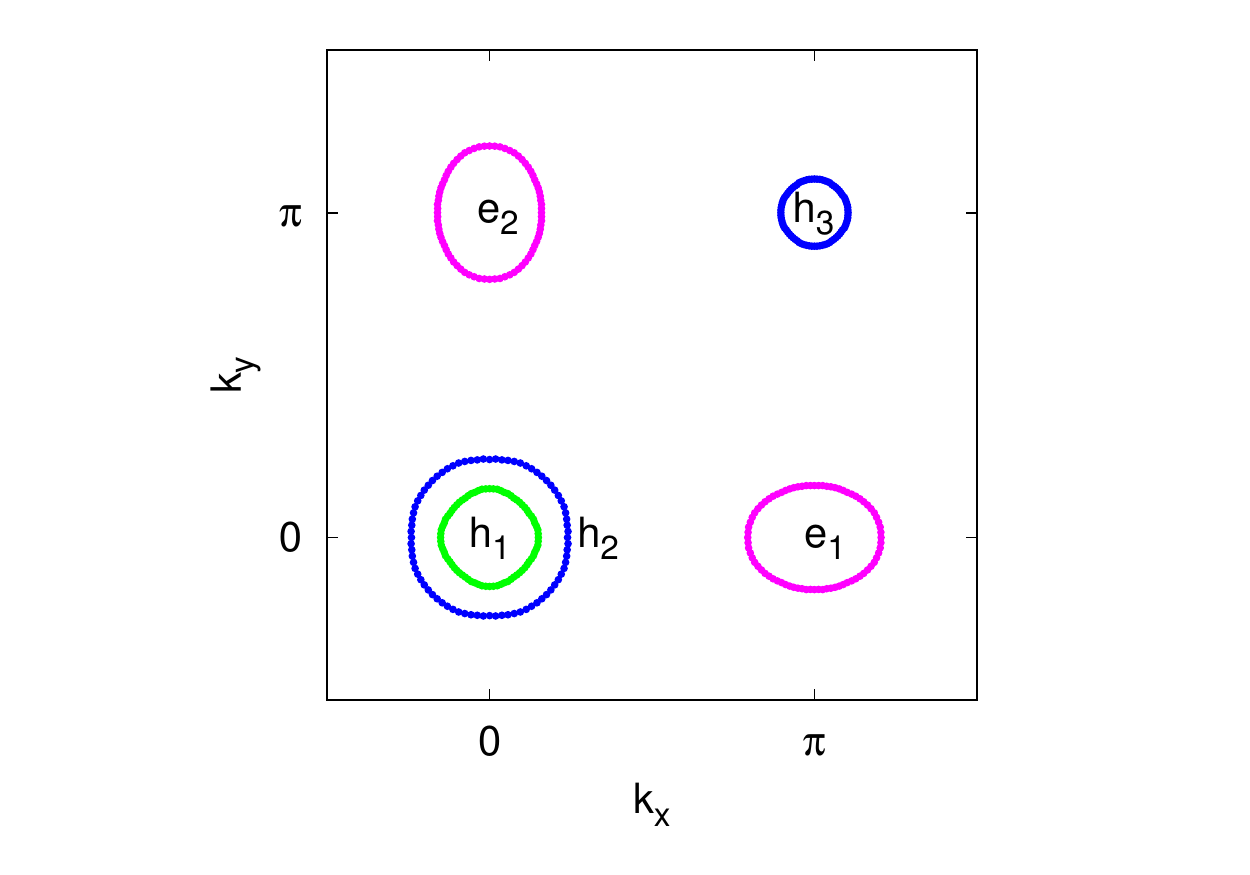}&
\includegraphics[width=0.45\columnwidth]{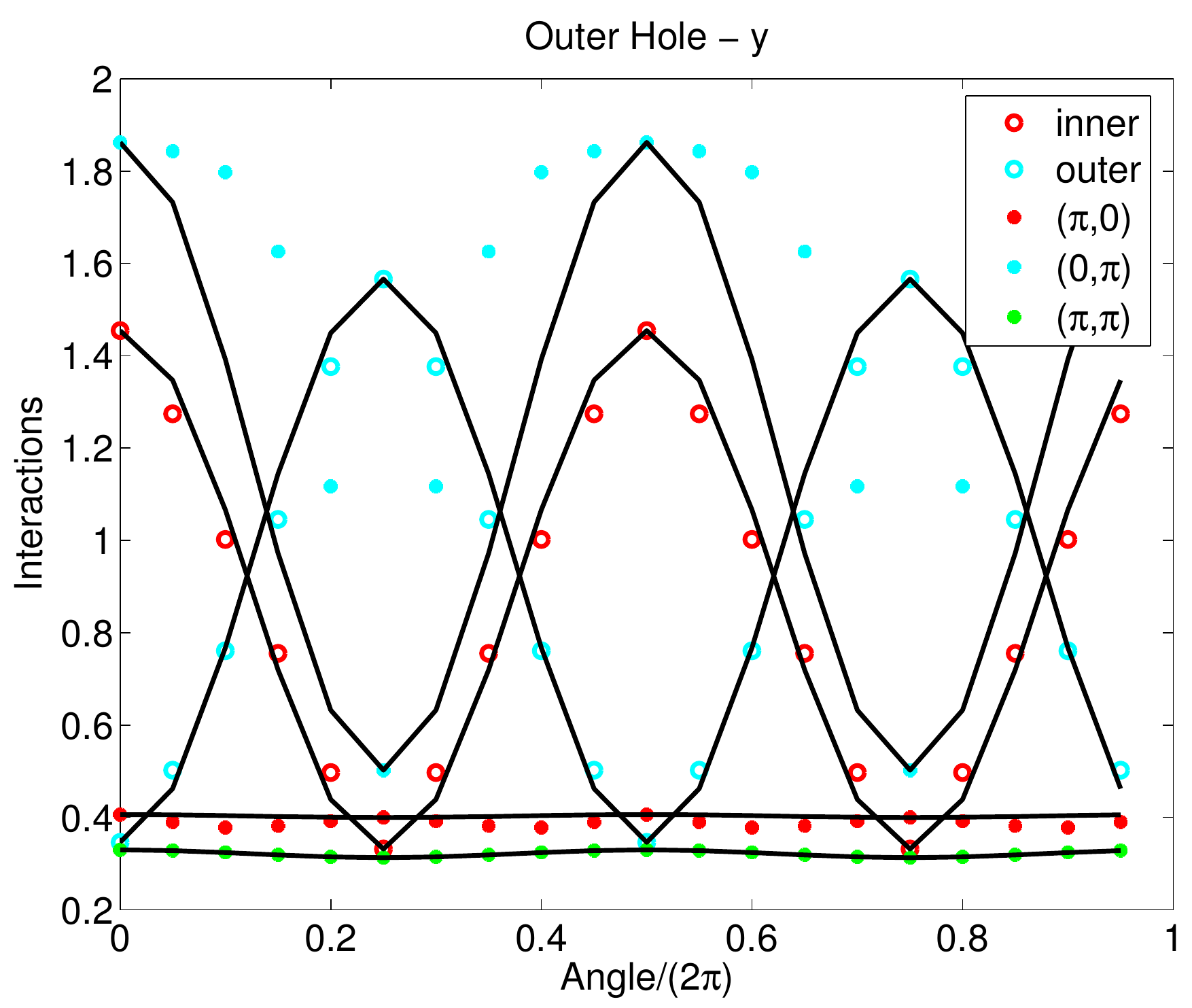}\\
\includegraphics[width=0.45\columnwidth]{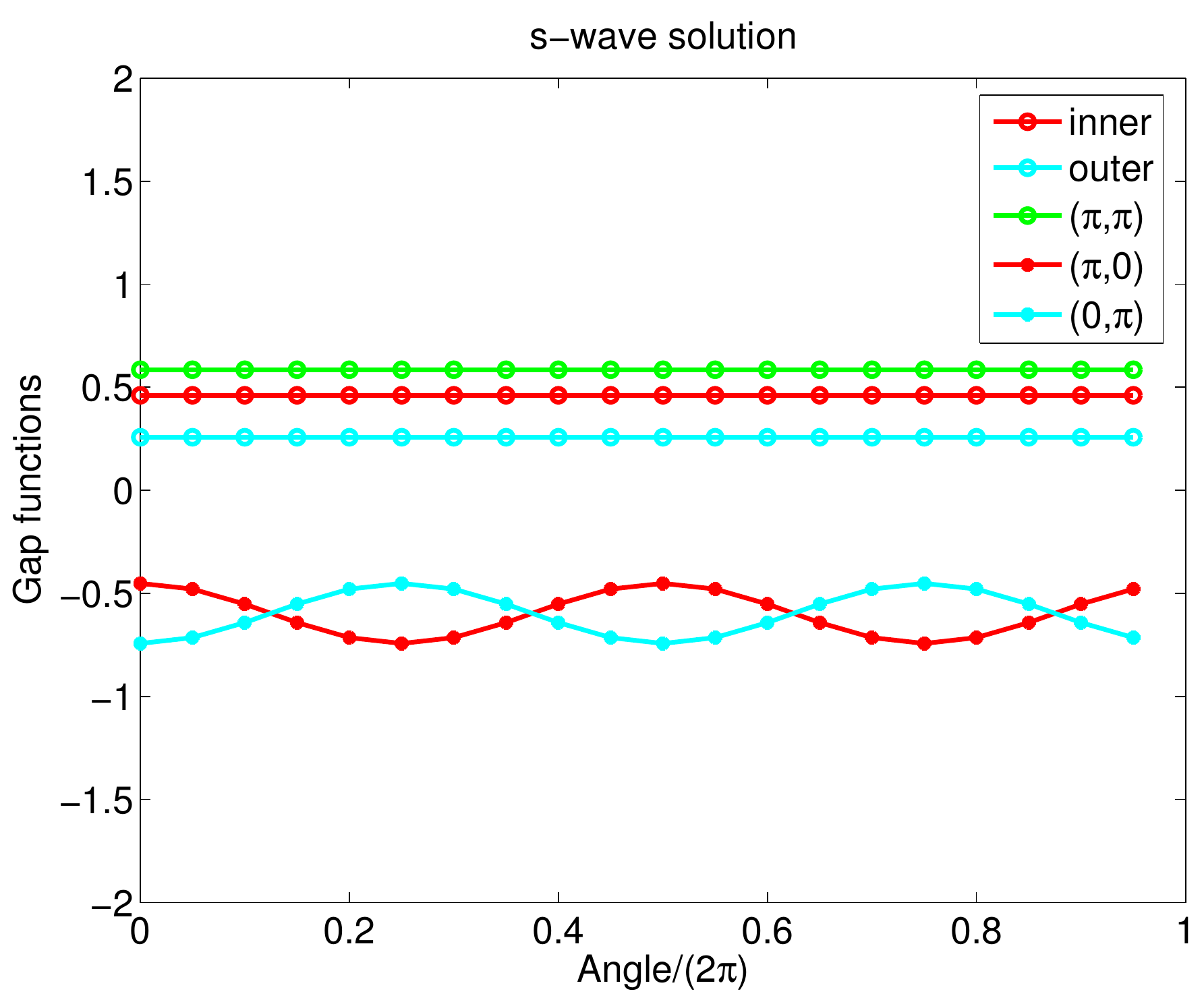}&
\includegraphics[width=0.45\columnwidth]{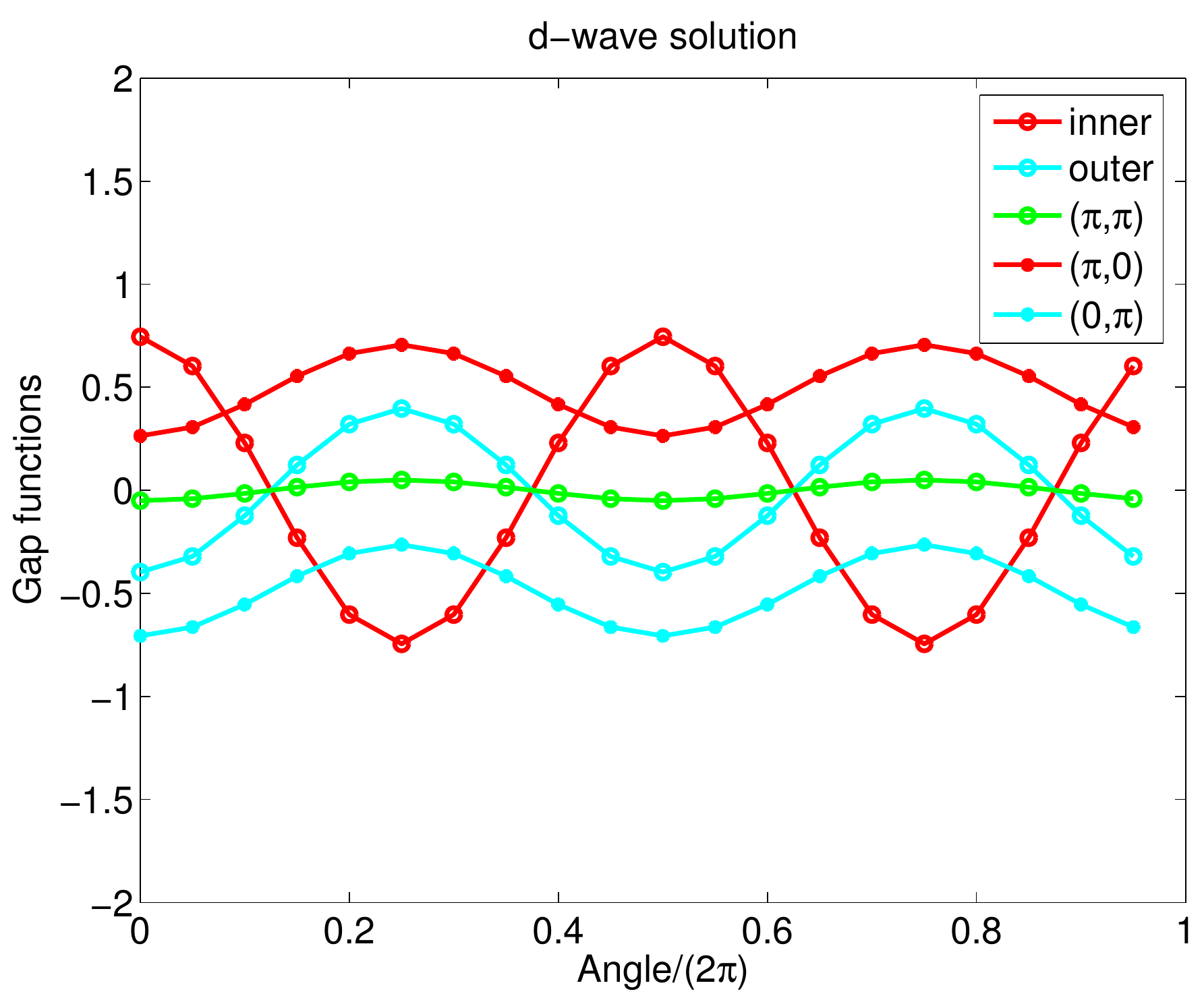}\\
\end{array}$
\caption{\label{fig:rev_4}
Representative case of small/moderate hole doping, when both hole and electron pockets are present.
 Panel a -- the FS, panel b --  representative fits
 of the interactions by LAHA (the dots are RPA results, the lines are LAHA expressions,  Eqs (\ref{s_4})-(\ref{s_5_d})). Panels c and d --
  the eigenfunctions in $s-$wave and $d-$wave channels for the largest $\lambda^s$ and $\lambda^d$.  From Ref.~[\onlinecite{maiti_11}].}
\end{figure}

\begin{figure}[htp]
$
\begin{array}{cc}
\includegraphics[width=0.45\columnwidth]{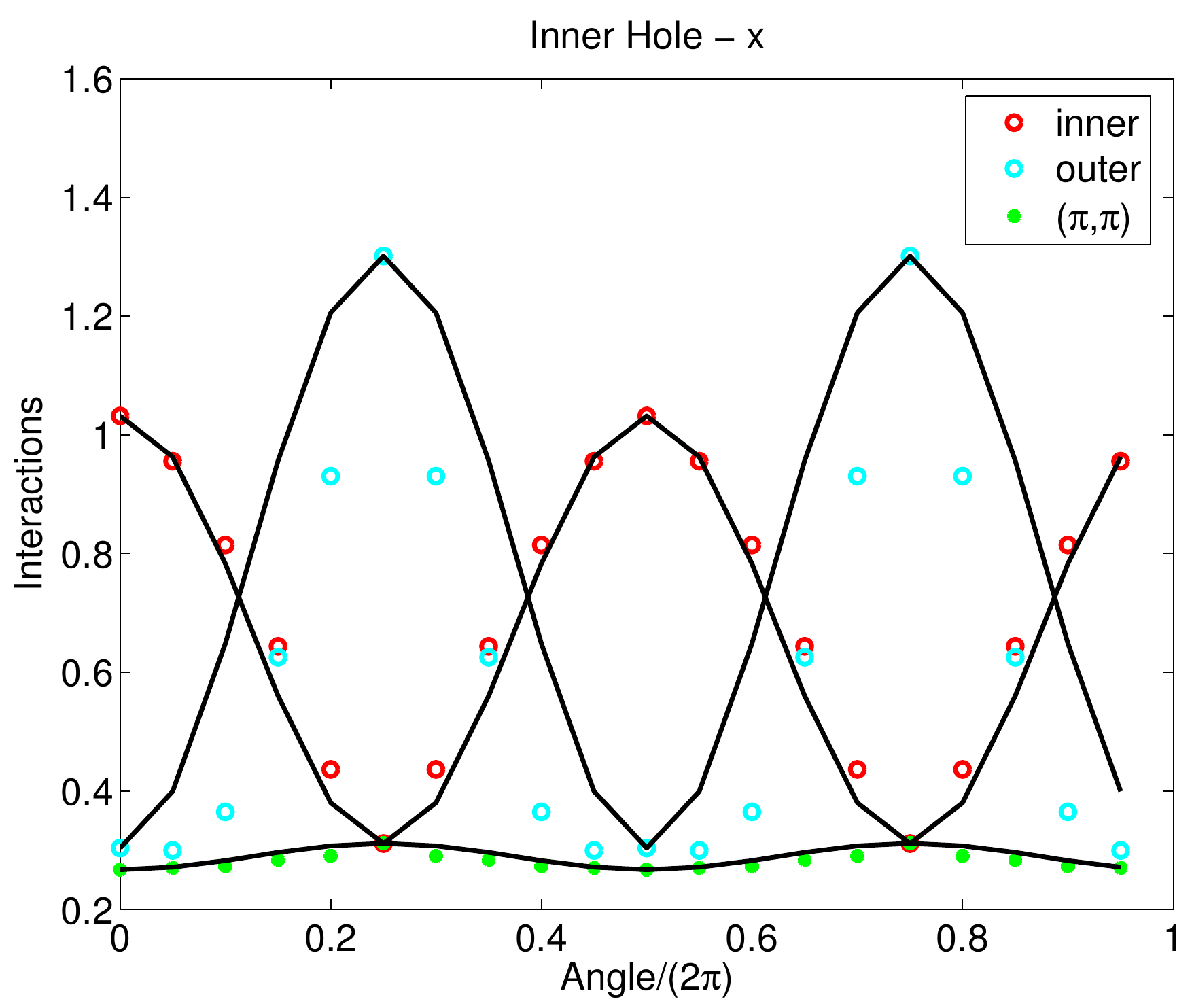}&
\includegraphics[width=0.45\columnwidth]{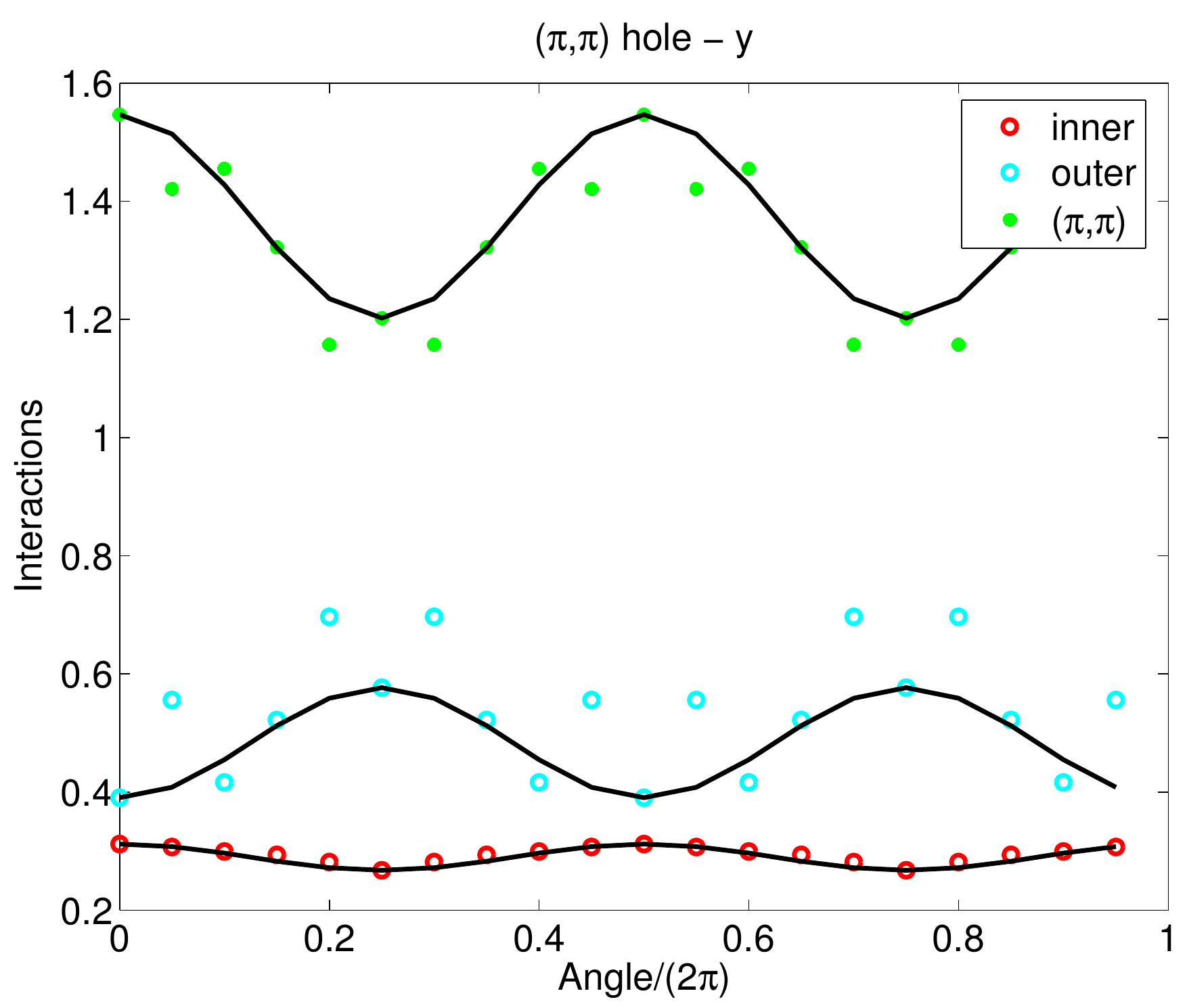}\\
\includegraphics[width=0.45\columnwidth]{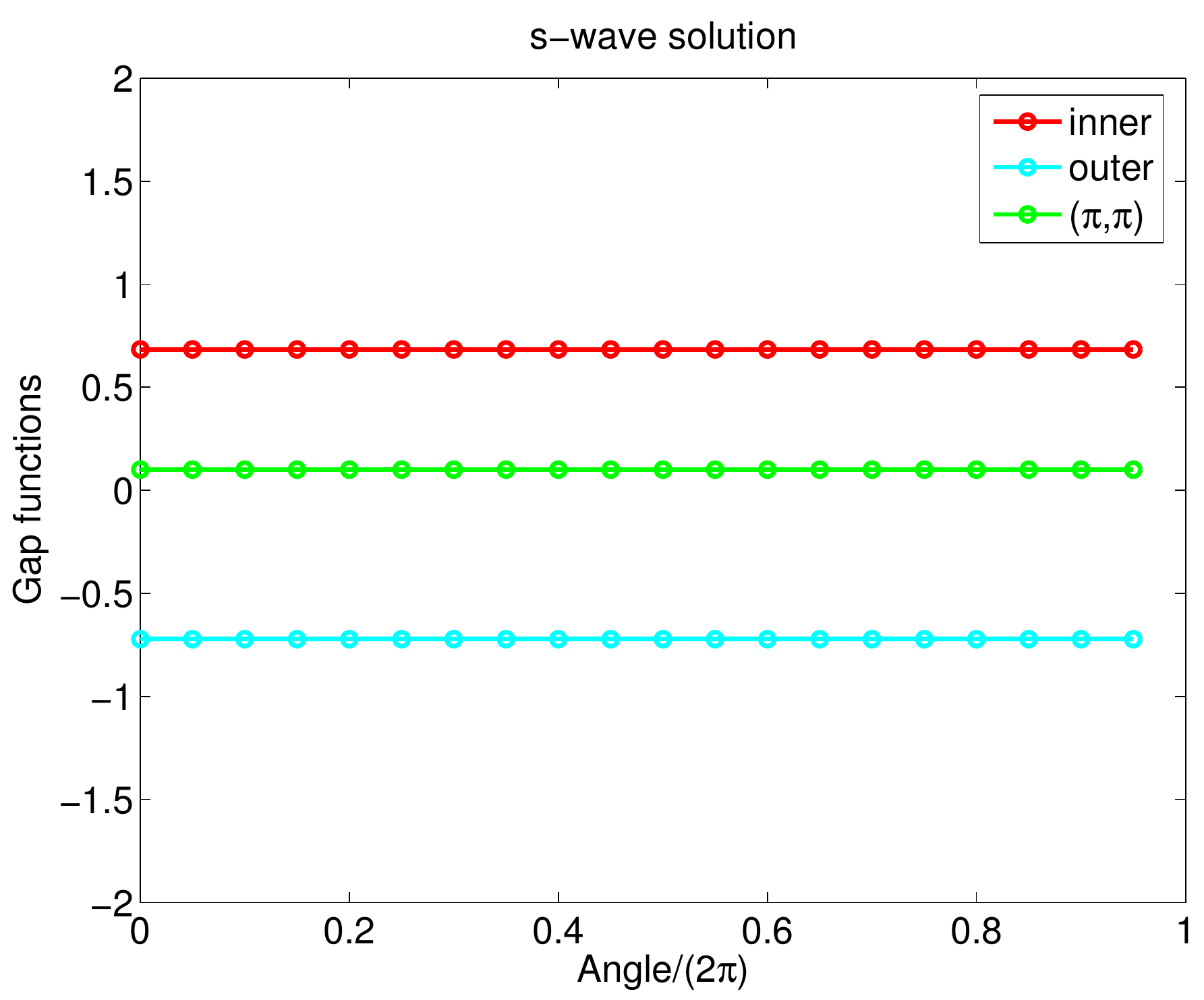}&
\includegraphics[width=0.45\columnwidth]{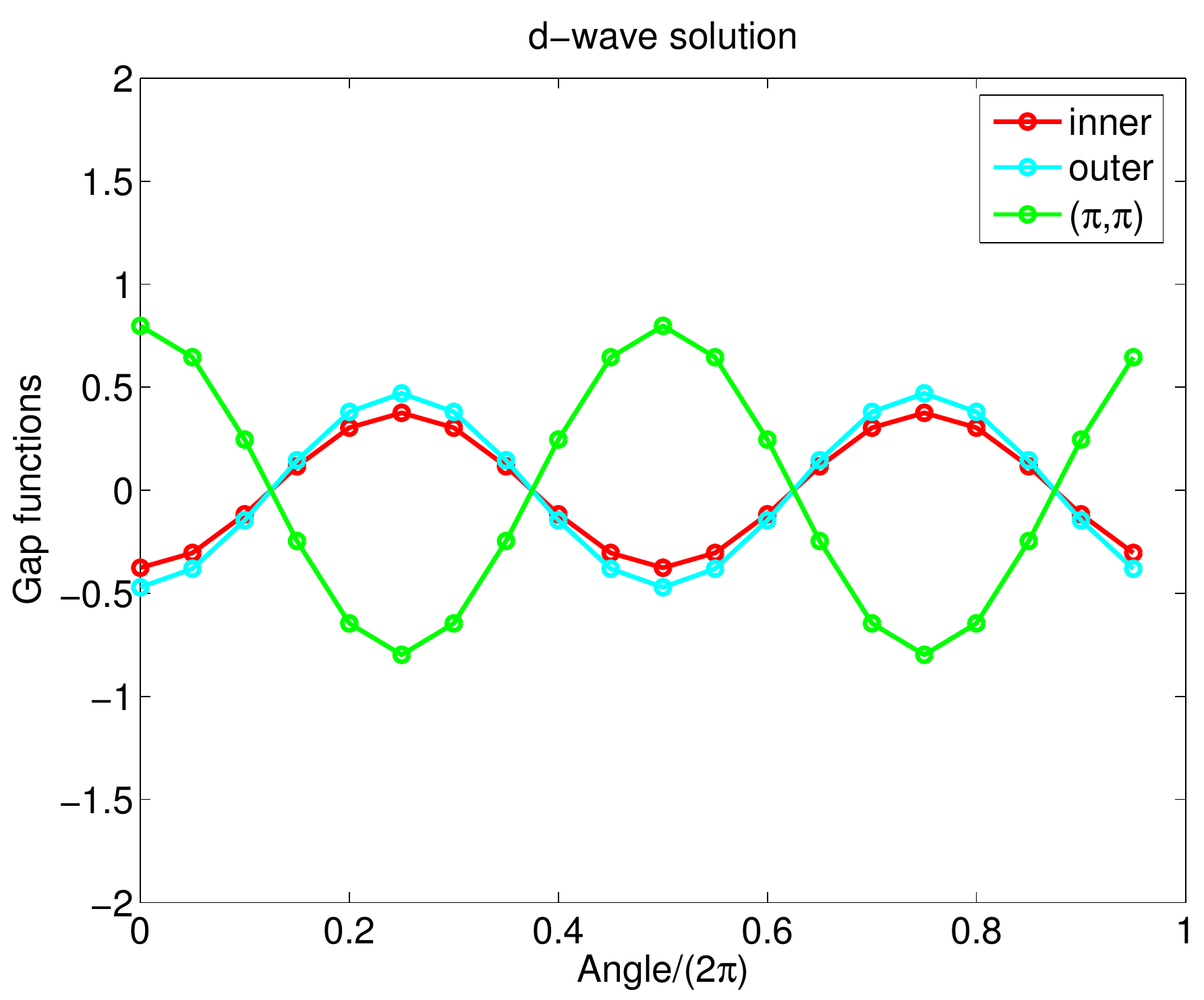}\\
\end{array}$
\caption{\label{fig:rev_5} The fits of the RPA
interactions by LAHA and the structure of $s-$wave and $d-$wave gaps in
 for strong hole doping ($\mu =-0.30eV$), when only hole
FSs are present.  From Ref.~[\onlinecite{maiti_11}].}
\end{figure}

\begin{table*}[htp]
\caption{\label{tab:2} Some of LAHA parameters  extracted from the fits in Figs.~\protect\ref{fig:rev_4} and \protect\ref{fig:rev_5} for hole doping. Block (i) corresponds to Fig. \protect\ref{fig:rev_4} (hole and electron pockets are present),  block (ii) corresponds to Fig. \protect\ref{fig:rev_5} ( no electron pockets).}
\begin{ruledtabular}
\begin{tabular}{lccccclcccccrccccc}
& \multicolumn{5}{c}{(i)}  & & \multicolumn{5}{c}{(iii)} \\
 \cline{2-6} \cline{8-12}
 $s-wave $&$u_{h_1h_1}$&$u_{h_1e}$&$\alpha_{h_1 e}$&$u_{ee}$&$\lambda_s$& & $u_{h_1h_1}$&$u_{h_1h_2}$&$u_{h_1
h_3}$&$u_{h_3 h_3}$&$\lambda_s$\\
&0.0.86&0.92&-0.18&1.00&0.58& &0.67&0.8&0.29&1.37&0.13\\
\cline{2-6} \cline{8-12}
$d-wave $&$\tilde{u}_{h_1h_1}$&$\tilde{u}_{h_1e}$&$\tilde{\alpha}_{h_1 e}$&$\tilde{u}_{e
e}$&$\lambda_d$& &$\tilde{u}_{h_1h_1}$&$\tilde{u}_{h_1 h_2}$&$\tilde{u}_{h_1 h_3}$&$\tilde{u}_{h_3 h_3}$&$\lambda_d$\\
&0.51&-0.45&-0.48&0.07&0.31& &0.36&-0.5&-0.02&-0.17&0.11\\
\end{tabular}
\end{ruledtabular}
\end{table*}

\subsubsection{Hole doping}

For small and moderate hole doping, the  FS
  contains 5 pockets --two hole pockets at $(0,0)$,
 two electron pockets at $(0,\pi)$ and $(\pi,0)$,
and one more hole pocket at $(\pi,\pi)$.
Representative FSs for hole doping,  typical fits by LAHA, the parameters extracted from the fit,
 and the solutions in s-wave and d-wave channels are shown in Fig. \ref{fig:rev_4} and in Table~\ref{tab:2}.
Just like for electron doping, there
 are universal and parameter-sensitive features.
The parameter-sensitive property is again the presence or absence of
accidental nodes in the $s$-wave gap along the electron FSs, although for
most of the parameters, the gap does not have nodes (see Fig.~\ref{fig:rev_4})
because the total $u_{he}$ increases once it acquires an
additional contribution $u_{h_3 e}$.

 There are two universal features. First,  the
$s$-wave eigenvalue is enhanced relative to a $d-$wave one and
 becomes  the leading instability  as long as both hole and electron pockets are present. Second,  the driving force for the
attraction in both $s$- and $d$- channels is again strong inter-pocket
electron-hole interaction ($u_{he}$ and ${\tilde u}_{he}$
terms), {\it no matter how small electron pockets are}.\\

{\it Extreme hole doping}

The situation again  changes rapidly once electron pockets disappear, see Fig.~\ref{fig:rev_5}.  Now electron-hole interaction becomes irrelevant, and the attractive pairing interaction may only be due to intra and inter-pocket interactions involving  hole pockets.
LAHA analysis shows [\onlinecite{maiti_11,maiti_kor}] that, at least for in some range of parameters,  there is an attraction in both
 $s-$wave and $d-$wave channels, and furthermore $\lambda_d \approx \lambda_s$, see Fig.~\ref{fig:rev_5}
The  near-equivalence of $s-$wave and $d-$wave eigenvalues was also found in recent unrestrictive RPA study~[\onlinecite{kuroki_11a}].
Within LAHA, the attractive
$\lambda_s$ is due to strong intra-pocket interaction between the two hole pockets centered at $(0,0)$.  The $s-$wave gap then changes sign
 between these two hole pockets.    The gap along $(\pi,\pi)$ pocket is induced  by a weaker inter-pocket interaction and is much smaller.
 LAHA neglects $\cos 4n \phi$  gap variations along the hole FSs (i.e., $s-$wave gaps are treated as angle-independent), but the theory can indeed be extended to include these terms.
The attractive $\lambda_d$  emerges by two reasons.
 First, the $d$-wave intra-pocket interaction ${\tilde
u}_{h_3 h_3}$ becomes negative, second, the
inter-pocket interaction ${\tilde u}_{h_1 h_2}$ between the two pockets at $(0,0)$ becomes larger in
magnitude than repulsive ${\tilde u}_{h_1 h_1}$ and ${\tilde
u}_{h_2 h_2}$ (see Table~\ref{tab:2}). The solutions with $\lambda_d>0$ then exist
separately for FSs $h_{1,2}$ and for $h_3$, the residual
inter-pocket interaction just sets the relative magnitudes and
phases between the (larger) gap at $h_3$ and (smaller) gaps at $h_{1,2}$.
The  $d$-wave gap with the same structure  has been obtain in  the fRG analysis at large hole doping~[\onlinecite{KFeAs_fRG}].

\subsection{LiFeAs}

\begin{figure}[htbp]
\includegraphics[width=0.45\textwidth]{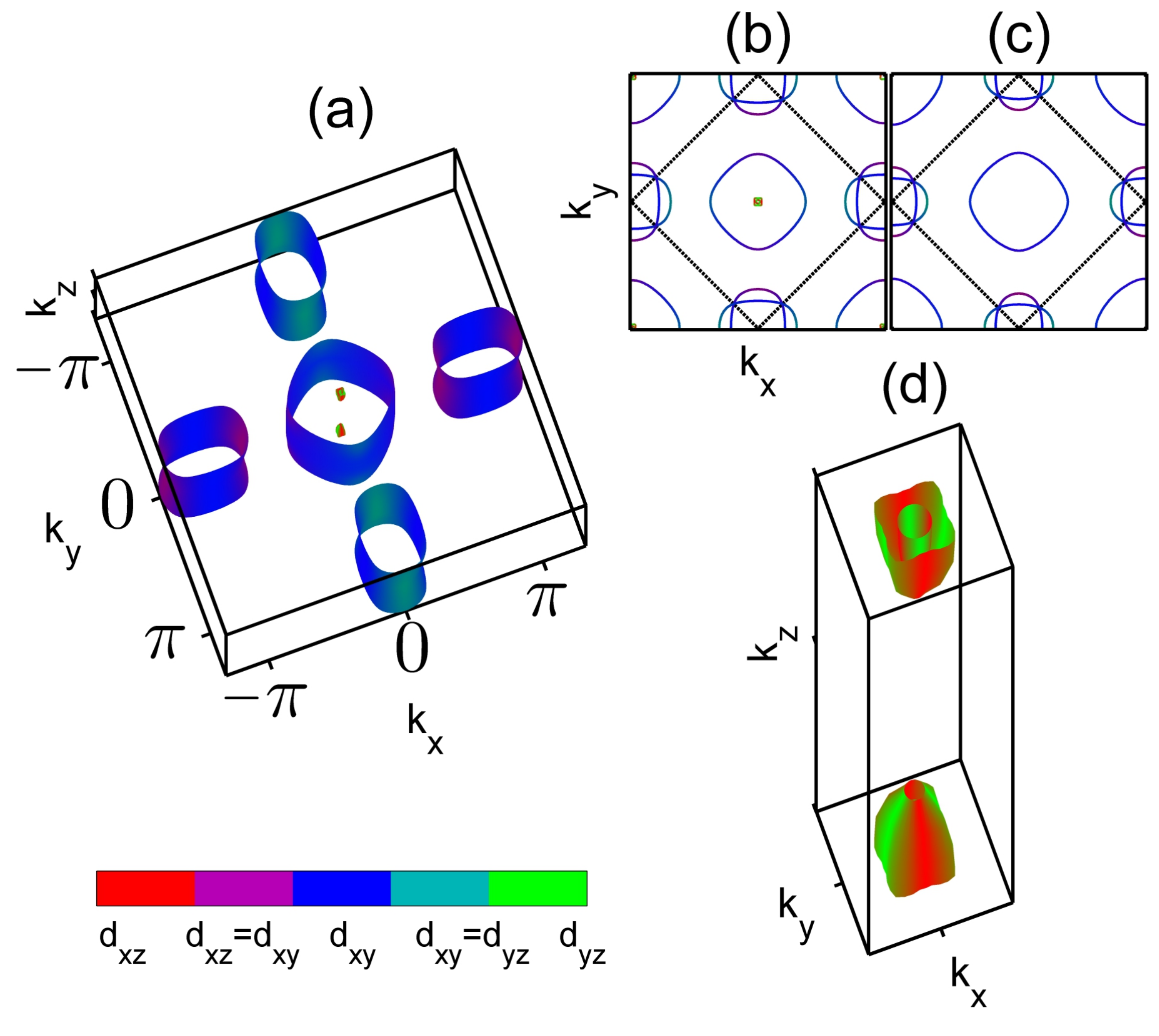}
    \caption{ Fermi surface of LiFeAs as deduced from the ARPES experiments: (a) shows the three-dimensional version of the Fermi surface and (b) and (c) refer to the two-dimensional cuts at $|k_z|=\pi$ (left) and $k_z=0$, respectively. Hole pockets are located a $(0,0)$ and $(\pi,\pi)$ and electron pockets are at $(\pm\pi,0)$ and $(0,\pm\pi)$. In case of $k_z=0$, the two tiny hole pockets $h_{1,2}$ vanish just below the FS and only $h_3$ and $e_{1,2}$ remain. (d) shows the zoomed region of the first BZ around the $\Gamma-$point of the BZ with tiny $\alpha$ hole pockets.}\label{fig:lifeas_fs}
\end{figure}

There is a possibility to obtain a more complex behavior even in systems which contain both hole and electron pockets.
One such example is LiFeAs.  Its electronic structure contains three hole and two electron pockets, however two $\Gamma-$centered hole pockets
 have strong 3D dispersion and  exist only near $k_z =\pi$ (see Fig.\ref{fig:lifeas_fs}
As a result the FS in the cross-sections at small $k_z$  consists of one hole and two electron pockets (hole $\gamma$ pocket and electron $\beta$ pockets),
 while in the cross-section at $k_z$ near $\pi$ the FS consists of three hole pockets and two electron pockets. The orbital content of the FSs for the two $\alpha$ FSs is very different from that for other three FSs.  Namely, the two $\Gamma-$centered hole pockets ($\alpha$ pockets) are made chiefly of $d_{xz}$ and $d_{yz}$ orbitals.  The other three FSs are made primarily of $d_{xy}$ orbital, with rather small admixture of $d_{xz}$ and $d_{yz}$ orbitals (Refs.\cite{peter_lifeas,ilya_lifeas}.
 These features indicate that the low-energy electronic structure of LiFeAs consists of two very different subsets. One is made out of quasi-2D $\gamma$ and $\beta$ pockets with primarily $d_{xy}$ orbital content, and the other is made out of $\alpha$ pockets, which are highly anisotropic along $k_z$ and are  made primarily out of $d_{xz}$ and $d_{yz}$ orbitals.

\begin{widetext}
\begin{center}
\begin{table}[hbtp]
\begin{center}
\tabcolsep=0.10cm
\begin{tabular}{cccccccccccccccc}
s-wave & $U_{h_1h_1}$ & $U_{h_2h_2}$ & $U_{h_3h_3}$ & $U_{h_1h_2}$ & $U_{h_1h_3}$ & $U_{h_2h_3}$ & $U_{h_1e}$ & $\alpha_{h_1e}$ & $U_{h_2e}$ & $\alpha_{h_2e}$ & $U_{h_3e}$ & $\alpha_{h_3e}$ & $U_{ee}$ & $\alpha_{ee}$ & $\beta_{ee}$ \\
\hline
$J=0.0U$&	 0.92	&	 0.99	&	 1.21	&	 0.95	&	 0.29	&	 0.23	&	 0.28	&	 -0.34	&	 0.22	&	 -0.49	&	 1.20	&	 -0.12	&	 1.20	 &	 -0.12	&	 0.03\\
$J=0.1U$&	 0.99	&	 1.09	&	 1.14	&	 1.03	&	 0.16	&	 0.10	&	 0.15	&	 -0.85	&	 0.08	&	 -1.65	&	 1.14	&	 -0.13	&	 1.14	 &	 -0.13	&	 0.04\\
$J=0.3U$&	 1.14	&	 1.28	&	 1.02	&	 1.20	&	 -0.09	&	 -0.15	&	 -0.12	&	 1.58	&	 -0.18	&	 1.10	&	 1.02	&	 -0.16	&	 1.03	 &	 -0.17	&	 0.05\\
\hline
\end{tabular}\caption{LAHA projected interactions in the $s-$wave channel for  \mbox{k$_z =\pi$}. The energies are in units of $U$.}
\end{center}
\begin{center}
\tabcolsep=0.10cm
\begin{tabular}{ccccccc}
s-wave & $U_{h_3h_3}$ & $U_{h_3e}$ & $\alpha_{h_3e}$ & $U_{ee}$ & $\alpha_{ee}$ & $\beta_{ee}$ \\
\hline
$J=0.0U$&	 1.53	&	 1.36	&	 -0.14	&	 1.22	&	 -0.13	&	 0.04\\
$J=0.1U$&	 1.48	&	 1.30	&	 -0.16	&	 1.16	&	 -0.14	&	 0.04\\
$J=0.3U$&	 1.40	&	 1.19	&	 -0.20	&	 1.04	&	 -0.18	&	 0.06\\
\hline
\end{tabular}\caption{LAHA projected interactions in the $s-$wave channel for \mbox{k$_z =0$}. The energies are in units of $U$.}
\end{center}
\label{table2_a}
\end{table}
\end{center}
\end{widetext}

\begin{figure}
\includegraphics[angle=0,width=\linewidth]{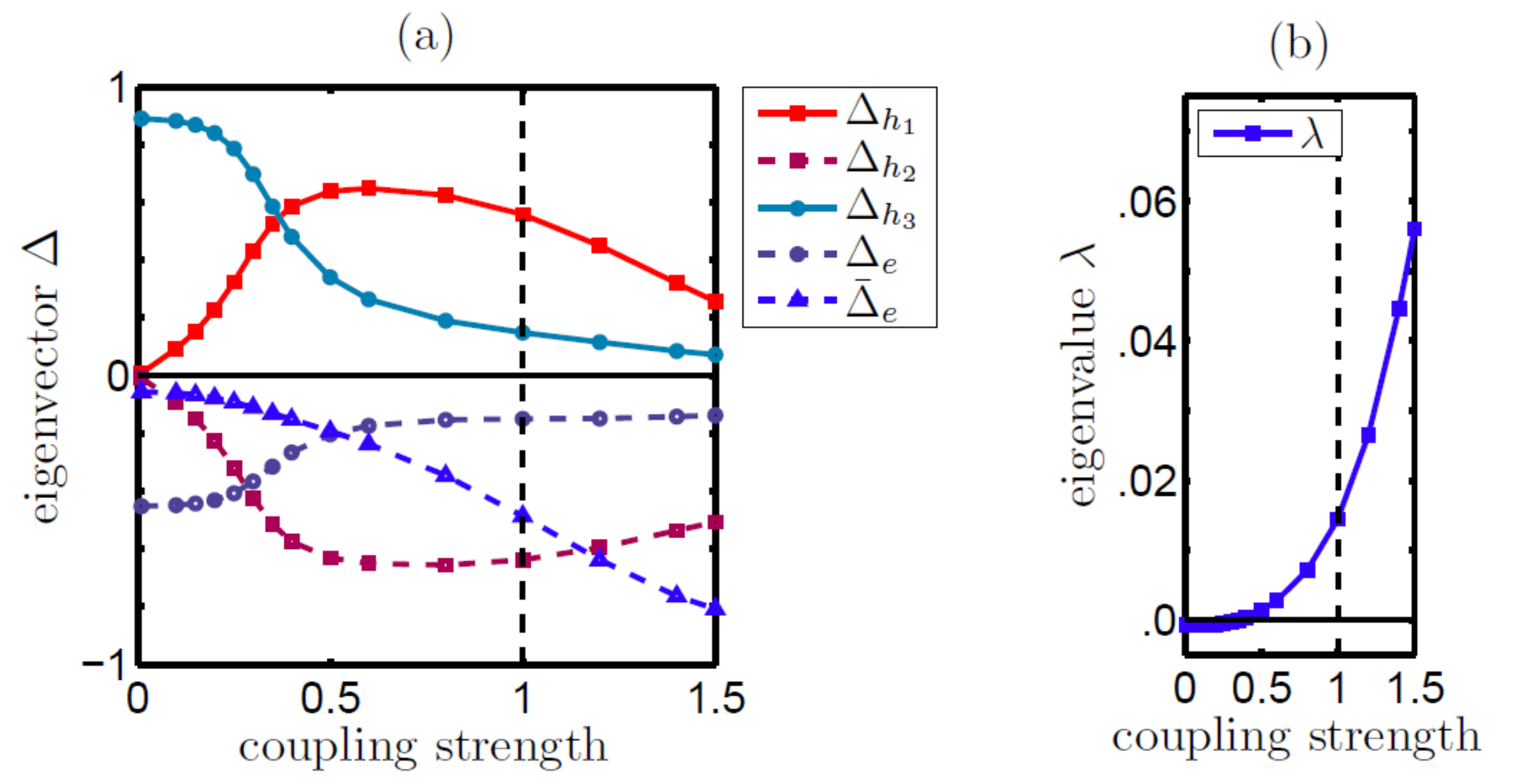}
\caption{Evolution of the gaps (a) and the largest eigenvalue (b) for the $s-$wave solution, with the
 coupling between the subset A (the two hole $\alpha$-pockets) and the subset B (the outer hole $\gamma$-pocket and the two electron $\beta$-pockets). From Ref. [\onlinecite{ilya_lifeas}]}
\label{fig:lifeas_sol}
\end{figure}

 The results~\cite{ilya_lifeas} for the interactions in $s-$wave channels within LAHA are shown in Tables IV and V.
 For the model described by Eq. (\ref{eq:multiorb_Hubbard_1}) with $U' = U-2J$ and $J'=J$, we clearly see
 that the two subsets are nearly separated for all $J/U$. This near-separation opens up a novel possibility for the structure of $s^{+-}$ gap. Namely, superconducting gaps consistent with the structure of interactions in Table IV
 are
\begin{align}
\begin{split}
\Delta_{h_1}(\phi)&=\Delta_{h_1}	\\
\Delta_{h_2}(\phi)&=\Delta_{h_2}	\\
\Delta_{h_3}(\phi)&=\Delta_{h_3}	\\
\Delta_{e_1}(\theta)&=\Delta_{e}+\bar{\Delta}_{e}\cos{2\theta}	\\
\Delta_{e_2}(\theta)&=\Delta_{e}-\bar{\Delta}_{e}\cos{2\theta}	\\
\end{split}\label{eq:swaveansatz}
\end{align}
 In a "conventional" $s^{+-}$  gap structure, the gaps on the three hole pockets are of the same sign.  Here, superconductivity within the subset
  of the two $\alpha$ pockets is primary due to inter-pocket repulsion between fermions near these pockets. When this repulsion exceeds inter-pocket repulsion, it gives rise to sign-changing $s^{+-}$ superconductivity between these two pockets.  In Fig. \ref{fig:lifeas_sol} I show the gap structure
   obtained for the parameters from Table IV
    and how it evolves when I artificially increase the interaction between $\alpha$ and $\beta$ pockets.
    When inter-subset interaction is strong, I obtain a conventional $s^{+-}$ superconductivity, with the same sign of the gap on all three hole pockets~\cite{peter_lifeas}. However, for small/moderate coupling between the $\alpha$ and $\beta-\gamma$ subsets, we see from Fig. \ref{fig:lifeas_sol}
     that the gaps on the two $\alpha$ pockets are of opposite sign.

 Another novel structure of an $s-$wave gap (termed as "orbital antiphase state") has been suggested in Ref. \cite{gabi_lifeas}. In this state, superconducting gap has the same sign on the two $\alpha$ pockets but changes sign between $\alpha$ pockets and $\gamma$ pocket. Such a state occurs if the coupling between the two subsets is strong and predominantly involves $\alpha-\gamma$ interaction.

\subsection{Superconductivity which breaks time-reversal symmetry}

 Several groups argued recently that multi-orbital character of FeSCs is an ideal playground to search for a truly novel spin-singlet superconductivity
  which breaks time-reversal symmetry (TRS).
   Spin-triplet superconductivity with broken time-reversal symmetry  ($p_x \pm i p_y$ state) has likely been found in  $Sr_2RuO_4$ \cite{Mackenzie},  which represents a solid-state analog of superfluid  ${}^3$He \cite{Volovikold,3he2}, but the \emph{spin-singlet} $d+id$ state has not yet been observed experimentally.
Such a state was once proposed as a candidate state for high $T_c$ cuprate superconductors \cite{laughlin}, but later gave way to
 a more-conventional TRS-preserving $d-$wave state.  A TRS breaking $d+id$ superconductivity has been recently predicted for  fermions on a hexagonal lattice (e.g., graphene) near van-Hove doping~\cite{rahul,ronny_gr}.

 For FeSCs, one proposal is to explore the region where $s-$wave and $d-$wave pairing channels are competitive in strength, and there is a transition from
   one pairing symmetry to the other, as one varies the parameters. In the intermediate regime, the system very likely falls into an intermediate $s+id$ state, with a broken TRS~\cite{khodas,ronny_review,s+id}.

\begin{figure*}[htp]
$
\begin{array}{cc}
\includegraphics[width=0.45\columnwidth]{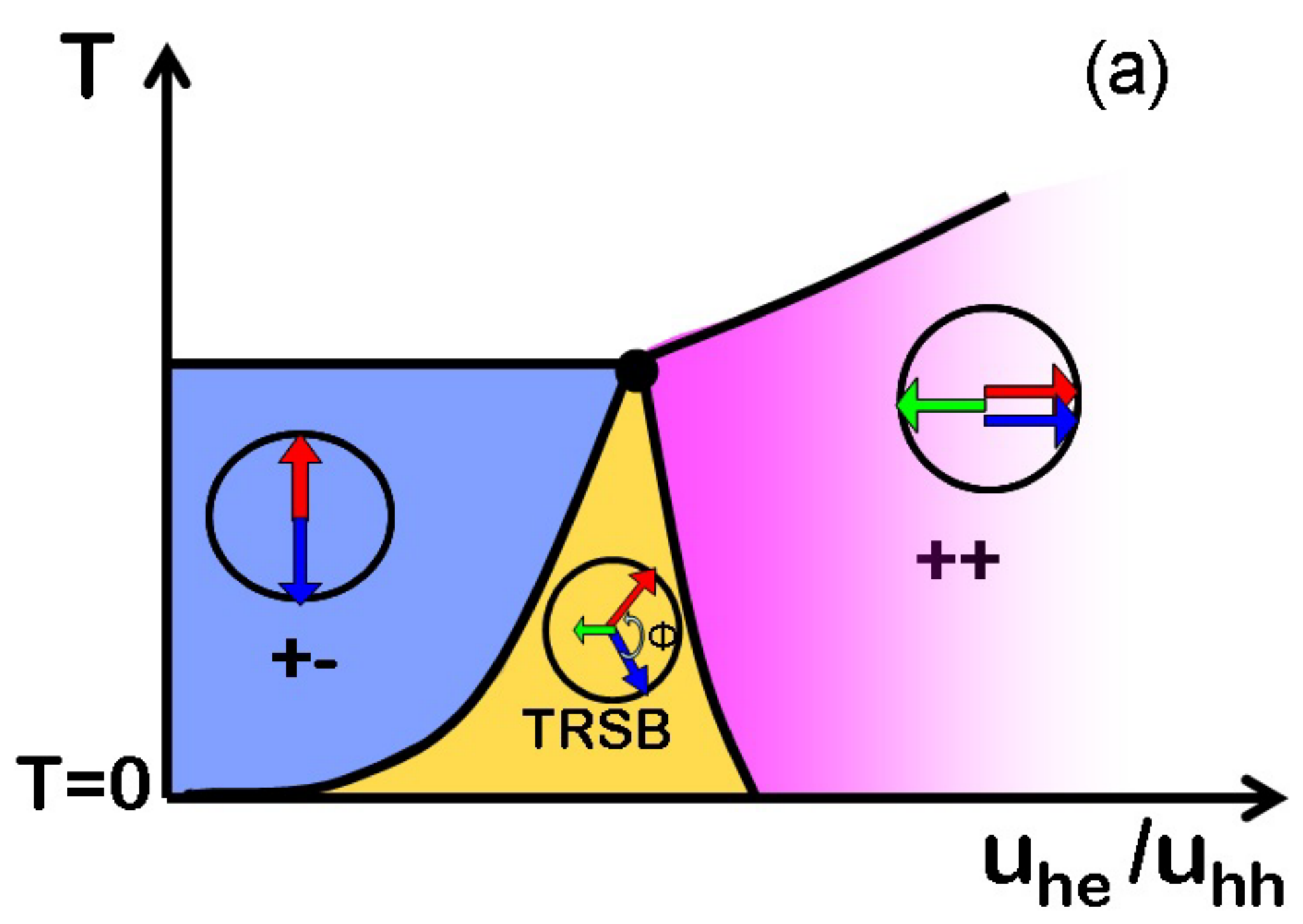}&
\includegraphics[width=0.45\columnwidth]{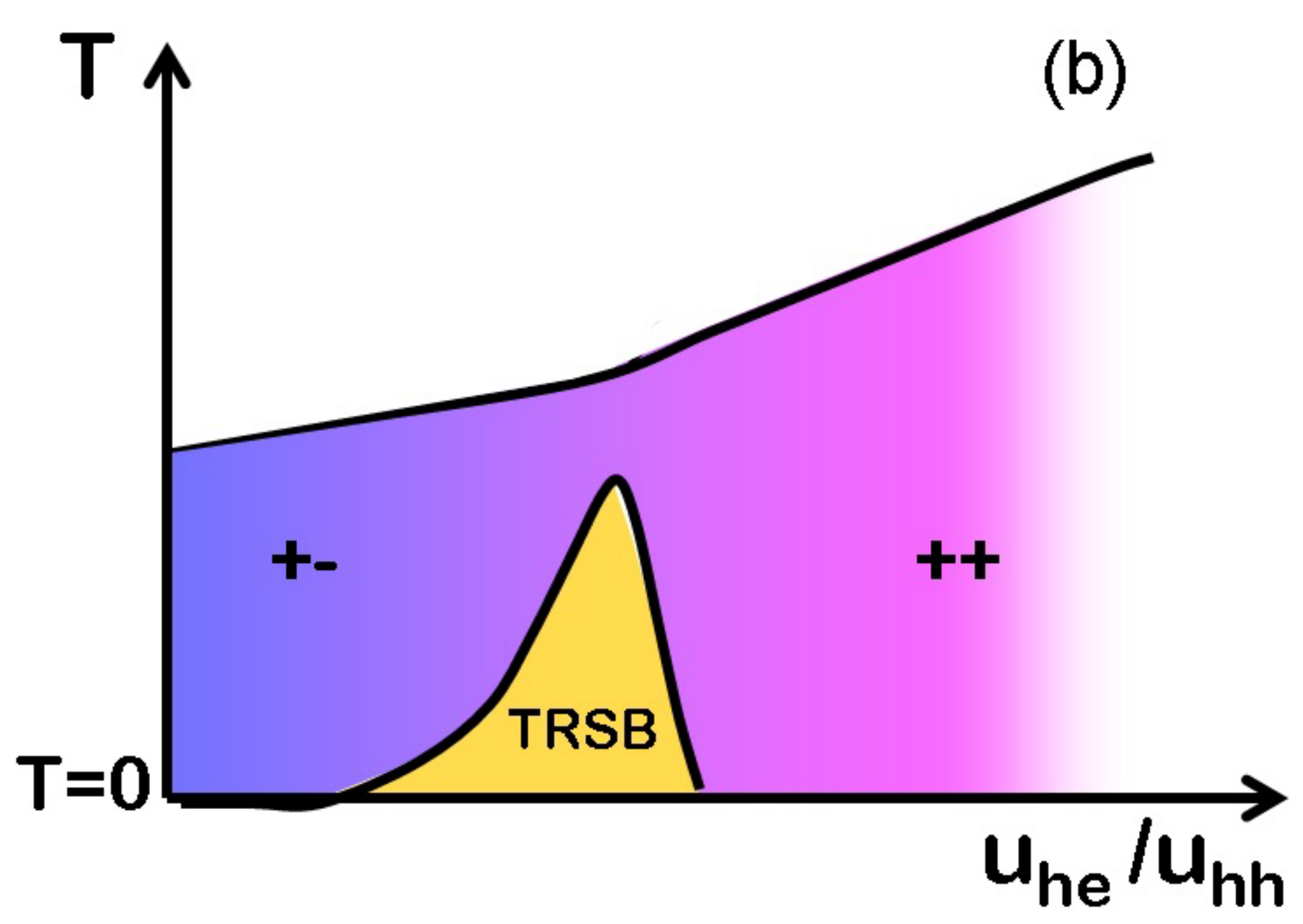}
\end{array}
$
\caption{\label{fig:trs} Qualitative phase diagram for
strongly hole-doped FeSC. I model the doping
dependence by varying the ratio of inter-pocket electron-hole and
hole-hole interactions $u_{he}/u_{hh}$. The $+-$ state has gaps of opposite signs on the two GCP's
and no gap on electron pockets, the $++$ state is an ordinary
$s\pm$ state, in which the gaps have opposite signs on hole and
electron pockets. The state with broken TRS is in between the two.
The gap structures are pictorially presented inside each region by vectors placed inside the circles.
 The magnitudes of the vectors represent $|\Delta_i|$ and the angles
represent the  phases.
Cases $(a)$ and $(b)$ are for equal and non-equal intra-pocket
interactions ($u_{h_1}$ and $u_{h_2}$) for the two hole pockets,
respectively . For ($a$), the state with broken TRS starts right at $T_c$ and
extends into a finite range at $T=0$. For ($b$), this state
splits off from the $T_c$ line and is only accessible at lower
temperatures, while immediately below $T_c$ the $+-$ state
gradually evolves into the $++$ state as $u_{he}/u_{hh}$
increases.}
\end{figure*}

   There is an even more exotic possibility to get a broken TRS state in an s-wave superconductor~\cite{zlatko_1,maiti_trs,lara,egor}.
 Consider as an example a system at extreme hole doping, like Ba$_{1-x}$K$_x$Fe$_2$As$_2$ at $x \approx 1$   and assume that the superconducting order is $s-$wave, with the sign change of the gap between the two $\Gamma-$centered hole pockets. Once hole doping gets smaller and electron pockets appear, the system eventually develops a "conventional" $s^{+-}$  superconductivity in which the gaps on the  two $\Gamma-$centered hole pockets have the same sign.  According to theory~\cite{maiti_trs}, the system evolution  with decreasing $x$  from a novel to a conventional $s^{+-}$ order may go through an intermediate state in which the relative phase $\phi$ between the gaps on the $\Gamma$-centered hole pockets gradually evolves from $\phi=\pm \pi$ in the novel s-wave state to $\phi =0$ in a conventional $s^{+-}$  state 
 (Fig. \ref{fig:trs}).  In between, the system selects  either $\phi$ or $-\phi$, which  are related by time-reversal transformation, i.e. it breaks time-reversal symmetry (an $s \pm i s$ state).  
 
 An intermediate state with broken time-reversal symmetry is also expected in Ba$_{1-x}$K$_x$Fe$_2$As$_2$ at $x \leq 1$ if the superconducting state in KFe$_2$As$_2$ is $d-$wave. In this situation, the system must transform from a $d-$wave at $x=1$ to an $s$-wave at a smaller $x$, and, like I just said, this normally involves an intermediate $s \pm id$ phase.  Another recent proposal for TRS broken superconducting state in FeSCs is $s+it$ superconductivity in the co-existence phase with SDW
  (see Ref. [\onlinecite{e_1}]). 

\section{Experimental situation on superconductivity in FeSCs}
\label{sec:7}

 As of today, there is no ``smoking gun'' experiment
which would carry the same weight as  phase-sensitive measurements of
$d_{x^2-y^2}$ gap symmetry in the cuprates~~[\onlinecite{d-wave}]. Still, there is enough experimental data to minimize the number of possible gap structures.

As we discussed in the previous section,  theoretically proposed  gap symmetry and structure  can be different for weakly/moderately doped systems with hole and electron FSs and for strongly doped  systems where FSs of only one type are present. It is then instructive to consider weak/moderate and strong doping separately.

\subsection{Moderate doping, gap symmetry}

The candidates are s-wave (either $s^\pm$ or $s^{++}$) or $d_{x^2-y^2}$ gap.
 The two behave very differently along the hole FSs centered at $(0,0)$ -- s-wave gap is nodeless with  $\cos 4 \phi$ variations, while d-wave gap has nodes along $k_x = \pm k_y$. ARPES measurements, both from synchrotron~~[\onlinecite{Evtush,122hole_ARPES,ARPES_1111,122electron_ARPES}] and using laser light~~[\onlinecite{laser_arpes}], show quite convincingly that the gap along hole FSs is nodeless in both hole and electron-doped FeSCs.  This unambiguously selects an s-wave.  Additional evidence in support of s-wave pairing comes from
 very flat low-T behavior of the penetration depth  in
 the highest $T_c$ 1111 FeSCs systems~~[\onlinecite{martin_09}].

\subsection{Moderate doping, $s^\pm$ vs $s^{++}$}

The distinction between  $s^\pm$ and $s^{++}$
 gaps is a more subtle issue, particularly given that both belong to the same $A_{1g}$ representation and also because in general $A_{1g}$
 gap on electron pockets may have strong oscillating component.
In general, the gaps on electron and hole FSs have non-equal magnitudes, and
 the issue whether the gap is $s^\pm$ or $s^{++}$ reduces to whether
 the gap averaged over an electron FS has the same sign
or opposite sign than the gap averaged over a hole FS.  This is not
 a fundamental  symmetry issue and, moreover, when $\cos 2 \phi$ oscillations are strong, one may switch from equal to opposite signs of the averaged gaps
 by a small change of parameters~~[\onlinecite{RG_SM}] or by adding impurities.~~[\onlinecite{efremov}] Still, when oscillations are not very strong, whether the eigenfunction has $s^\pm$ or $s^{++}$ character
  is essential because it determines, to a large extent, whether  the pairing is driven by spin  or by orbital fluctuations (see Sec.\ref{sec:4}).

The experimental data most frequently cited in support of $s^\pm$ gap is
 the observation of a magnetic  resonance in neutron scattering~~[\onlinecite{Cruz,resonance}].
If, as many researchers believe, the resonance is a spin exciton, it exists
 at a momentum $Q$ if the gaps at FS momenta $k_F$ and $k_F +Q$ are of opposite sign. Experimentally, in most FeSCs the resonance is observed~~[\onlinecite{Cruz,resonance}] near $Q = (\pi,\pi)$ in the
folded BZ, which in this zone is precisely the distance between electron and hole FSs. The excitonic resonance then exists if the gap changes sign between hole and electron pockets and does not exist if the gap doesn't change sign.  A similar reasoning has been used in identifying the
  the resonance seen in the cuprates with a fingerprint of
 $d_{x^2-y^2}$ gap symmetry~~[\onlinecite{eschrig}]

 The neutron peak is the resonance
 if it is narrow and is located below twice the gap value. The argument made by the supporters of $s^{++}$ scenario~~[\onlinecite{jap}] is that the observed neutron peak is more broad than the resonance seen in the cuprates, and that there is no firm evidence that the peak energy is below $2\Delta$ for the minimum gap.  For $s^{++}$ gap structure, there is no resonance, but there is a redistribution of the neutron spectral weight immediately above 2$\Delta$ what gives rise to a local maximum in the magnetic structure factor~~[\onlinecite{jap,kuroki_11,eremin_11}].
  Still, the majority of researchers do believe that the
 observed neutron peak is a resonance, and the fact that it is quite broad is at least partly due to $\cos 2 \phi$ gap variations along the electron FSs [\onlinecite{eremin_11}].

Another rather strong evidence in support of  $s^\pm$ gap
is the observed variation of the quasiparticle interference pattern in a magnetic field~~[\onlinecite{hanaguri}] although the interpretation of the data has been subject of debates~~[\onlinecite{debates}]. It  was also argued~~[\onlinecite{co-existence_Fe}] that the very presence of the co-existence region
between SC and stripe magnetism in FeSCs is a fingerprint of an $s^\pm$ gap, because for $s^{++}$ gap a first order transition between a pure magnetic and a pure SC state is a much more likely scenario.

\subsection{Moderate doping, nodal vs no-nodal $s^\pm$ gap}

Let's assume  that the pairing is driven by spin fluctuations and the
gap has $s^\pm$ structure. In 2D scenario, such
 gap has $\cos 2 \phi$ variations along electron FSs, which, according to theory, can be rather strong, particularly in electron-doped FeSCs.  Experimental data show that, whether or not the gap is nodeless or has nodes, depends on the material, on the doping, and on whether SC co-exists with SDW order.

\subsubsection {Hole doping}

 For hole-doped FeSCs (e.g. for Ba$_{1-x}$K$_x$Fe$_2$As$_2$) the data indicate
 that the gap is nodeless, away from the co-existence region.
  This is consistent with the theory (see Sec. \ref{sec:5}).
ARPES experiments do not show any angular variation of the gap along both hole and electron FSs~~[\onlinecite{Evtush,122hole_ARPES}], but it is not entirely clear whether ARPES can at present distinguish between the gaps  on the two electron FSs which in folded zone are both centered at $(\pi,\pi)$. Thermal conductivity data show that $\kappa/T$ tends to zero in the limit of $T=0$, in line with what one should expect for a nodeless SC~~[\onlinecite{reid_11}]. Specific heat data also show non-nodal behavior~~[\onlinecite{heat_hole}]. The interpretation of the penetration depth data requires more care as the data do show a power-law behavior $\lambda (T) - \lambda (0) \propto T^a$ with $a \sim 2$ (Refs. ~[\onlinecite{122hole_pen_depth}]).
 Such a behavior is expected for a SC with point nodes, but it is also
 expected in a wide range of $T$
 for a nodeless $s^\pm$ SC in the presence of modest inter-band scattering by non-magnetic impurities~~[\onlinecite{impurities}]. Penetration depth measurements on artificially irradiated samples~~[\onlinecite{kim_10}]  support the idea that the gap is nodeless and power-law
 $T^a$ behavior of   $\lambda (T) - \lambda (0)$ is due to impurities.

\subsubsection{Electron doping}

For electron-doped FeSCs, e.g., 122 materials like Ba(Fe$_{1-x}$Co$_x$)$_2$As$_2$
or 1111 materials like  NdFeAsO$_{1-x}$F$_x$,   ARPES shows no-nodal gap along hole FS~~[\onlinecite{ARPES_1111,122electron_ARPES}],
 but there are no data on the gap along each of
 the two electron FSs. At optimal doping, the data on both thermal conductivity~~[\onlinecite{thermal_opt,hashimoto}] and penetration depth~~[\onlinecite{hashimoto,pen_opt}] are consistent with no-nodal gap  However, the data for overdoped  Ba(Fe$_{1-x}$Co$_x$)$_2$As$_2$ indicate that gap nodes may develop: the behavior of $\lambda (T)$ becomes more steep, and $\kappa/T$ now tends to a finite value~~[\onlinecite{thermal_opt}], expected  for a SC with line nodes. The data also show  $\sqrt{H}$ behavior of $\kappa$ in a magnetic field~~[\onlinecite{thermal_opt}]
 expected for a SC with line nodes~~[\onlinecite{volovik}], but it was argued that the behavior resembling $\sqrt{H}$ can be obtained even if $s^\pm$ gap has no nodes~~[\onlinecite{bang}].
  There is also clear anisotropy between in-plane
conductivity and conductivity along $z$ direction, what was interpreted~~[\onlinecite{thermal_opt}] as an indication that the nodes may be located near particular $k_z$.
Specific heat data in  overdoped   Ba(Fe$_{0.9}$Co$_{0.1}$)$_{2}$As$_{2}$ were also interpreted as evidence for the nodes.~~[\onlinecite{heat_el_over}]

The development of the nodes in $s^\pm$ gap upon electron doping is in line with the theory. The farther the system moves away from the
SDW phase, the weaker is the increase of  intra-band electron-hole interaction and hence the stronger is the competition from intra-band repulsion. As I discussed in Sec.\ref{sec:3}), the gap adjusts to this change by increasing its $\cos 2 \phi$ component in order to effectively
reduce the effect of the intra-band repulsion in the gap equation.

There is also experimental evidence for $\cos 2 \phi$ gap oscillations from
 the observed oscillations~~[\onlinecite{wen_10}] of the field-induced component of the specific heat $C(H,T)$ in superconducting FeTe$_{1-x}$Se$_x$ ($x \sim 0.5$).
  The measured $C(H,T)$ oscillates with the direction of the applied field as $\cos 4 \phi$. In theory, such an oscillation is related to the behavior of $\Delta^2 (\phi)$ (Ref. ~[\onlinecite{vekhter_10}]), hence $\cos 2 \phi$ gap oscillations in $\Delta$
lead to $\cos 4 \phi$ oscillations in $C(H,T)$.
The observed field and temperature dependence of the prefactor for  $\cos 4 \phi$ term are consistent with the idea that the oscillations are caused by
$\cos 2 \phi$ term in $\Delta$.  These data were also interpreted as evidence for no-nodal gap because if $\cos 2 \phi$ gap oscillations were strong and the gap had nodes at accidental points, the behavior  of $\Delta^2$ would be
 more complex than the observed $a+b \cos 4 \phi$.

For  LiFeAs, which is undoped but  has FS structure similar to electron-doped FeSCs, no-nodal behavior has  been observed in ARPES~~[\onlinecite{LiFeAs_ARPES2}],
 specific heat~~[\onlinecite{LiFeAs_ARPES1_nutron_scat}], penetration depth~~[\onlinecite{thermal_pen_LiFeAs}] and NMR~~[\onlinecite{buechner,ma_10}] measurements. An $a + b \cos 4\phi$ variation of the gap on the hole $\gamma$ pocket, consistent with $s-$wave superconductivity, has been observed~\cite{sergei_b}, together with
  $c + d|\cos 2\phi|$ variations of the gap on electron pockets~\cite{ding_lifeas}. The latter is precisely what is expected theoretically for an s-wave superconductor when the hybridization between the two $\beta$ pockets is weak. The gap on the $\alpha$ pocket probably also has  angle dependence, but the pocket is too small to detect it in ARPES measurements.

\subsubsection{Co-existence region with SDW}

Taken at a face value, thermal conductivity and penetration depth data
 indicate that the gap becomes nodal deep in the co-existence regime in both
 hole-doped and in electron-doped FeSCs. The most striking evidence comes from thermal conductivity~~[\onlinecite{thermal_opt,reid_11}] -- in the co-existence regime $\kappa/T$ tends to a finite value at $T \to 0$ and shows $\sqrt{H}$ behavior, both typical for a SC with line nodes. From theory perspective, the gap 
  remains nodeless near the onset of co-existence with SDW if it was nodeless outside f the co-existence phase~\cite{parker},
   however, deep in the co-existence phase angular variation of the gap increase due to FS reconstruction (Ref. [\onlinecite{e_2}]). 

 \subsubsection{Isovalent doping}

Electron or hole doping is not the only way to change the properties of FeSCs. Another route is to replace one pnictide with the other. The most common replacement is As $\to$ P.  P-containing materials include the very first
FeSC -- LaFeOP, with $T_c \leq 5K$ (Ref. ~[\onlinecite{lafop}]), the family BaFe$_2$ As$_{1-x}$P$_{x}$ with the highest $T_c$ around 30K (Ref. ~[\onlinecite{matsuda}]), and  LiFeP~~[\onlinecite{private}].  Penetration depth, thermal conductivity, specific heat, and NMR
 data~~[\onlinecite{thermal_p}] in these materials all show the behavior consistent with line nodes. In particular, $\kappa$ scales linearly with $T$ at low $T$ and displays $\sqrt{H}$ behavior in a magnetic field, and $\lambda (T) - \lambda (0)$ is also linear in $T$ down to very low $T$.  Laser ARPES data
show~~[\onlinecite{laser_arpes}] that the gap along FS is nodeless, so the
 nodes likely are located on electron FSs.

On general grounds,  the existence of the nodes on electron FSs is in line with theory predictions particularly as BaFe$_2$ As$_{1-x}$P$_{x}$ has the same structure of 4 cylindrical FSs as electron-doped FeSCs for which nodes are most likely.
 It has been argued~~[\onlinecite{Kuroki}] that a replacement of As by P changes the hight of a pnictide with respect to Fe plane, what effectively reduces inter-pocket
electron-hole interaction, in which case the gap develops nodes to reduce
 the effect of intra-pocket repulsion.  However, this argument is only suggestive, and it is not entirely clear at the moment why all P-based FeSCs have
nodes. One way to analyze this semi-quantitatively is to study the correlation between $2\Delta/T_c$ on the hole FS and the presence of the nodes on electron FSs. This study shows~~[\onlinecite{maiti_delta}] that from this perspective P-based FeSCs are indeed the ``best case'' for the gap nodes.

Another open issue is the location of the nodes along z- direction.
 Oscillations of thermal conductivity
 with the direction of a magnetic field have been measured recently~~[\onlinecite{vekhter_11}], and
  $\cos 4 \phi$ component of these oscillations has
 been interpreted using the modified 2D form of the gap on an electron pocket $\Delta_e (k_z) = \Delta_0 (1 + \alpha (k_z) \cos 2 \phi)$. The best fit to the data yields $\alpha(k_z) >1$ for some $k_z$ and $\alpha (k_z) <1$ for others,  in which case the nodes form
 patches along $k_z$. This gap structure has been reproduced in microscopic calculations~\cite{khodas_3}, but whether this is the only explanation of the data
 is unclear.

It is  still possible, though, that the nodes are
 located on a hole FS, near particular $k_z$, as some of 3D theories suggest~~[\onlinecite{3D}]. Another possibility, which is also not entirely ruled out,
is that the system behavior near the surface, probed  by ARPES, is not the same as in the bulk. The
 probability that this is the case is not high, though, because some ARPES data have been obtained using a laser light which probes states located farther from the surface than in conventional synchrotron-based ARPES.

\subsection{Strongly doped FeSCs}

\subsubsection{Electron doping}

Strongly electron doped  materials are represented by a family of
 A$_x$Fe$_{2-y}$Se$_2$
($A = K, Rb, Cs$)~~[\onlinecite{exp:AFESE,exp:AFESE_ARPES}] of which  K$_0.8$Fe$_{1.7}$Se$_2$ is the most studied material. $T_c$ in  A$_x$Fe$_{2-y}$Se$_2$ is rather high, almost 40K.  ARPES shows~~[\onlinecite{exp:AFESE_ARPES}] that
 only electron FSs are present in  A$_x$Fe$_{2-y}$Se$_2$, while
hole pockets are at least $60 meV$ from the FS, although hole dispersion above $60 meV$ is still clearly visible in ARPES.
 Two electron FSs are at $(0,\pi)$ and $(\pi,0)$, like in other FeSCs,
 and there is, possibly, another electron FS at $(0,0)$.
RPA, LAHA and fRG calculations for these systems predict~\cite{Maier2011a,Das2011,Das2011,Wang2011c,maiti_11,dhl} that the gap should
 have a d-wave symmetry, at least for the case when the FSs are only at $(0,\pi)$ and $(\pi,0)$. A d-wave symmetry in this situation means that the gaps on the two electron FSs behave as $\Delta_0 (\pm 1 + \alpha \cos 2 \phi)$, and all calculations yield $\alpha <1$, i.e., no nodes (neglecting 3D effects). One theoretical  alternative is  $s^{++}$ symmetry by one reason~~[\onlinecite{kontani_se}] or the other~~[\onlinecite{yu_11,bernevig}], another is $s^{+-}$ state
   between electron pockets~\cite{mazin_a,khodas,khodas_2}.
   At present, both ARPES~\cite{exp:AFESE_ARPES} and specific heat data~\cite{Hardy_13}
   point that the gap is nodeless, at least for most of $k_z$ values.
 Of particular relevance here are ARPES data on a small electron pocket centered at $k_z =\pi$ and $k_x=k_y=0$. These data show~\cite{kfeas_pi} that the gap has no nodes, and, taken at a face value, rule out $d-$wave. At the same time, neutron data clearly show~\cite{AFESE_neu} spin resonance, which, if interpreted as spin exciton~\cite{eschrig},  requires a sign change of the gap. Both ARPRS and neutron data and recent Raman data~\cite{e_3}  
  are consistent with the novel $s^{+-}$ gap, but more studies are needed to verify whether this state is the right one for  A$_x$Fe$_{2-y}$Se$_2$

\subsubsection{Hole doping}

The case of extreme hole doping is represented by  KFe$_2$As$_2$ ($T_c =3K$), which is at the opposite end from parent  BaFe$_2$As$_2$ in the family of  K$_x$Ba$_{1-x}$Fe$_2$As$_2$.
  According to ARPES~~[\onlinecite{KFeAs_ARPES_QO}], this system has  no electron pockets. It contains hole pockets at $(0,0)$ and 
  additional hole pockets around $(\pi,\pi)$, but whether the latter play any role for superconductivity is not clear at the moment.

   Both thermal conductivity and penetration depth measurements clearly point to nodal behavior [\onlinecite{KFeAs_exp_nodal}].
  There is, however, no ``smoking gun''  symmetry-sensitive measurement, so whether the gap is a d-wave or an $s$-wave with nodes due to strong $\cos 4 \phi$ gap component on one of the FSs remains an open issue.  Recent study of $T_c$ under pressure has found~\cite{louis_last} that
  $T_c$ initially decreases with pressure initially, and then suddenly changes trend  above a critical
pressure $P_c$ and start increasing. This is a strong indication of the near-degeneracy between different pairing states in KFe$_2$As$_2$.
These states can be $s$ and $d$, or different $s-$wave states. Like I said before, some theorists suggested mixed states, like $s+id$ or $s+is$, which break TRS. Zero-field $\mu$SR measurements so far have not detected spontaneous internal magnetic fields, expected for $s+id$ state~\cite{sonier}.  Whether such fields are generated for $s+is$ state is less clear.
 This is an active field of research and I refrain from discussing it in more detail.

\subsubsection{FeTe$_{1-x}$Se$_{x}$}

There has been high interest recently in the properties of Fe-chalcogenide FeTe$_{1-x}$Se$_{x}$. The parent compound
 FeTe$_x$ is a SDW metal, but with high magnetic moment and magnetic order different from that in Fe-pnictides (see \cite{dukatman} for details).
  Superconductivity emerges around $x =0.5$, and magnetic fluctuations and FS-geometry at these $x$ do not differ substantially from Fe-pnictides,
   and it is reasonable to expect that magnetic fluctuations may mediate $s^{+-}$ supeconductivity. This, however, has not been studied in detail yet.
  On the other end, at $x=1$, the system behavior is quite unusual --  the structural transition sets in at around $80K$ (see \cite{dresden}), well
   before magnetic fluctuations develop.  Structural order increases as $T$ decreases, but changes trend below superconducting $T_c \sim 9$, what clearly shows that
    structural order competes with superconductivity. The existence of structural transition without strong magnetic fluctuations fueled speculations  that
     structural order in FeSe may reflect spontaneous orbital order, i.e., orbital fluctuations are "in the driver's seat". If this is the case, one should
      expect $s^{++}$ superconductivity in this material.  Experimental studies in the superconducting state of  FeTe$_{1-x}$Se$_{x}$ are called for to resolve this issue. There is also an interesting and highly unusual system behavior under pressure -- structural transition temperature goes down and magnetic fluctuations rapidly develop~\cite{pressure}.

   A highly unusual behavior has been detected in thin films of FeSe.  ARPES measurements detected only electron pockets~\cite{arpes_filmsfese}, like in K$_x$Fe$_{2-y}$Se$_2$. Like in   K$_x$Fe$_{2-y}$Se$_2$, $T_c$ is rather high, $T_c \sim 60K$ (even higher $T_c$ have been reported~\cite{tcspecul}).
   Whether FeSe (and, more generally, FeTe$_{1-x}$Se$_{x}$ for $x \geq 0.5$) falls into the same category as Fe-pnictides remains to be seen.

\subsection{Summary}

Overall, the agreement between itinerant theory and experiment with respect to the type of SDW order, the interplay between the nematic order and magnetism
 (e.g., the normal state phase diagram as a function of doping) and the symmetry and structure of the superconducting gap  is reasonably good.
With respect to superconductivity, itinerant approach  predicts that the gap is $s^{+-}$  in most systems, with angular variation, chiefly on electron pockets, and with accidental nodes in some systems. A more complex gap structure emerges in systems with strong hole or strong electron doping.

\section{Conclusion}
\label{sec:8}

he analysis of the normal state behavior and superconductivity in FeSCs is a fascinating subject
 because of multi-orbital/multi-band nature of these materials.
This review is an attempt to present a coherent picture of itinerant scenario for FeSCs.  I discussed the SDW magnetism,  which in most FeSCs corresponds to stripe order, the pre-emptive nematic phase, and the origin of superconductivity and the symmetry and structure of the superconducting gap in different classes of FeSCs and at different doping levels.  It is safe to say that there is no major disagreement between theory predictions within the itinerant scenario and the experiments.  This by no means implies that FeSCs fall into a class of "weakly coupled Fermi liquids". There are numerous indications that the coupling is strong and is not that much different from that in the cuprates, where, we know, Mott state does develop near half-filling. Nevertheless, most of parent compounds of FeSCs are metals, and, in my view, that low-energy physics of FeSCs is adequately captured within a moderate coupling itinerant scenario. Up to what temperature/energy one can extent the itinerant approach is another question.  This scale varies from material to material, but, still, is larger than
  the scales associated with superconducting $T_c$ and is likely larger that SDW transition temperature $T_{sdw}$ and the temperature $T_n$ associated with the development of a nematic order.  The physics at higher temperature/energies is outside the validity of the itinerant approach.

\section*{Acknowledgements}

I acknowledge helpful discussions with a large number of colleagues, including  E. Abrahams, L. Bascones, L. Benfatto, A. Bernevig, S. Borisenko, B. Buechner, S. Budko,
 P. Canfield,  A. Carrington, P. Coleman, A. Coldea,
 V. Cvetkovic, L. Digiorgi,  I. Eremin, L. Fanfarillo,
 R. Fernandes, S. Graser, H. Ding, W. Hanke, P. Hirschfeld, K. Honerkamp, D. Efremov, I. Eremin, J. Kang, A. Kemper, S. Kivelson, M. Khodas,
 J. Knolle, H. Kontani, G. Kotliar, M. Korshunov, K. Kuroki, D-H. Lee, T. Maier, S. Maiti, D. Maslov, Y. Matsuda, I. Mazin, A. Millis, K. Moller, M. Norman, 
  S. Pandey, R. Prozorov, J-Ph. Reid, D. Scalapino, T. Shibauchi, Q. Si, J. Schmalian, J. Sonier, V. Stanev,  L. Taillefer, H. Takagi, M. Tanatar, Z.~Tesanovic, R. Thomale, O. Vafek, M. Vavilov,  A. Vorontsov, and H.H. Wen.  This work was supported by the Office of Basic Energy Sciences U.S. Department
of Energy under the grant \#DE-FG02-ER46900.

\end{document}